\documentclass[final,3p]{elsarticle/elsarticle}

\usepackage{graphicx}
\usepackage{epsfig}
\usepackage{epstopdf}
\usepackage{subfig}

\usepackage{amssymb}
\usepackage{amsthm}
\usepackage{amsmath}
\usepackage{setspace}
\usepackage{lineno}
\usepackage{appendix}
\usepackage{hyperref}
\usepackage{verbatim}
\usepackage{filecontents}
\usepackage[all]{hypcap}

\journal{Nucl. Instrum. Meth. A}

\begin{document}

\onecolumn


\begin{frontmatter}



\title{The design and performance of a prototype water Cherenkov optical time-projection chamber}

\author[efi]{Eric Oberla}\corref{cor1}\ead{ejo@uchicago.edu}
\author[efi]{Henry J. Frisch}
\address[efi]{Enrico Fermi Institute, University of Chicago; 5640
  S. Ellis Ave., Chicago IL, 60637}

\begin{abstract}

A first experimental test of tracking relativistic charged
particles by `drifting' Cherenkov photons in a water-based optical
time-projection chamber (OTPC) has been performed at the Fermilab
Test Beam Facility. The prototype OTPC detector consists of a
77~cm long, 28~cm diameter, 40~kg cylindrical water mass
instrumented with a combination of commercial
$5.1\times5.1$~cm$^2$ micro-channel plate photo-multipliers
(MCP-PMT) and $6.7\times6.7$~cm$^2$ mirrors.  Five MCP-PMTs are
installed in two columns along the OTPC cylinder in a small-angle
stereo configuration. A mirror is mounted opposite each MCP-PMT on
the far side of the detector cylinder,  effectively doubling the
photo-detection efficiency and providing a time-resolved image of
the Cherenkov light on the opposing wall. Each MCP-PMT is coupled
to an anode readout consisting of thirty 50$\Omega$ microstrips. A
180-channel data acquisition system digitizes the MCP-PMT signals
on one end of the microstrips using the PSEC4 waveform
sampling-and-digitizing chip operating at a sampling rate of
10.24~Gigasamples-per-second. The single-ended microstrip readout
determines the time and position of a photon arrival at the face
of the MCP-PMT by recording both the direct signal and the pulse
reflected from the unterminated far end of the strip . The
detector was installed on the Fermilab MCenter secondary beam-line
behind a steel absorber where the primary flux is multi-GeV muons.
Approximately 80 Cherenkov photons are detected for a
through-going muon track in a total event duration of $\sim$2~ns.
By measuring the time-of-arrival and the position of individual
photons at the surface of the detector to $\le$100~ps and a
few~mm, respectively, we have measured a spatial resolution of
$\sim$ 15~mm for each MCP-PMT track segment, and, from linear fits
over the entire track length of $\sim40$~cm, an angular resolution
on the track direction of $\sim60$~mrad.
\end{abstract}


\end{frontmatter}



\section{Introduction}
\label{sec:intro}

We report the design, construction, and performance of a small
prototype `Optical Time Projection Chamber (OTPC) consisting of 40
kg of water viewed by five Micro-channel Plate Photomultipliers
(MCP-PMTs).  The principle of the OTPC is that with photodetectors
with adequate time and space resolution, measuring the time and
position of arrival of each individual photon emitted by Cherenkov
radiation from charged particles traversing a volume allows the
measurement of the photon drift times. Given the arrival positions
and the drift times, the track position and direction can be
reconstructed, in analogy with reconstructing tracks from the
electron drift times in a conventional gaseous or noble liquid
TPC.

Due to the faster drift velocity in a medium of photons compared
to that of electrons, the expected spatial resolution measured by
timing will be poorer. However, the scale for the required spatial
resolution for a large water neutrino detector, for example, is
set by the radiation length in water, which is approximately
40~cm~\cite{pdg}. A resolution of less than a few cm per optical
sensor may then be adequate to do event tracking for isolated or
separated tracks.

 The OTPC rests on three enabling technologies:
\begin{enumerate}
\setlength{\itemsep}{-0.03in} \item the use of MCP-based
photodetectors with a single photo-electron transit-time
resolution measured in 10's of picoseconds and a correlated
position measurement with a resolution measured in
mm~\cite{lappdtiming, milnes, ossystrip, mcplifetime}, able to
cover a large surface area economically. In a conventional large
water Cherenkov detector the size of the PMT precludes knowing the
position of arrival to within the size of the cathode.

\item The use of microstrip anodes~\cite{tang, lappdanode}, with
bandwidths matched to the photodetector pulse, to cover large
areas while preserving the temporal and spatial information
encoded by the incident photon. The microstrips provide spatial
resolution along the strip direction from the difference in pulse
arrival times at the two ends, and in the transverse direction by
charge sharing among strips, allowing mm resolution in two
dimensions while requiring only a one-dimensional array of
electronics readout channels~\cite{andrey}.

\item The development of an economical (in quantity) low-power
waveform sampling Application Specific Integrated Circuit (ASIC)
capable of multiple samples on the rising edge of the fast pulse
from an MCP-PMT~\cite{milnes, mcplifetime}, and intrinsic intra-chip jitter
measured in picoseconds. In this case we have developed the PSEC4 ASIC,
operated at 10.24 Gigasamples-per-second, with an effective dynamic range of 10.5
bits~\cite{psec4}.
\end{enumerate}

The prototype whose design and performance is described below
differs from a future large detector. In particular, a small path
length allows ignoring dispersion of the photon velocities and
scattering in the water. On the other hand, the much larger
LAPPD{\small $\texttrademark$} detectors have demonstrated a gain 10-times larger
than the commercial MCP-PMTs used here, with the potential for
better time resolution\cite{lappdcollab, andrey}.
In a larger detector, measuring
tracks over a longer length and with a larger photo-detection coverage would provide more
constraints on the 3D event topology (including the potential for an extended and optimized use of optical mirrors).

\subsection{A prototype OTPC}

\begin{figure}[]
\centering
\includegraphics[trim=.1cm .6cm .1cm .4cm, scale=.4, clip=true]{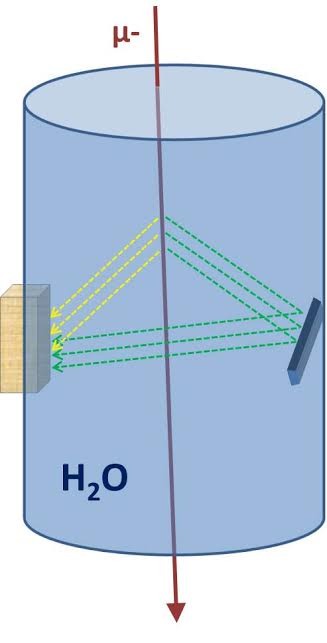}
\caption[The OTPC concept]{The concept of the prototype
water-based OTPC. A charged particle emits Cherenkov light in the
water volume. For each time and spatial-resolving MCP-PMT (left),
a mirror is mounted on the opposing side. The mirror adds a
discrete set of reflected photons (green) that arrive at the
MCP-PMT after a delay due to the extra path length, and can be
time-resolved from the direct photons (yellow). } \label{fig:otpc}
\end{figure}

To demonstrate this technique, we have built a 40~kg water
Cherenkov detector using a combination of commercial MCP-PMTs and
optical mirrors. A concept drawing of the optical setup of one of
the five MCP-PMT's of  this detector is shown in
Figure~\ref{fig:otpc}. For each planar MCP-PMT mounted on the
water volume, a mirror is mounted on the opposing side creating an
image of the Cherenkov light hitting the opposing wall. For a
given charged particle track, the mirror adds a discrete set of
reflected photons that can be time-resolved from the direct
photons, thus economically doubling the photo-detection efficiency
over the angular acceptance of the detector.

As the LAPPD MCP-PMTs are not yet available for use outside of a
test setup~\cite{andrey, rsimatt}, this prototype relies on
commercial MCP-PMTs that exhibit approximately the same detection
resolutions, but with 1/16 the photo-active area per unit. The
commercial photo-detector used is the Planacon XP85022 MCP-PMT
device from PHOTONIS~\cite{photonis}. The total photo-cathode
coverage in the prototype OTPC is 125~cm$^2$ on the surface area
of 1700~cm$^2$ that encloses the fiducial volume.

\subsection{Organization of the Paper}
\label{organization}

The OTPC prototype detector is described in
$\S$\ref{sec:detector}, including the design of the optics in
$\S$\ref{sec:Detoptics} and details on the photodetector modules
(PM) in $\S$\ref{sec:pmodule}. The readout system, consisting of
the fast waveform sampling digitizers of the front-end and the
data-acquisition system (DAQ), is described in
$\S$\ref{sec:dataacq}. Section~\ref{sec:exp} describes the
experimental setup in the Fermilab MCenter Test Beam area, and the
trigger used in the data-taking follows in
$\S$\ref{sec:exttrigger}. The reduction of the raw data to times
and positions is presented in $\S$\ref{sec:reduce}.
Section~\ref{sec:lasertests} describes the response of the
detector to single photons. The OTPC photo-detection efficiency
and gain are derived in Sections~\ref{sec:PDeff}
and~\ref{subsec:gain}. The test-beam results are shown in the
final four sections, including the number of photo-electrons along
the track~($\S$\ref{sec:charge}), resolving the direct and
mirror-reflected Cherenkov photons and the time-projection along
the beam axis~($\S$\ref{sec:timing}), the angular
resolution~($\S$\ref{sec:angle}), and the spatial resolution of
the reconstructed tracks~($\S$\ref{sec:3d}). The conclusions are
given in $\S$\ref{conclude}.

\section{Detector}
\label{sec:detector}

\begin{figure}
\centering
\includegraphics[trim=9cm 5.5cm 3cm 4cm, scale=.35, clip=true]{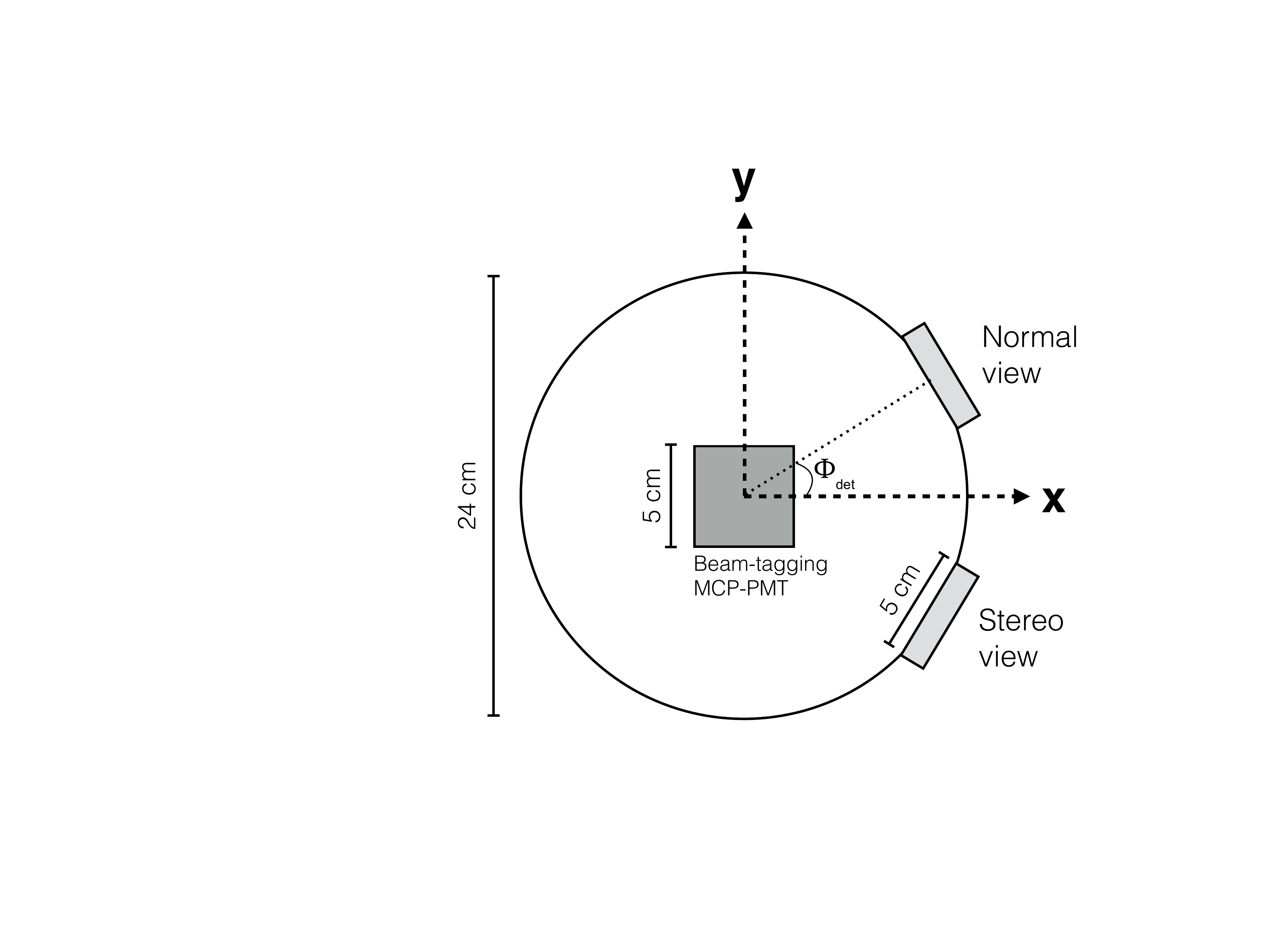}
\caption[Rear-view scale drawing of the OTPC detector]{Rear-view
scale drawing of OTPC detector showing the normal and stereo
MCP-PMT mounting positions. The normal and stereo mounted PMs are
symmetrical about the x-axis in the OTPC coordinate system, and
separated by an azimuthal angle of $\Phi$$_{det} = 32.5^{\circ}$.
The beam comes out of the page in the +z direction}
\label{fig:otpcdraw}
\end{figure}

The prototype OTPC detector is constructed from a 24~cm
inner-diameter Poly-Vinyl Chloride (PVC) cylindrical pipe cut to a
length of 77~cm~\cite{porterpipe}. Six 11~cm diameter ports were
machined in the tube arranged in 2 columns of three ports each
along the cylindrical axis, as shown in Figure~\ref{fig:otpcdraw}.
These columns have an azimuthal separation of 65$^{\circ}$. The
photodetector modules (PMs), described in $\S$\ref{sec:pmodule},
are mounted on five of these ports (the 6th port was occupied by a
conventional PMT used as a diagnostic trigger). The column with 3
PMs installed is denoted the `normal' view and the other the
`stereo' view.

\begin{figure}[]
\centering
\includegraphics[trim=8cm 9cm 3cm 1.2cm, scale=.45, clip=true]{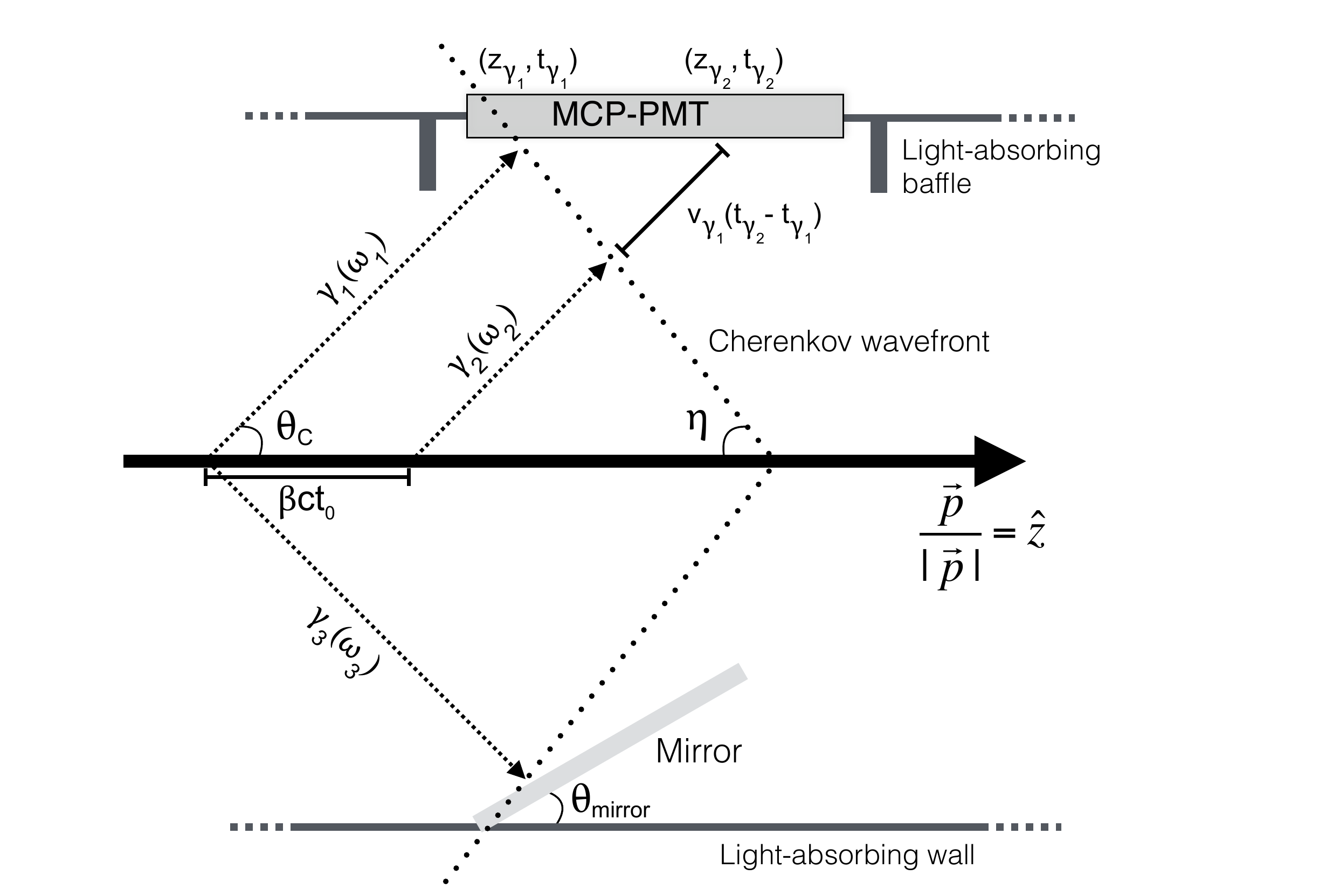}
\caption[Cherenkov radiation and the OTPC optics]{Two-dimensional
projection of the Cherenkov light generated by a relativistic
charged particle (heavy, black arrow) and the OTPC optics. The
drawing shows the optical paths of three photons: $\gamma$$_1$ and
$\gamma$$_2$ are directly detected, $\gamma$$_3$ is reflected at
the mirror. The mirror is mounted at an angle,
$\theta$$_{mirror}$, of 48$^{\circ}$.} \label{fig:otpcray}
\end{figure}

For each PM, a first-surface broadband optical mirror is mounted
on the opposing wall facing the PM port. The 7.6~cm square mirrors
are installed on the interior wall of the cylinder at an angle of
48$^{\circ}$, denoted as $\theta$$_{mirror}$ in
Figure~\ref{fig:otpcray}. The remaining exposed PVC surfaces of
the interior wall are coated with light absorbing paint~\cite{opticsdata}.

The detector is filled with 40~L of deionized water for the target
volume. Because of the detector angular acceptance, the Cherenkov
light directionality, and since only the first five the ports are
instrumented, the effective fiducial volume is limited to roughly
the top two-thirds of the total volume, corresponding to a
fiducial mass of $\sim25$~kg.

\subsection{Water quality}
\begin{figure}[]
\centering
\includegraphics[trim=0cm 0cm 0cm 0cm, scale=.45, clip=true]{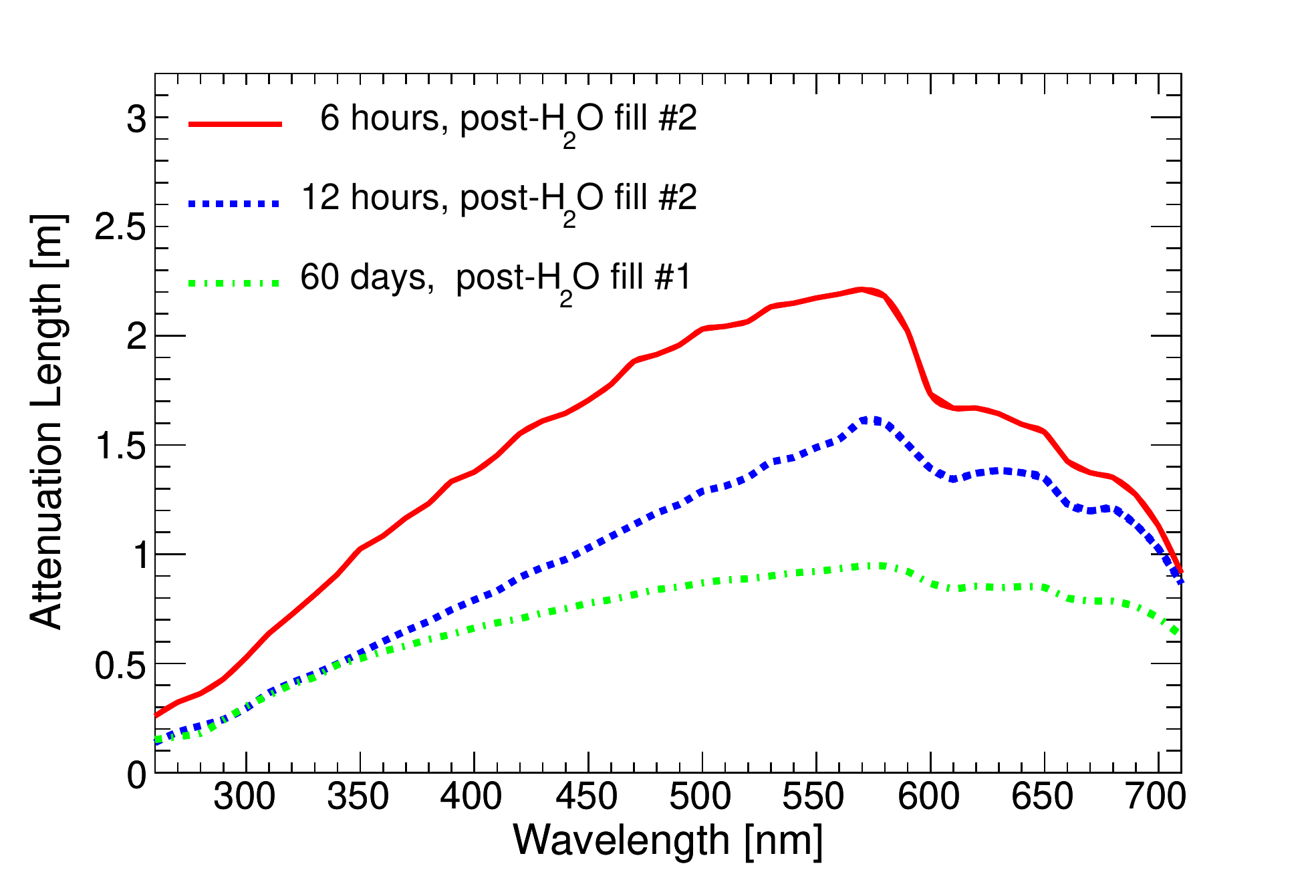}
\caption[Optical attenuation length in water]{
Calculated attenuation length of the OTPC water samples over a wavelength range of 250 to 720 nm using data obtained from dual-beam spectrophotometer measurements.
}
\label{fig:waterattn}
\end{figure}

The optical quality of the  water  in the OTPC was measured to be
notably poor, although it was not a limitation for the short
photon path lengths in this detector. We found the water
attenuation length by measuring 1-cm-deep samples with a dual-beam
spectrophotometer~\cite{spectro} at various post-fill intervals,
as shown in Figure~\ref{fig:waterattn}. Compared to a pure water
standard, the absolute absorbance at over the 300-500~nm range is
50 to 100 times worse. This can be attributed to several factors.
One, there is no filtration system to remove particulates in the
volume. Second, in the 12~hr and 60~day measurements a noticeable
degradation in the near-UV part of the spectrum was observed. This
is likely due to UV stabilizers in the PVC enclosure leaching into
the water~\cite{dazeley}.
Data taken within the first 6~hours of a water fill have a $1/e$
attenuation length \textgreater ~0.5~m at~300~nm. The longest path
lengths in the OTPC, taken by the mirror-reflected photons, are
approximately 0.3~m.

\subsection{Detector Optics}
\label{sec:Detoptics}
\begin{figure}[]
\centering
\includegraphics[trim=.6cm 0cm 0cm 0cm, scale=.4, clip=true]{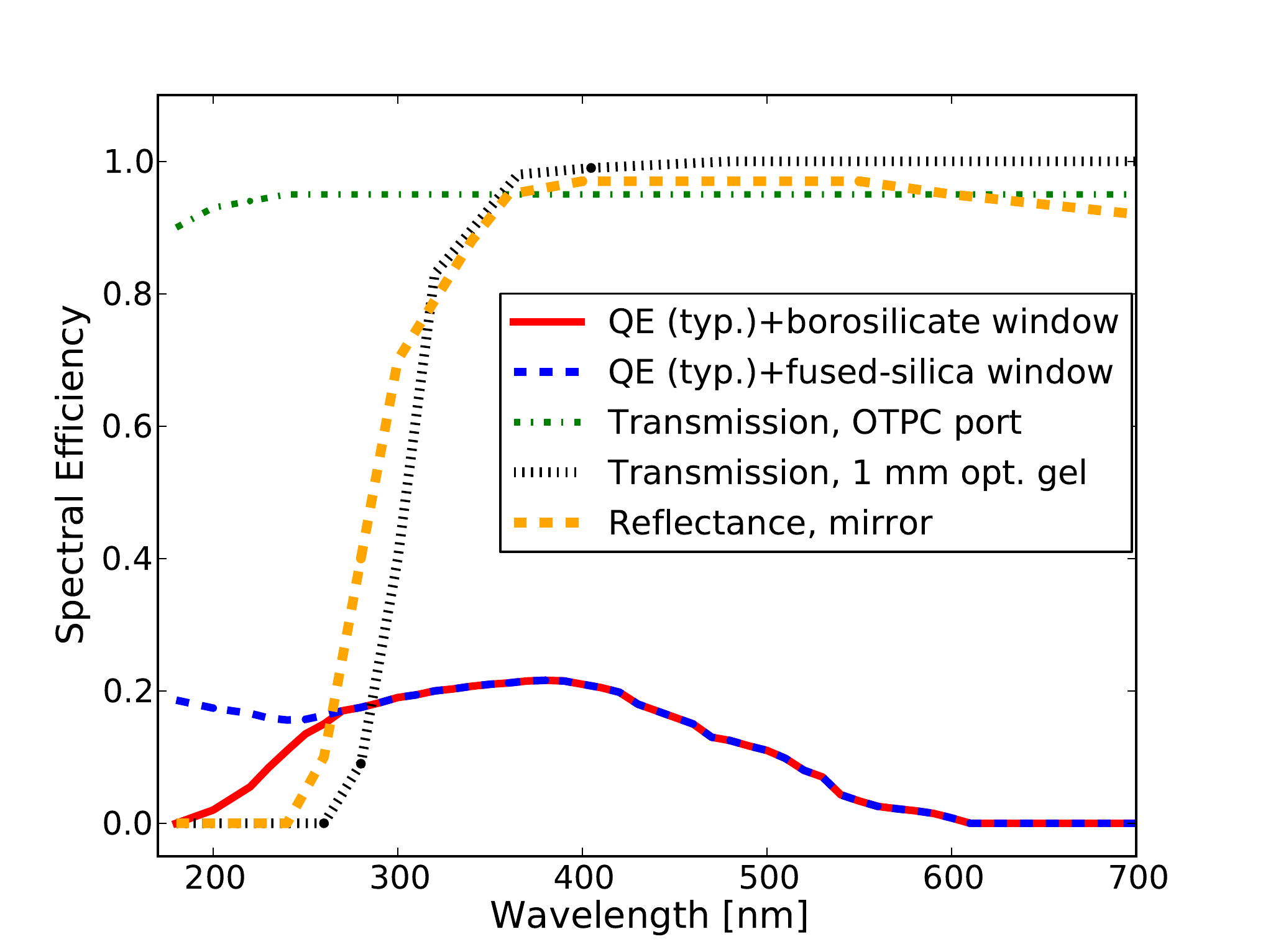}
\caption[Spectral efficiency of detector components]{ Optical
properties of the OTPC optical components over wavelength range of
200 to 700~nm. These parameterizations are used to model the
optical properties in a simulation of the detector using the
Chroma toolkit~\cite{chroma2011}. } \label{fig:spectrals}
\end{figure}
The optical properties of the detector components in the light
path, as quoted by the manufacturers~\cite{photonis, opticsdata},
are shown versus wavelength in Fig.~\ref{fig:spectrals}. Despite
efforts to capture as much near-UV Cherenkov light as possible by
using fused-silica windows, we found that these wavelengths are
cut off by the optical coupling gel between the window of the port
and the MCP-PMT. The characteristics of the light-absorbing paint
were modeled with a broadband 95\% absorbance and a 5\% diffuse
reflectance.

\begin{figure}[]
\centering
\subfloat[]{\includegraphics[height=6.1cm]{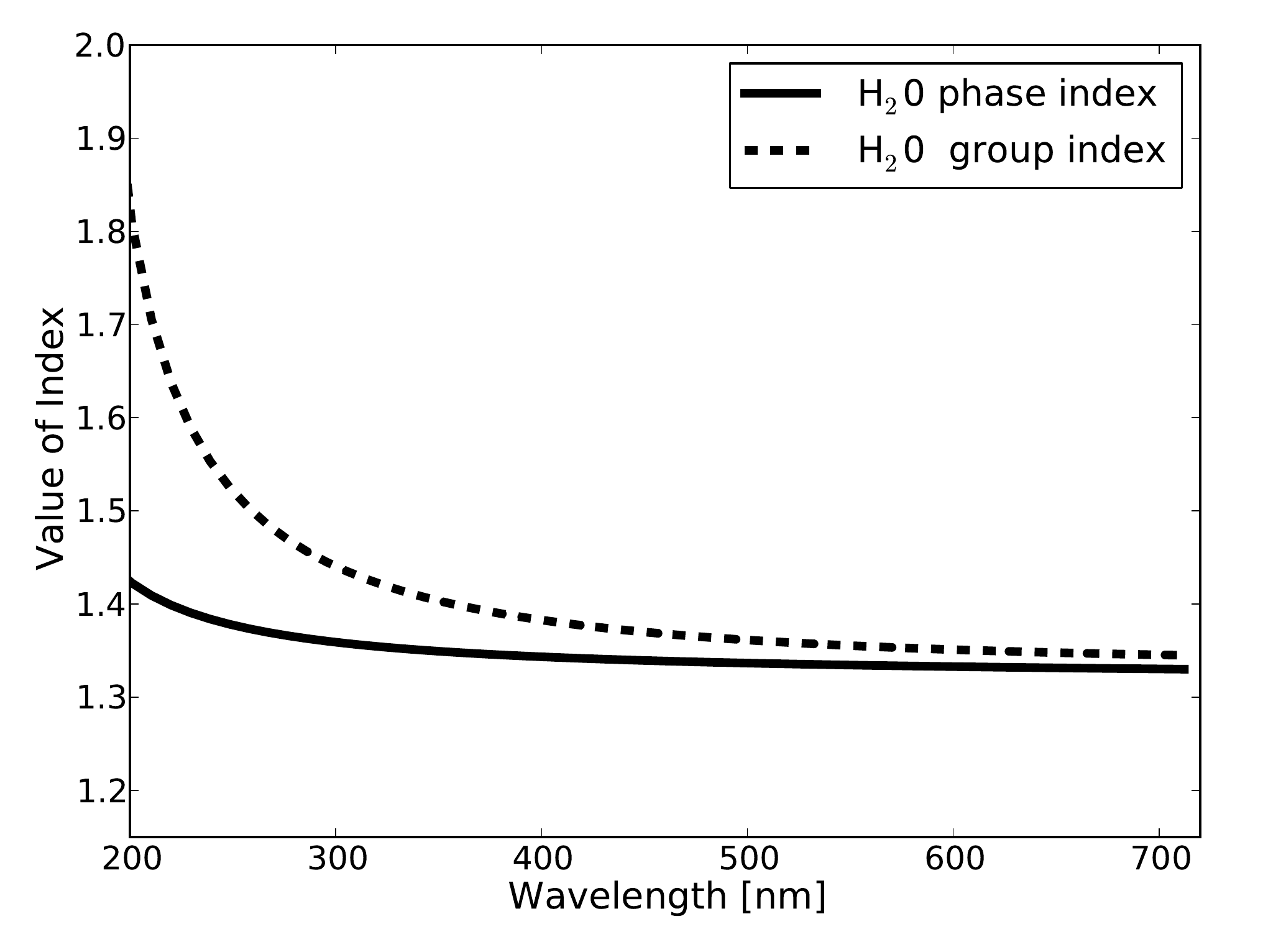}}
\subfloat[]{\includegraphics[height=6.3cm]{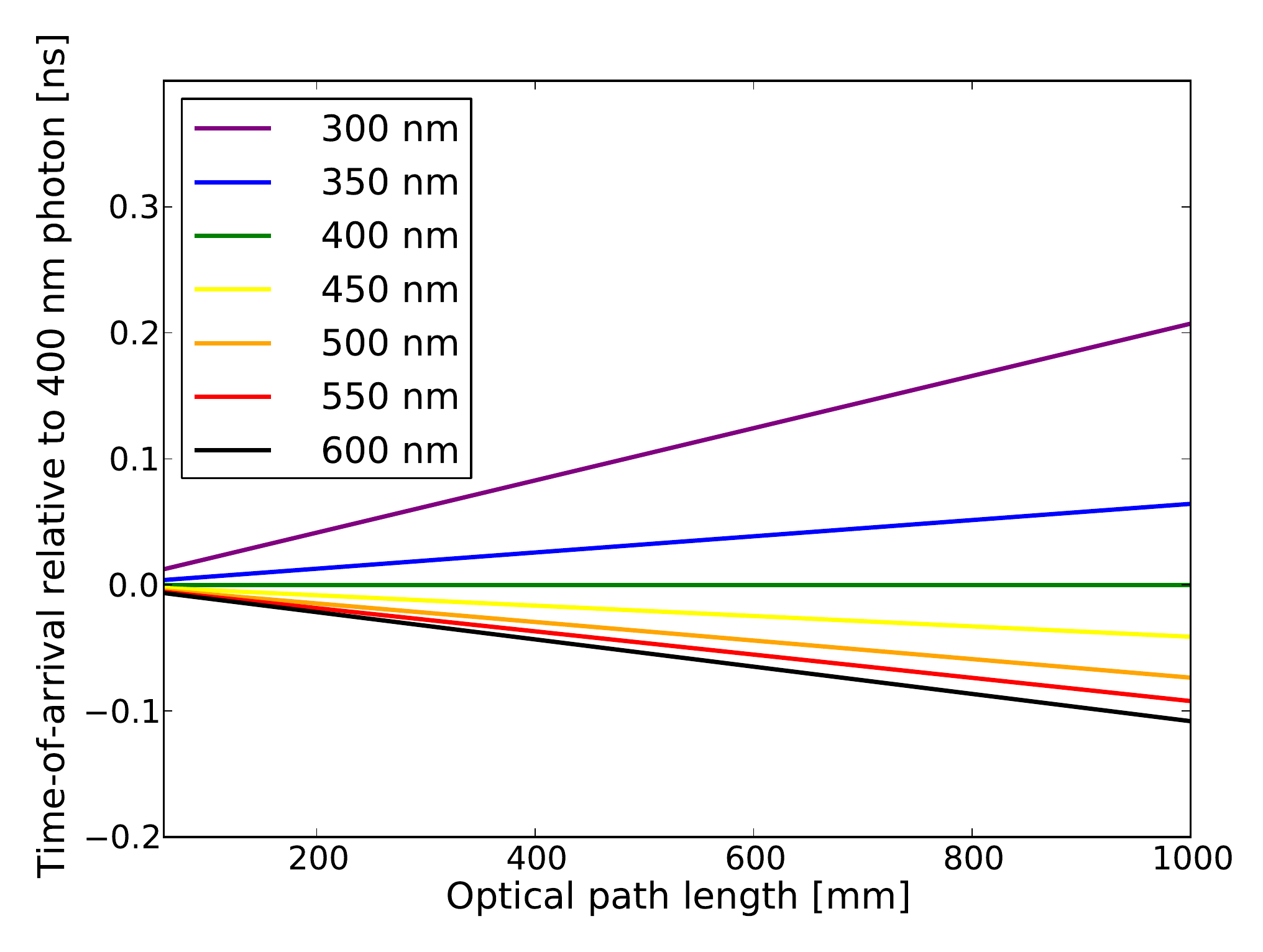}}

\caption[Phase and group indices of refraction]{
(a) Phase and group indices of refraction in water over the optical wavelength range.
The water phase-index of refraction data were taken from~\cite{waterindex}.
(b) Timing errors due to chromatic dispersion of the group velocity, referenced to a 400~nm photon, versus optical path length.
}
\label{fig:indices}
\end{figure}

With low-timing jitter and high granularity MCP-PMTs, the water
chromaticity and scattering properties become important as the
distance from the particle track to the photo-detector increases.
The Cherenkov photons, which are generated at an angle specified
by the phase velocity of the dielectric medium, propagate at the
speed at which energy is transported in the dielectric: the group
velocity. As shown by Jackson~\cite{jackson}, the group velocity
is expressed as
\begin{equation}
v_g(\omega) = \frac{c}{n_g(\omega)} = \frac{c}{n(\omega)\:+\:\omega\:(dn/d\omega)}
\end{equation}
where $n(\omega)$ and $n_g(\omega)$ are the phase and group index of the medium, respectively.
The group index is both greater in magnitude and more dispersive than $n(\omega)$,
as shown in Figure~\ref{fig:indices}a.

The spectrum of detected photons is the convolution of the OTPC
spectral response and the Cherenkov spectrum~\cite{franktamm}. A
full detector simulation, which includes the OTPC geometry, water
volume, and optics, was built using the Chroma
toolkit~\cite{chroma2011}. A simulation of $10^4$ muons with a
uniform distribution of energies between 1 and 10 GeV gives a
spectrum of detected photons of 370$^{+110}_{-50}$~nm, with the
spread covering about 95$\%$ of the spectrum around the mean
wavelength. Over this wavelength range, \begin{math} \left< n/n_g
\right> \approx 0.97 \end{math}.

The relative timing errors due to chromatic dispersion of the group velocity are shown in Figure~\ref{fig:indices}b.
At the maximum photon drift-length in the prototype OTPC ($\approx$30~cm), the expected timing error is about $\pm$25~ps,
which is small compared to the measured single photon timing resolution of 75~ps shown in $\S$\ref{sec:lasertests}.
Averaging over the dispersion effects,
the effective photon drift-speed in the OTPC is given by the weighted average of the group
velocity over the detected photon spectrum, $<v_{group}> =  218 \: \:  {\rm mm \: ns^{-1}}$.

\subsection{Track reconstruction equations}
\label{subsec:trackeqn} Consider a particle traveling through the
OTPC as shown in Figure~\ref{fig:otpcray}, but further
generalizing the track by allowing a polar angle, $\theta_i$, with
respect to the z-axis (see the caption of
Figure~\ref{fig:otpcdraw}).
Two Cherenkov photons generated along the particle path at $t=0$ and $t=t_{0}$ are detected with a relative time and longitudinal position of $\Delta t_{\gamma_{21}} = t_{\gamma_2}-t_{\gamma_1}$ and $\Delta z_{\gamma_{21}} = z_{\gamma_2}-z_{\gamma_1}$.  The relative timing is given by 
\begin{equation}
\Delta t_{\gamma_{21}} = t_o + (\frac{L_{\gamma_2}}{v_{\gamma_2}} - \frac{L_{\gamma_1}}{v_{\gamma_1}}) = t_o + \frac{\Delta L_{\gamma_{21}}}{<v_{group}>}
\label{eqn:time}
\end{equation}
where $\Delta L_{\gamma_{21}}$ is the path length difference and $<v_{group}>$ is the weighted average of the group velocity over the optical range of the detector. The experimentally measured relative times and z-positions in terms of the track variables are
\begin{subequations}
\begin{equation}
\Delta t_{\gamma_{21}} = t_o \: ( 1 - \frac{\beta\:c}{<v_{group}>}\:\tan\theta_i)
\end{equation}
\begin{equation}
\Delta z_{\gamma_{21}} = \beta \: c \:  t_o \: \cos\theta_i
\end{equation}
\end{subequations}
A useful relation is the time-projection on the z coordinates. The projection over an infinitesimal track length is
\begin{equation}
\frac{dt}{dz} \approx \frac{1}{\beta \:c} - \frac{\tan\theta_i}{<v_{group}>}
\label{eqn:tproj}
\end{equation}
for small angles along a path parallel to the detector plane. The OTPC photon drift speed in Eqn.~\ref{eqn:tproj} is given by $<v_{group}>$.

An additional set of reflected photons is detected at each MCP-PMT
from a mirror mounted on the opposite side of the inner cylinder
as shown by $\gamma_3$ in Figure~\ref{fig:otpcray}. This is not a
necessary ingredient of an OTPC, but is an effective and
economical method of increasing limited photocathode coverage.
Each mirror is mounted at an angle, $\theta_{mirror}$ of
90$^{\circ}$-$\theta_{c}$. For a small range of particle angles,
the mirror creates a time-delayed image of the Cherenkov light on
the opposing wall at the MCP-PMT.

The discrete mirror positions provide a technique to measure the
longitudinal track location across a single MCP-PMT by using the
measured time difference between the direct and mirror-reflected
Cherenkov light. The displacement, $r$, from the OTPC center-line
is given by
\begin{equation}
r = \frac{1}{2}\:(\Delta t\:<v_{group}> - \:d)  \:
\bigg(\:\frac{1}{\sin\theta_c} - \frac{<v_{group}>}{\beta c
\tan(\theta_c)}\:\bigg)^{-1} \label{eqn:mirror}
\end{equation}
where $\Delta t_{mirror} $ is the measured time difference at the MCP-PMT
and $d$ is the inner cylinder diameter.
Averaging over the dispersion effects and using $\beta$=1, we insert values in Eqn.~\ref{eqn:mirror}:
\begin{equation}
r =  0.737 \:\: (\Delta t\:<v_{group}> - \:d)
\label{eqn:mirror2}
\end{equation}

The scaling factor accounts for relative optical path lengths when
the particle is off the OTPC center-line. For example, for tracks
closer to the mirror, the reflected Cherenkov photons detected at
the PM are generated along the charged particle track at a later
time than the direct photons. The distance, $d$, from the angled
mirror center-point to the photo-detector module is 18$\pm$1~cm.

\section{Photodetector Module}
\label{sec:pmodule}

The OTPC is instrumented with five Photodetector Modules (PM),
which provide the photon detection using  the commercial
`Planacon' XP85022 MCP-PMT device from PHOTONIS~\cite{photonis}.
These square devices have a $5\times5$ cm$^2$ photo-active area
and an anode comprised of a $32\times32$ square pad array with a
1.6~mm pad pitch.  The three MCP-PMTs mounted in the normal view
are 25~$\mu$m MCP-pore-size tubes with fused silica windows
manufactured in 2014-15. On the stereo view are 2 MCP-PMTs; one
from the same manufacturing run, and another considerably older
MCP-PMT with a borosilicate window. The latter MCP-PMT exhibited
lower gain and detection efficiency.

\begin{figure}[]
\centering
\includegraphics[scale=.35]{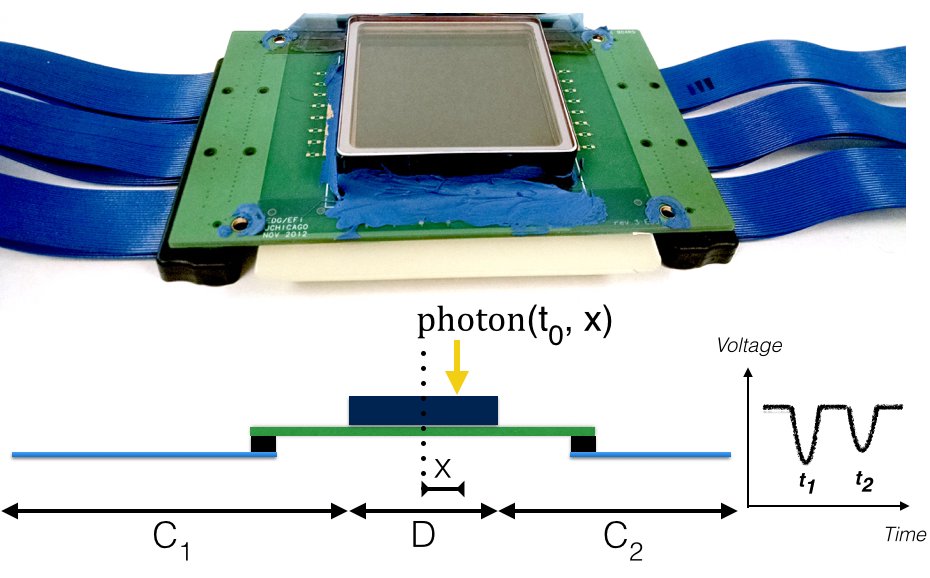}
\caption[PHOTONIS Planacon MCP-PMT mounted on transmission line
anode board]{A PHOTONIS XP85022 Planacon MCP-PMT~\cite{photonis}
mounted on a custom thirty-two channel, 50~$\Omega$ transmission
line anode board is shown in the top photograph. The Planacon has
an anode of 1024 pads in a $32\times32$ array over the $5\times5$
cm$^2$ active area. These are mapped to the microstrip anode
readout card in rows of 32. The bottom diagram outlines the
technique for using the single-channel waveform timing to extract
the incident photon position along the transmission line. The
times of the direct and open-end reflected waves on the anode
microstrip are denoted by {\it t$_1$} and {\it t$_2$}.}
\label{fig:planacon_photo}
\end{figure}

While reading out each of the Planacon's 1024 anodes might provide
the best-case granularity and pile-up rejection per channel, it
presents a challenge when instrumenting a detector with many such
photo-detectors. Instead, we adopt a similar  microstrip readout
scheme as the first-generation glass package
LAPPDs~\cite{lappdanode, tang}. A thirty-two channel microstrip
anode PCB board was designed such that a column of 32 anode pads
is mapped to a single 50~$\Omega$ transmission line. Each
microstrip line serves as both the DC~return and the high fidelity
signal path for
\begin{math}\frac{1}{32}\end{math} of the MCP-PMT. The central 30 strips
are digitized while keeping the two edge strips terminated at 50~$\Omega$.
The board-mounted photo-tube is shown in
Figure~\ref{fig:planacon_photo}.

The accelerated electron shower  from the MCP will induce a
quasi-TEM wave between the microstrip's signal and reference
plane~\cite{russermicrowave}. Once created, this wave will
propagate equally along both directions on the 1D microstrip.
Two high-density multi-channel 50$\Omega$ coaxial cables per
microstrip array end bring these waveforms off the
board~\cite{samtec}. The readout strategy adopted for the OTPC is
to digitize the waveforms on the microstrips at only one end,
leaving the multi-channel cables at the other end unterminated.
Thus, the wave that initially propagates along a transmission-line
microstrip toward the open end will be reflected with the same
polarity back towards the digitizer for that strip, which then
captures the waveforms caused by the initial pulse in both its
direct and reflected manifestations (see Fig.~\ref{fig:planacon_photo} and
$\S$\ref{sec:dataacq}).

For a single photo-electron signal, the digitizer will record two transient waveforms along the microstrip.
The times of the direct and open-end reflected waves, {\it t$_1$} and {\it t$_2$} can be extracted
from these transients, as diagrammed in Figure~\ref{fig:planacon_photo}.
Using the measured times we can solve for the photon longitudinal position, {\it x},
and time-of-arrival, {\it t$_o$}:
\begin{subequations}
\begin{equation}
\label{eqn:tdiff}
x = v_{prop}\:\frac{t_2 - t_1}{2} - \frac{D + 2\:C_1}{2}
\end{equation}
\begin{equation}
\label{eqn:t0}
t_0 = \frac{t_2 + t_1}{2} - \frac{1}{v_{prop}}(D + C_2 + C_1)
\end{equation}
\label{eqn:tline}
\end{subequations}
where {\it D}, {\it C$_1$}, and {\it C$_2$} are the microstrip
lengths of the MCP-PMT footprint, the reflected wave cable length,
and the direct wave cable length respectively. The wave speed
along the microstrip is {\it {v}$_{prop}$}, and is derived in
$\S$\ref{subsec:singlephoton}. A 10'' (25.4~cm) high density
coaxial cable provides a delay for the reflected signal and an
identical 6'' (15.2~cm) cable connects the microstrip card to the
waveform digitizer board~\cite{samtec}.

\section{Front-end data acquisition}
\label{sec:dataacq} The waveforms on the anode microstrip of each
PM are digitized at 10.24~Gigasamples-per-second (GSPS) using the
ACquisition and Digitization with pseC4 (ACDC) printed circuit
card. This card uses 30~channels of the PSEC4 ASIC with an analog
bandwidth~$\ge$~1~GHz~\cite{psec4}. The signal input to the PSEC4
is AC-coupled and the on-chip signal voltage range is from
0.1-1.1~V. A pedestal voltage-level of 0.8~V is set for the OTPC
data runs, allowing ample headroom for negative polarity pulses.
The digitized waveforms from the PSEC4 chips are sent to an
on-board field-programmable gate array (FPGA), which buffers the
digital data. Serial communication and data readout to a system
data acquisition card are performed over a dual low-voltage
differential signaling (LVDS) link.

The ACDC card runs off a 40~MHz clock that is distributed to the
five PSEC4 ASICs and FPGA using an on-board  jitter-cleaning PLL
chip~\cite{pllchip}. This clock is up-converted in the PSEC4 ASIC
using a voltage-controlled delay line (VCDL) to achieve a
10.24~GSa/s sampling rate over 256 sample cells~\cite{psec4}.
Though an enabling and power-efficient means of garnering high
sampling rates, the time-steps in a VCDL-controlled switch
capacitor array are non-uniform due to process variations in the
integrated circuit fabrication. Methods of calibrating of the
time-base of PSEC4 and related CMOS chips are discussed in
\cite{psec4, blab, kurtis, rittsnr}.

There are two sources of error in the PSEC4 times-base: the
differential non-linearity (DNL) and the integral non-linearity
(INL)~\cite{psec4}. The DNL is the cell-to-cell time-base
variation from the nominal time-step and the INL is a measure of
the aggregate deviation over many sampling cells. We apply a crude
calibration that considers the average INL over the PSEC4
time-base, as shown in Table~\ref{tab:timebase}. The time-steps
between sampling cells are defined as $\Delta$$t$$_{n}$ :=
$t$$_{n+1}$- $t$$_{n}$. With this basic calibration the sampling
step is assumed to be uniform over the first 255 cells, which
leaves  DNL and local-INL errors uncorrected. The PSEC4
`wraparound' is considered separately and has a much larger
differential time-step.

\begin{table}
  \caption[Time-base calibrations for the PSEC4 ASIC]{Coarse time-base
   calibrations for the PSEC4 ASIC. The time-step from sample 255
   to sample 0 is the `wraparound' of the circular sampling buffer,
   and is unique.} \vspace{5 pt}
  \centering
  \centering
  \begin{tabular}{ l  l }
    Time step $\Delta$t$_{n}$  & Value [ps]\\ \hline 
    [0,254] & 96.5 \\ 
    255   & 393.3 \\ 
  \end{tabular}
\label{tab:timebase}
\end{table}



\section{Experimental Setup}
\label{sec:exp}

The OTPC was installed on the beam line of the MCenter secondary
beam at the Fermilab National Laboratory Test-Beam
Facility~\cite{ftbf}, approximately 3~m behind a 1.09~m thick
steel absorber. The absorber served as both a beam-stop for the
secondary beam so that the dominant particle type seen by the OTPC
was muons, and a target station and collimator to generate a
tertiary beam utilized by the LArIAT experiment, to which we
operated parasitically~\cite{lariat2014}.

\begin{figure}[]
\centering
\includegraphics[scale=.082]{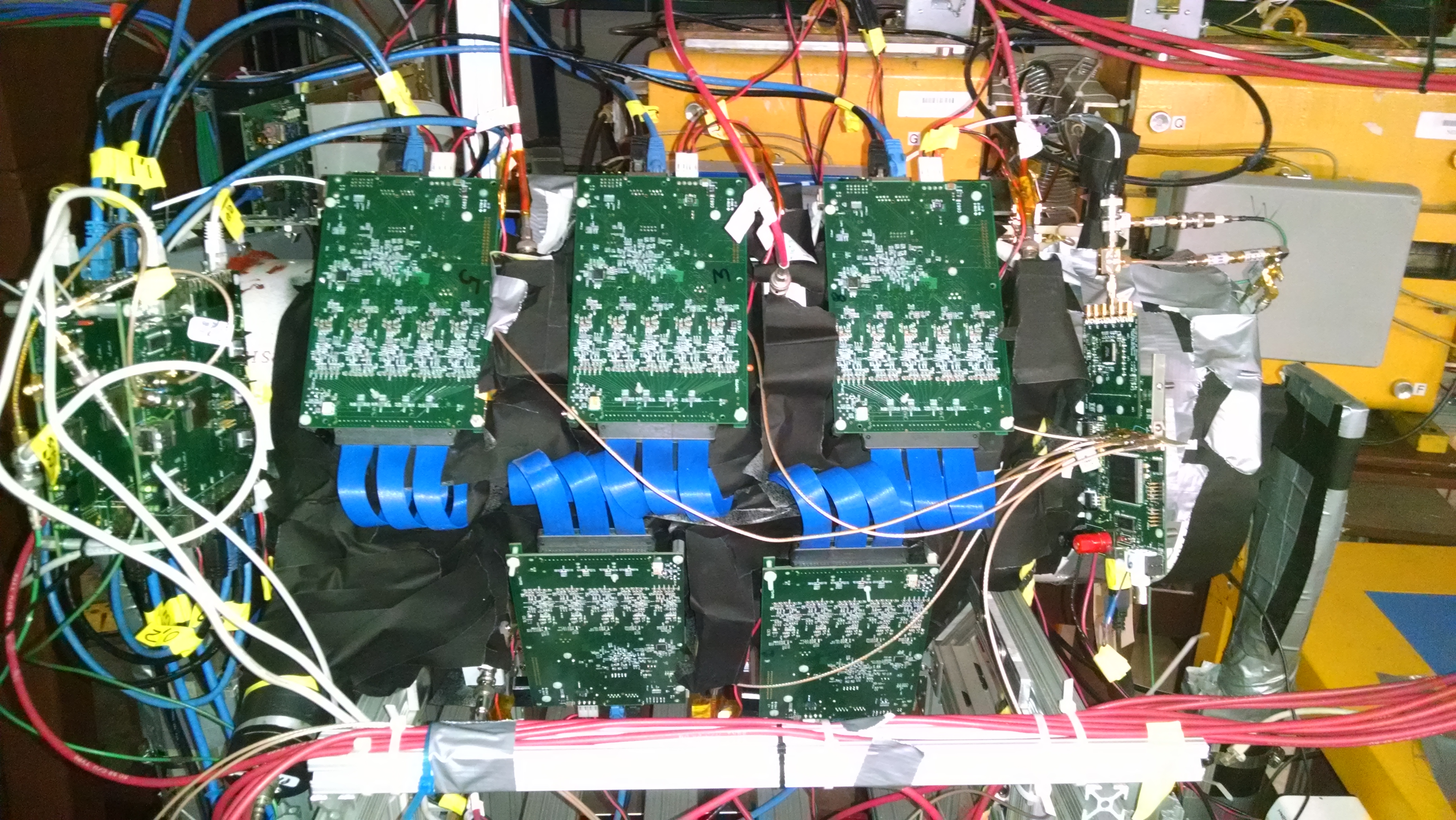}
\caption[The OTPC mounted in the MC7 enclosure hall]{The prototype
OTPC installed in the MC7 enclosure hall at the beam height of 208
cm. The upstream steel absorber block is out of the frame to the
right. Mounted facing inwards, the five PMs connect to the system
data acquisition (DAQ) cards that are mounted on the left. The DAQ
system connects to a remotely-controlled laptop on an adjacent
rack, which also houses the beam trigger coincidence and
discrimination logic. The yellow dipole magnets on the adjacent
tertiary beam-line used by the LArIAT experiment are visible in
the background.} \label{fig:OTPCphoto}
\end{figure}

A side view of the installed OTPC is shown in Figure \ref{fig:OTPCphoto}
where the upstream absorber is out of the frame.
The five OTPC PM modules are mounted facing inwards on the water volume.
Each of the PMs is connected to the system central cards over two~Category~6 (CAT6) network cables.
The system cards interface to a remotely-controlled Linux PC on an adjacent rack that also houses the
beam trigger coincidence and discrimination logic.
High voltage (HV) is supplied to the detector from the counting room over $\sim$200~m SHV cables;
the low voltage for the electronics is supplied within the enclosure.
The OTPC water volume corresponds to 2.1 radiation lengths along the cylindrical axis.

The beam is delivered to the Meson Lab during the 4.2-second
slow-spill at the beginning of the 60~second Fermilab accelerator
complex `super cycle'. During the OTPC test-beam runs, a secondary
beam of positive pions impinged on the MCenter target at momenta
of 8, 16, or 32~GeV/c. Because the OTPC was located behind the
secondary beam absorber, the particle flux was dominated by muons
(85$\%$), with a mixture of hadrons (10$\%$), electrons (1$\%$),
and photons (4$\%$) from punch-through and showering, as verified
in a G4beamline simulation at a beam momentum of
16~GeV/c~\cite{g4beam}.

\section{The Trigger}
\label{sec:exttrigger}

The OTPC trigger is created using two systems. The first is the
beam trigger, which provides a coincidence signal of a
through-going particle. Second, a level-0 (L0) trigger is made
locally on each PM by discriminating the analog signals on every
channel using the PSEC4 threshold-level trigger. This two-level
system is required when using the PSEC4 chip due to its relatively
short analog memory. When running the PSEC4 at a sampling rate of
10.24~GSPS, the waveforms on the PSEC4 capacitor array are
over-written in 25~ns intervals. The beam coincidence trigger
decision requires a few hundred nanoseconds due to cable and
electronic delays, which necessitates the fast L0 trigger to save
these signals.

\subsection{Beam trigger system}
\label{subsec:beamtrig} 
\begin{figure}[]
\centering
\includegraphics[scale=.5]{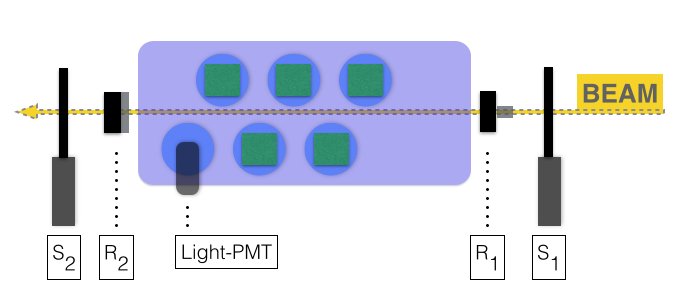}
\caption[Diagram of beam trigger]{Diagram of the external
beam-trigger (not to scale). The beam trigger coincidence can be
made between  four sub-detectors: the front and back
scintillators+PMTs (\textit{S$_{1}$} and \textit{S$_{2}$}), and
the front and back fused-silica Cherenkov radiators+MCP-PMTs
(\textit{R$_{1}$} and \textit{R$_{2}$}). An additional diagnostic
trigger is generated by a 1''~PMT (`Light-PMT') mounted on the
last OTPC port to observe Cherenkov light in the water volume. }
\label{fig:TrigDraw}
\end{figure}

The beam trigger is a set of scintillator
and Cherenkov detectors external to the OTPC water volume, as
shown in Figure~\ref{fig:TrigDraw}. A pair of $5\times5$~in$^2$
plastic scintillators mounted fore ($S_1$) and aft ($S_2$) provide
good coverage over the OTPC cross-sectional area. Another pair of
detectors, $R_1$ and $R_2$, are MCP-PMTs coupled to fused-silica
Cherenkov radiators that provide a prompt beam-trigger signal.
$R_1$ is a $1\times1$~in$^2$ active area, single-pixel MCP-PMT
mounted on the front cap of the OTPC\footnote{$R_1$ is a PHOTONIS
XP85011 MCP-PMT, which has an anode structure consisting of an
8-by-8 array of pixels. The $R_1$ signal is made by summing the
central 32 pixels through 10~$\Omega$ series resistors on each
pixel. $R_2$ is an additional PHOTONIS XP85022 MCP-PMT with an
anode structure of 32-by-32 pixels, read out by the 30-microstrip
anode board.}. In the central region of $R_1$ is a small
cylindrical (1~cm diameter) fused-silica radiator. $R_2$ is a
$2\times2$~in$^2$ MCP-PMT that is coupled to a matching 5~mm thick
square fused-silica radiator. It is mounted on the rear of the
detector as also shown in Figure~\ref{fig:otpcdraw}.

\begin{figure}[]
\centering
\includegraphics[scale=.31]{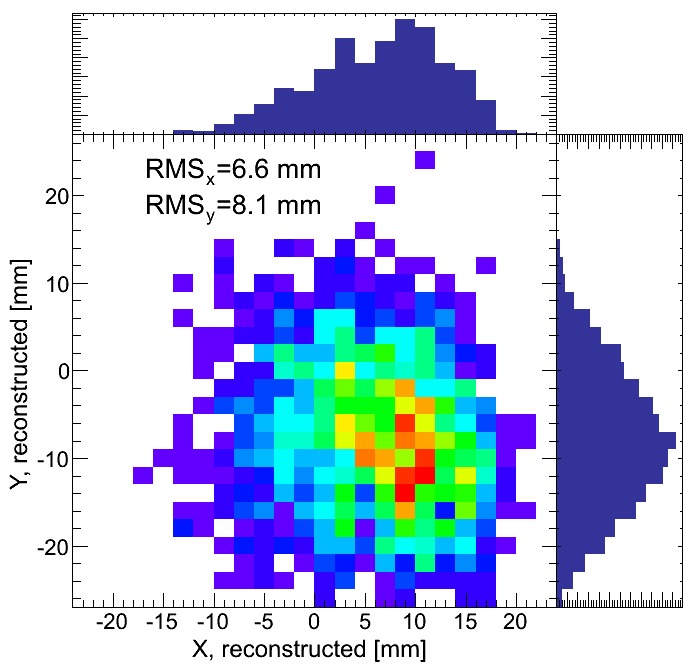}
\vspace{-.2cm} \caption[Output beam profile using position and
time-sensing MCP-PMT]{Reconstructed output particle position
through the rear MCP-PMT+Cherenkov radiator for a 16~GeV/c
dataset. This MCP-PMT is part of the beam trigger system and is
denoted as $R_2$ in Fig.~\ref{fig:TrigDraw} in the coordinate
system defined in Fig.~\ref{fig:otpcdraw}. The data are
histogrammed in $2\times2$~mm$^2$ bins, where the Y direction is
taken from the time difference along the anode transmission line
(in the same manner as the single-photon case,
$\S$\ref{subsec:singlephoton}) and the X position is reconstructed
using the integrated-charge centroid between microstrips.}
\label{fig:back-trig}
\end{figure}

An additional six channels of PSEC4 electronics are implemented on
the OTPC to monitor the analog signals of $S_1$ and $S_1$, as well
as to provide a threshold discriminator for the $R_1$ signal. This
is done with a PSEC4 Evaluation Card~\cite{psec4}, which also
serves to fan-out the $R_1$ trigger signal to the PMs as an
unbiased L0 trigger signal. The $R_2$ signals are read out and
digitized in the same manner as the PMs, allowing spatial and
time-tagging of the out-going particle, as shown in
Figure~\ref{fig:back-trig}. For each event, the digitized signal
from $R_1$ is recorded with those of $R_2$, allowing for a
time-of-flight measurement ($\approx$100~ps resolution).

The beam trigger signal is made with standard NIM electronics. The
signals $S_1$, $S_2$, and $R_1$ are discriminated and and fed to a
coincidence unit. Two configurations were used during data runs:
(1) $S_1$+$S_2$+$R_1$, and (2) $S_1$+$R_1$. Trigger configuration
(1) is the default through-going particle trigger. Configuration
(2) allowed the recording of particles that may scatter out, stop,
or shower within the water volume. Both digital trigger signals
are sent to the OTPC data acquisition system, which then sends
these trigger signals to the front-end ACDC cards.

\subsection{Fast level-0 trigger}
\label{subsec:l0trig}

\begin{figure}[]
\centering
\includegraphics[scale=.46]{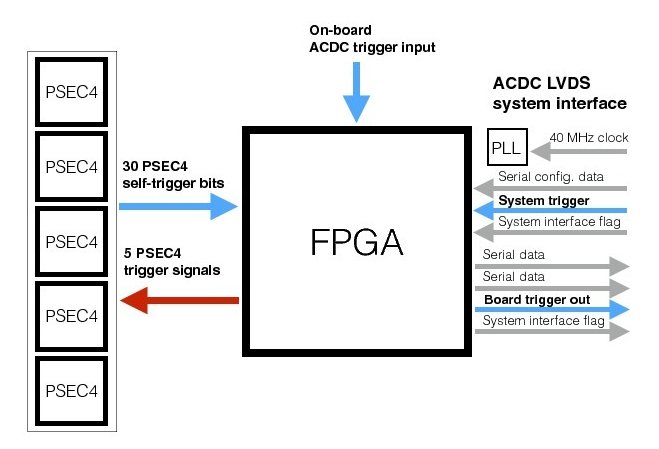}
\caption[Diagram of level-0 trigger]{Diagram of the ACDC level-0
(L0) trigger and low-voltage differential signaling (LVDS) system
interface. The blue lines (lighter shade) denote the ACDC trigger inputs and
outputs. The L0 trigger is made as a logical combination of the 30
PSEC4 self-trigger bits and the on-board fast trigger input. In
the configuration for which the L0 trigger is made in the FPGA,
the five PSEC4 trigger signals (red line) are sent to the PSEC4
ASICs in order to hold their analog values. If a beam-trigger
signal is received within a coincidence window of 175~ns, these
values are digitized.} \label{fig:boardtrig}
\end{figure}

The beam trigger signal is formed and returned to the OTPC electronics with a latency of approximately 130~ns due to logic and cable signal delays.
The L0 trigger is implemented to hold the analog waveforms on the PSEC4 switched capacitor array
if an event-of-interest is locally triggered. These waveforms are held for an adjustable length of
time before either registering a beam-trigger and digitizing or, if no beam-trigger is received,  releasing and re-starting the 10.24 GSPS sampling.

The L0 trigger is made within the field-programmable gate array (FPGA) on the ACDC card as
outlined in Figure~\ref{fig:boardtrig}.
Each of the six PSEC4 channels has a built-in threshold discriminator that asynchronously latches when an input signal crosses the threshold voltage. The signal polarity is set on the PSEC4 chip and the threshold levels are adjusted with an on-board digital-to-analog converter chip.
An ACDC card has 30 such self-trigger bits that are sent to the FPGA.

The PSEC4 self-trigger bits are used in the FPGA to form the L0
trigger signal. Once the L0 trigger is made on the FPGA, the
trigger signals are simultaneously sent to  each of the five PSEC4
chips on the board, which then freeze the analog levels on the
sampling capacitors.
The round-trip L0 trigger time is 20-25~ns, which is sufficient to
save the waveforms before the PSEC4 circular buffer begins to
over-write. The coincidence window between the L0- and
beam-trigger signals was set to 175~ns, after which time the L0
trigger signal is released, restarting the sampling.  The
coincidence window allows for a small rate of accidental triggers
on one or more ACDC boards, which are removed by requiring a
synchronized L0 trigger time-stamp across all boards for each
event.

\section{Data Reduction}
\label{sec:reduce} An OTPC event is recorded using 180~channels of
10.24~GSPS data in a time window of 25~ns. The first 150 channels
contain the data from the OTPC photodetector modules (PM) and the
remaining thirty channels, 150-179, contain the trigger
information from the $R_1$ and $R_2$ detectors(described in
$\S$\ref{sec:exttrigger}). The digitized signal from $R_1$ (see
Fig.~\ref{fig:TrigDraw}) is saved in channel 179 and is used as a
coarse zero-time reference. The events are recorded in the PSEC4
chip asynchronously to the 40~MHz system clock, which leaves the
location of an event in the PSEC4 buffer undetermined. The $R_1$
signal time is used as a reference to find the Cherenkov photon
signals in the digitized data.

The following procedures are performed on these data before analyzing an event:
\begin{enumerate}

\item A pedestal calibration is performed for each of the 256 sample cells in all channels \cite{psec4}.

\item The firmware time-stamps are checked on the front-end ACDC boards.
If the time-stamps are not synchronized, the event is discarded.

\item
The data on all channels are shifted in the PSEC4 buffer such that the $R_1$ peak-signal
time is located at sample cell 90. With this shift, the hits from the Cherenkov photons are
located in the first one-third of the 256 sample PSEC4 buffer.

\item A reference baseline level, created using the median value of sample cells 200-255, is subtracted from each channel.

\item A time-base is constructed from the sample-steps by applying the coarse calibration shown in Table~\ref{tab:timebase}.  The data points that cross the wraparound region ($\S$\ref{sec:dataacq}) are interpolated to have an evenly spaced event time-base of 97~ps per sample.
\end{enumerate}
At this point in the data reduction process, the event is aligned with respect to the $R_1$ trigger signal, baseline-subtracted, and referenced to a time-base. The event size is: 180 channels $\times$ 12 bit ADC counts $\times$ 260 time-steps over 25~ns (70~kB).

\begin{figure}[]
\centering
\includegraphics[trim=.13cm 14cm .35cm 1cm, clip=true, scale=.44]{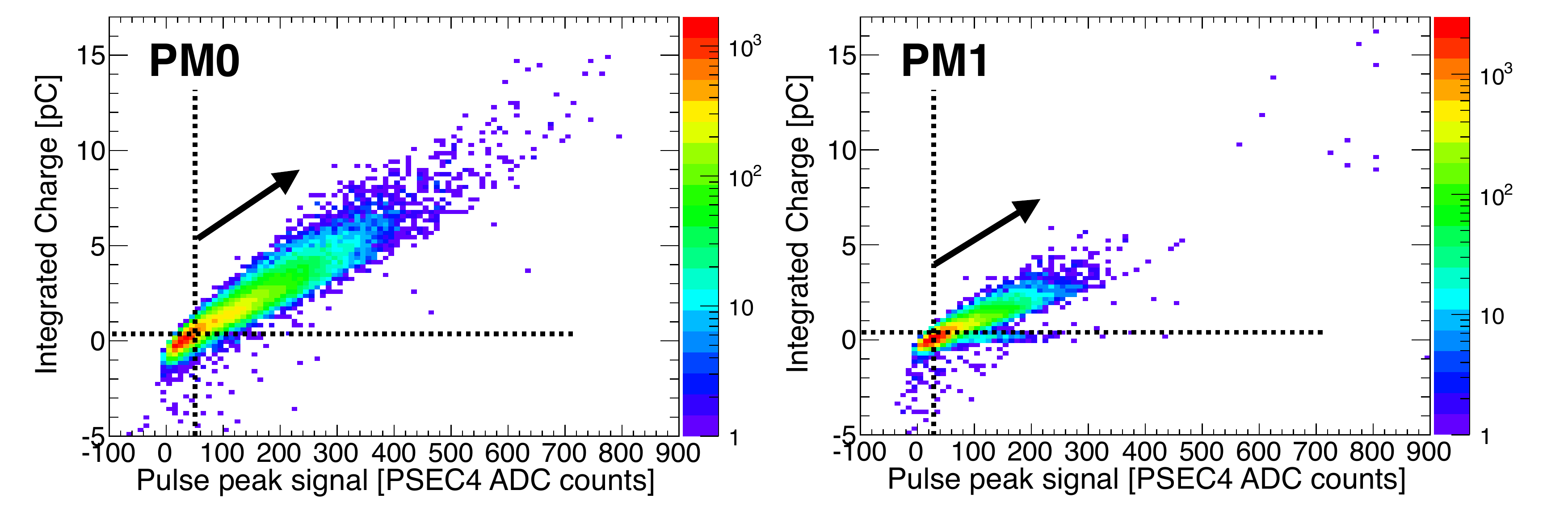}
\caption[Waveforms: integrated charge vs. signal peak]{Correlation plots of integrated charge vs. pulse peak signal for PMs~0~and~1. Channels with waveforms above a minimum integrated charge and ADC count threshold, shown as dashed lines, are saved and further processed to measure the photon time-of-arrival. }
\label{fig:pulsecuts}
\end{figure}

The event is further reduced by extracting the trigger information
in channels 150-179, narrowing the event time-window to hold only
the region of Cherenkov photon hits, and removing below-threshold
channels as shown in Figure~\ref{fig:pulsecuts}~\cite{mythesis}.
 After the data reduction, the event comprises the following trigger information:
 particle output x-position, particle output y-position, time-of-flight, $R_1$ trigger time,
$R_2$ trigger time, $R_1$ total charge, $R_2$ total charge, and the number of distinct hits in $R_2$.
The reduced size of the event time-stream data is: number of OTPC channels above threshold ($\le$150)
$\times$ 12 bit ADC counts $\times$ 110 time-steps over 10.67 ns (number of channels $\times$ 0.17~kB).

\subsection{Extracting the photon time-of-arrival}
\label{sec:features}

\begin{figure}[]
\centering
\includegraphics[trim=.13cm 0cm 0cm 1.4cm,clip=true,height=5.8cm]{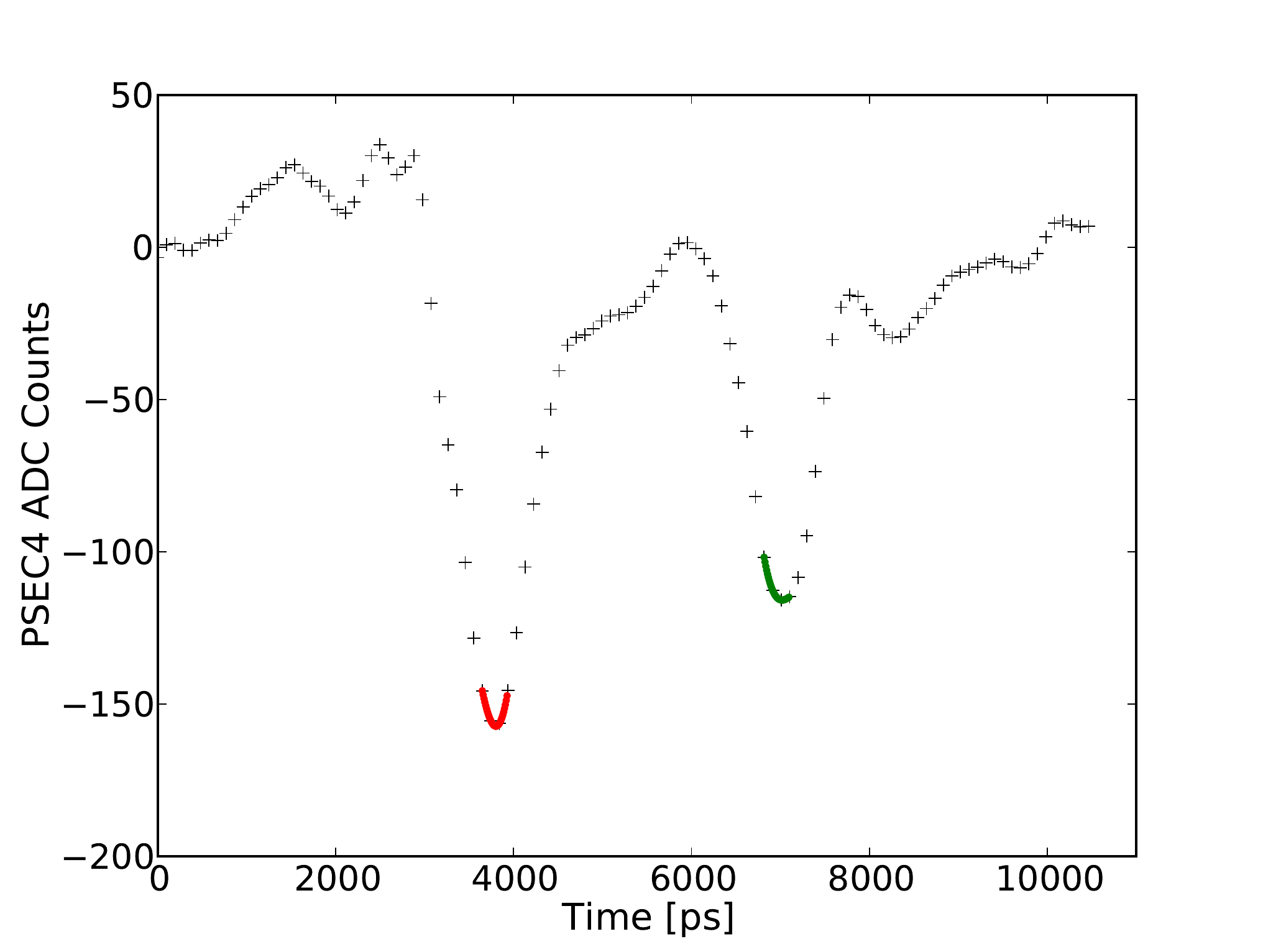}
\caption[Pulse feature extraction techniques]{OTPC digitized waveform from a Cherenkov photon. A single photo-electron waveform is shown with the peak-interpolation time-extraction method. The interpolated peak-signal regions on the waveform are shown in red (green), which represent the direct (anode-reflected) pulses.}
\label{fig:wfmfit}
\end{figure}

For each detected photon, we measure its 2D spatial position on
the OTPC cylinder ($z_{det}$, $\phi_{det}$) and its
time-of-arrival, $t_{det}$. By projecting the measured photon
arrival times on the two spatial coordinates we can reconstruct
the charged particle track in 3D. We take the detected photon
azimuth angle, $\phi_{det}$, to be the PM location about the OTPC
cylinder and its z-position, $z_{det}$, as the microstrip position
along the OTPC longitudinal axis. There are 2 discrete
$\phi_{det}$ angles and 150 discrete $z_{det}$ positions. This
section describes how we extract the time-of-arrival for each
photon from the digitized waveforms.

We found the most robust algorithm is to interpolate the PSEC4
waveform in a small region around the peak. A window of
4~time-steps (97~ps each) is interpolated with a cubic spline on a
9.7~ps grid and the pulse time is taken from the peak of the
interpolated region. An example of an OTPC channel hit by a
Cherenkov photon is shown in Figure~\ref{fig:wfmfit}.

For an event, the output of the peak-interpolation algorithm is a set of times ([$t_{1a}$,..], [$t_{2a}$,...]) from the direct and anode-reflected pulses. To get the photon hit time, we apply Eqn.~\ref{eqn:t0} and require causality using Eqn.~\ref{eqn:tdiff}.
That is, the time difference between the direct and anode-reflected pulse should be located on the microstrip line within the 5~cm length of the MCP-PMT.
The causality requirement is that $t_{2a}$~-~$t_{1a}$~$\in$[3.1, 3.9]~ns after solving Eqn.~\ref{eqn:tdiff} for the cable lengths in the PM.

\section{Single photo-electron response}
\label{sec:lasertests}

A laser test-stand was built to characterize the single photo-electron response of the PMs.
A 405~nm PiLas laser was used in pulsed mode, which can achieve pulse widths of 33~ps~FWHM~\cite{pilas}.

\subsection{Single-photon}
\label{subsec:singlephoton}

\begin{figure}[]
\centering
\includegraphics[scale=.32]{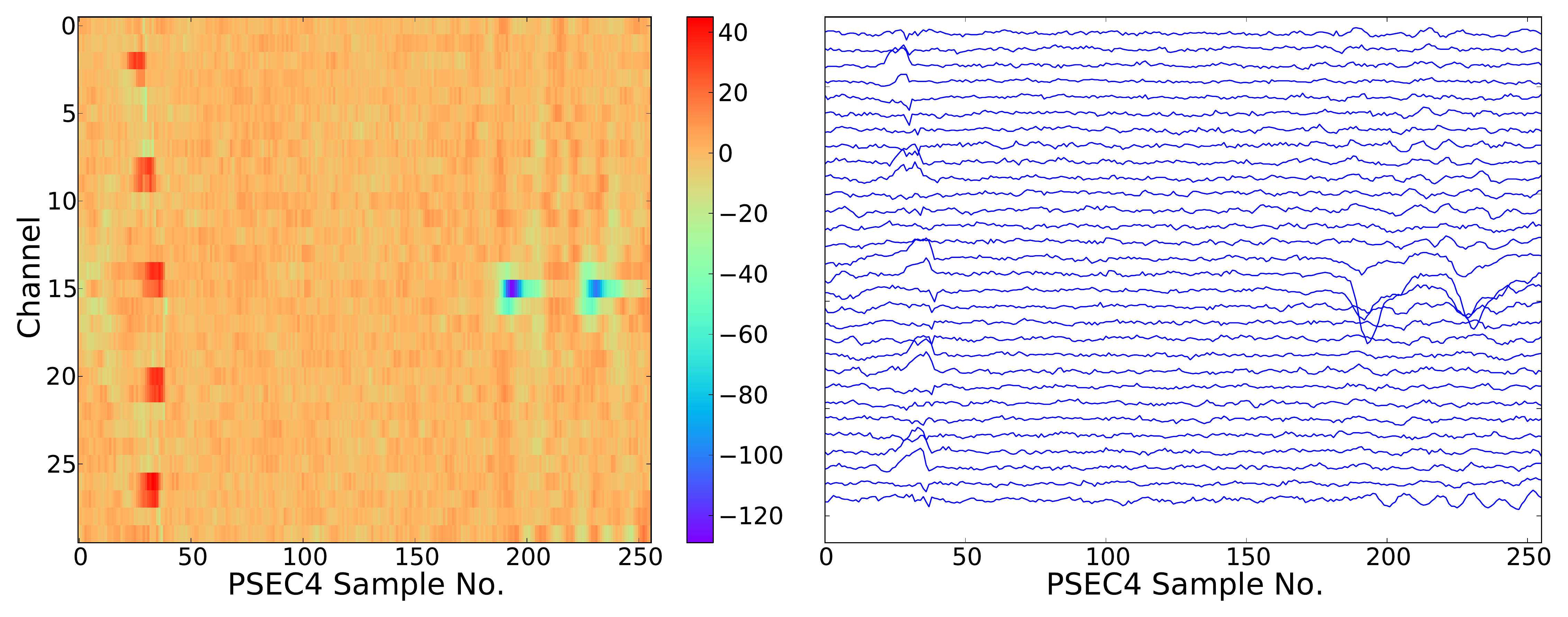}
\caption[Single photo-electron event]{A single photo-electron recorded by the PM using a pulsed laser.
The left panel shows the ACDC channel vs. PSEC4 sample number with the color indicative of the digitized value in PSEC4 ADC counts.
 The corresponding PSEC4 waveforms  are shown on the right, in which the two peaks in the waveform are the direct and reflected pulses on the anode microstrip.}
\label{fig:singlepe_event}
\end{figure}

In the single photon mode, an iris is placed on the free-space
output of the PiLas optical head, which picks off a 1~mm diameter
portion of the beam. The iris is placed a few mm off the incident
beam axis in order to produce an attenuation of approximately
50$\%$. Following the iris is a neutral density filter
(Transmittance = 10$^{-5}$) to attenuate the beam to the
single-photon level. About 5$\%$ of the events recorded in the
single-photon test setup have a laser-correlated signal.

\begin{figure}[]
\centering
\includegraphics[trim=0cm .5cm 1cm .5cm, scale=.46, clip=true]{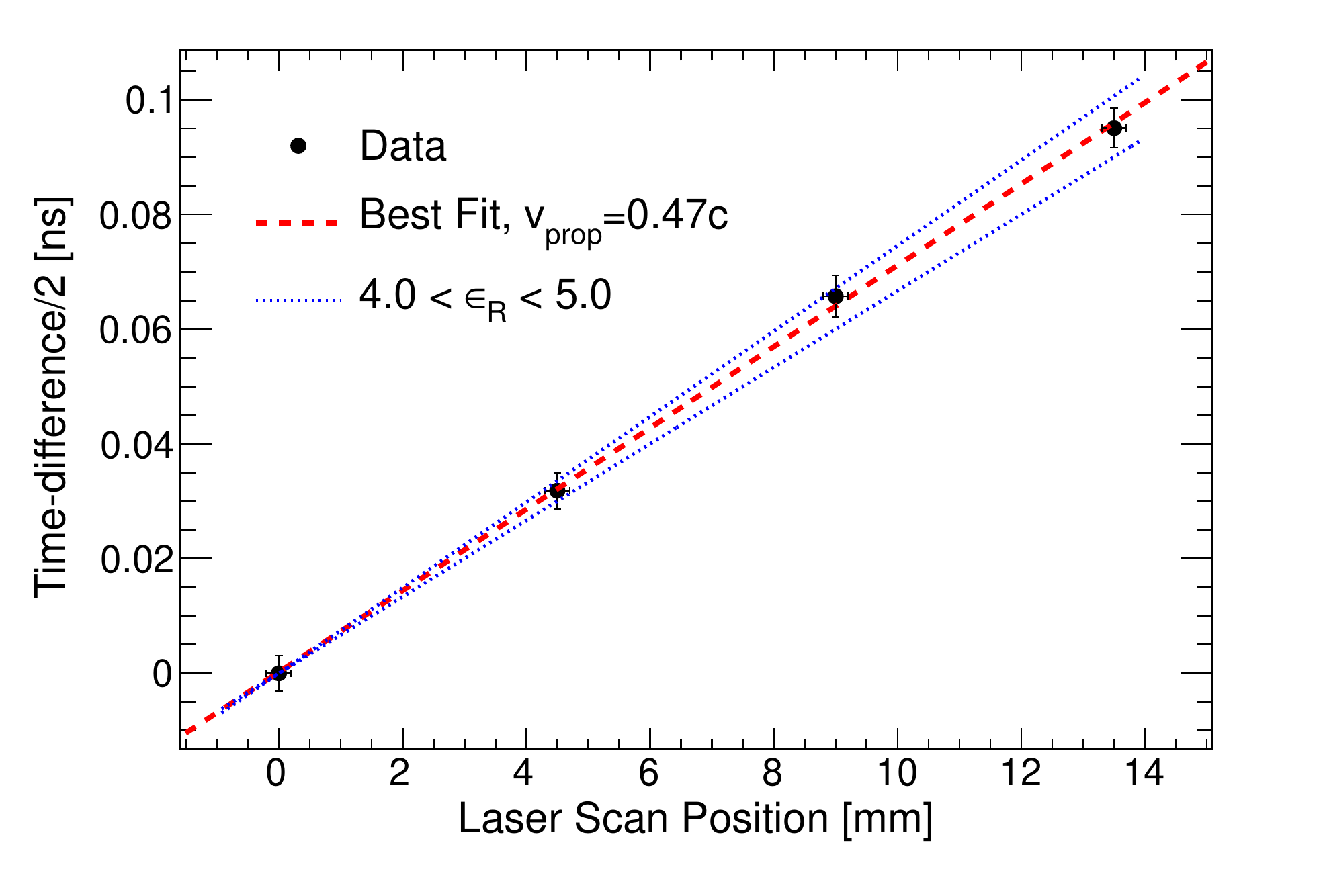}
\caption[Propagation delay along anode microstrip]{ Single
photo-electron laser scan to measure the propagation velocity
along the microstrip line. The vertical error bars are the
3$\sigma$ standard error on the time-difference measurement (the
single event statistical error on the time-difference is
$\le$35~ps for all laser positions). The data agree with the
expected velocity given the typical dielectric permittivity of an
FR4 printed-circuit board substrate ($\epsilon= 4.3\pm0.2$).}
\label{fig:anode_delay}
\end{figure}

Figure~\ref{fig:singlepe_event} shows the PM response to a single
photo-electron. Typical signals are on the order of 100 PSEC4-ADC
counts~($\sim$25~mV) for the first pulse and somewhat less for the
time-lagged pulse due to the additional resistive attenuation.
There is a 20~dB pre-amp board on the ACDC input so the
unconditioned single photo-electron MCP-PMT signals are typically
between $\sim$2-5~mV\footnote{Measurements with the LAPPD MCPs
have demonstrated gains $\ge$ $10^7$, which give single
photo-electron signals ten-times as large into the same
50~$\Omega$ anode~\cite{andrey}.}. An approximately synchronous
line in many channels near the beginning of the recording window
is also visible in the figure.  This is the time at which the
PSEC4 chips are triggered.

The single photo-electron position resolution is obtained by
scanning the laser beam on the active area of the PMs both
parallel and transverse to the microstrip direction. Results from
a scan along the microstrip direction are shown in
Figure~\ref{fig:anode_delay}. Four data points were taken at a
spatial separation of 4.5~mm with a time-difference resolution of
$\sim$35~ps measured for all laser positions. This is the inherent
electronics timing resolution for the single photo-electron pulses
digitized using one channel of the PSEC4 chip, with the basic (not
complete) time-base calibration described in
$\S$\ref{sec:dataacq}.

Figure~\ref{fig:anode_delay} shows the time difference along the microstrip vs. the laser beam position,
from which the signal propagation velocity, $v_{prop}$,
in Eqn.~\ref{eqn:tline} along the microstrip to be calculated.
The best fit $v_{prop}$ is found to be $0.47 \pm 0.02 \: c$.

\subsection{Multi-photon}
\label{subsec:multiphoton}

A similar test-stand setup is employed to measure the relative
timing between single photons within the same laser pulse. In this
setting, the iris is re-aligned with the laser beam axis, and a
10~mm focal length convex lens is put in the laser path. The
MCP-PMT sits 30~cm behind the lens, which projects an elliptical
laser beam spot over the MCP-PMT active area. The same neutral
density filter is installed after the lens to attenuate to an
average of a few photons per pulse.

\begin{figure}[]
\centering
\includegraphics[trim=.2cm 0cm 0cm 0cm, height=6cm, clip=true]{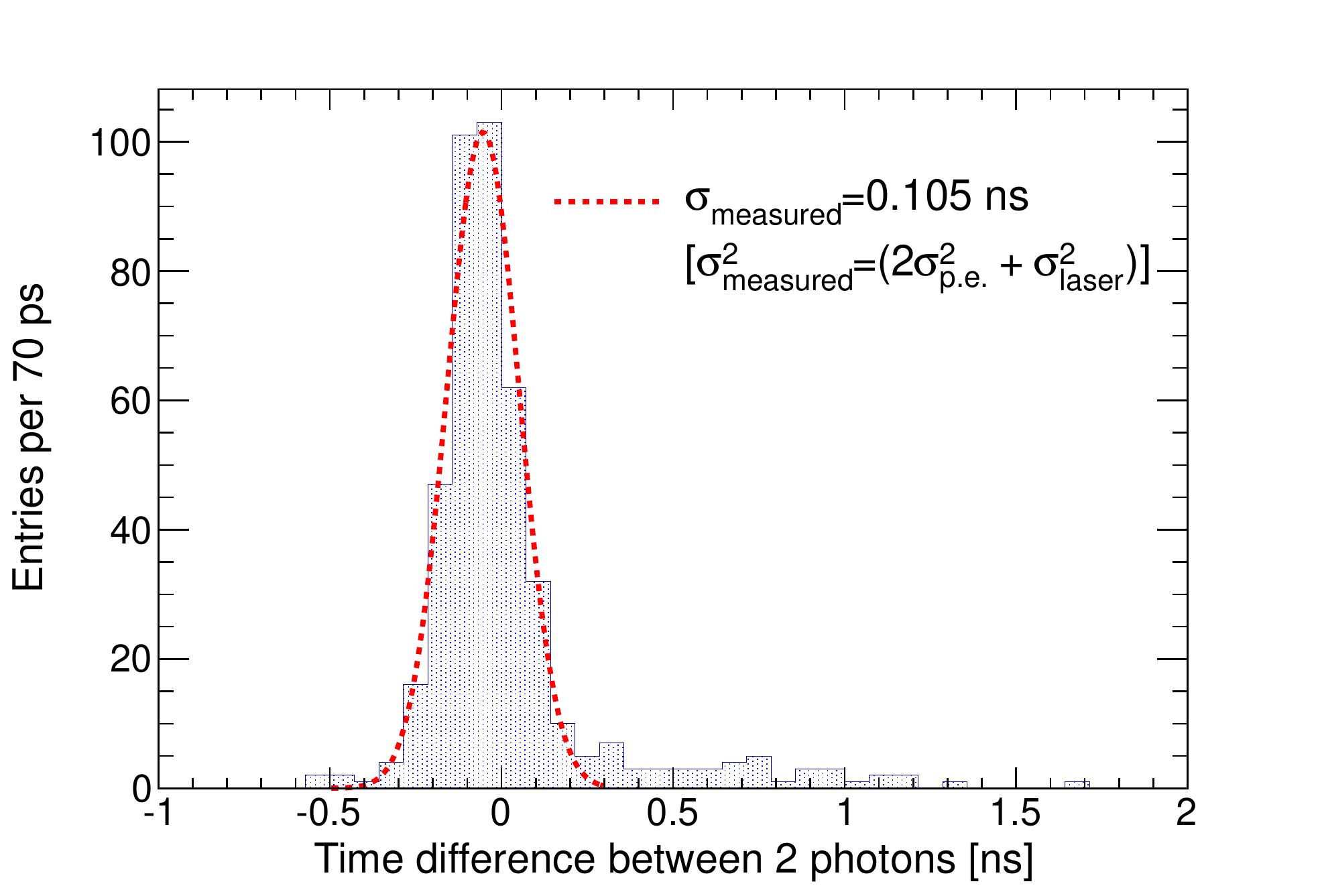}
\caption[Single photo-electron time resolution]{Relative timing measurement between 2 photons on the PM.
The underlying single photo-electron electronics + MCP-PMT timing resolution for this measurement is 75 picoseconds. }
\label{fig:timingpe}
\end{figure}

Events are analyzed that have two recorded photo-electrons, in
which the signals are clearly separated between channels above and
below the MCP-PMT center-line. The relative timing between photons
is shown in Figure~\ref{fig:timingpe}. A tail is seen in the
distribution, which is likely caused by optical reflections in the
test setup that are recorded late. The core resolution is shown to
be 105~ps, which corresponds to a single photo-electron PM
resolution of 75~ps.

From these measurements, we conclude that the 1-$\sigma$ single photo-electron
time-of-arrival and space statistical errors are
$(\sigma_t,\sigma_x,\sigma_y)$ = (75~ps, 2~mm, 3~mm).
The $\sigma_x$ resolution is taken from the microstrip pitch on the anode board.

\section{Photo-detection properties}
\label{sec:occupies}
The data presented are taken from 16~GeV/c secondary beam runs in which the optical water quality was at or above the
6-hour curve in Figure~\ref{fig:waterattn} and were taken using the through-going particle trigger
(the beam-trigger configuration 1 as described in $\S$\ref{subsec:beamtrig}).

\subsection{Photo-detection efficiency}
\label{sec:PDeff}

\begin{figure}[]
\centering
\includegraphics[scale=.45]{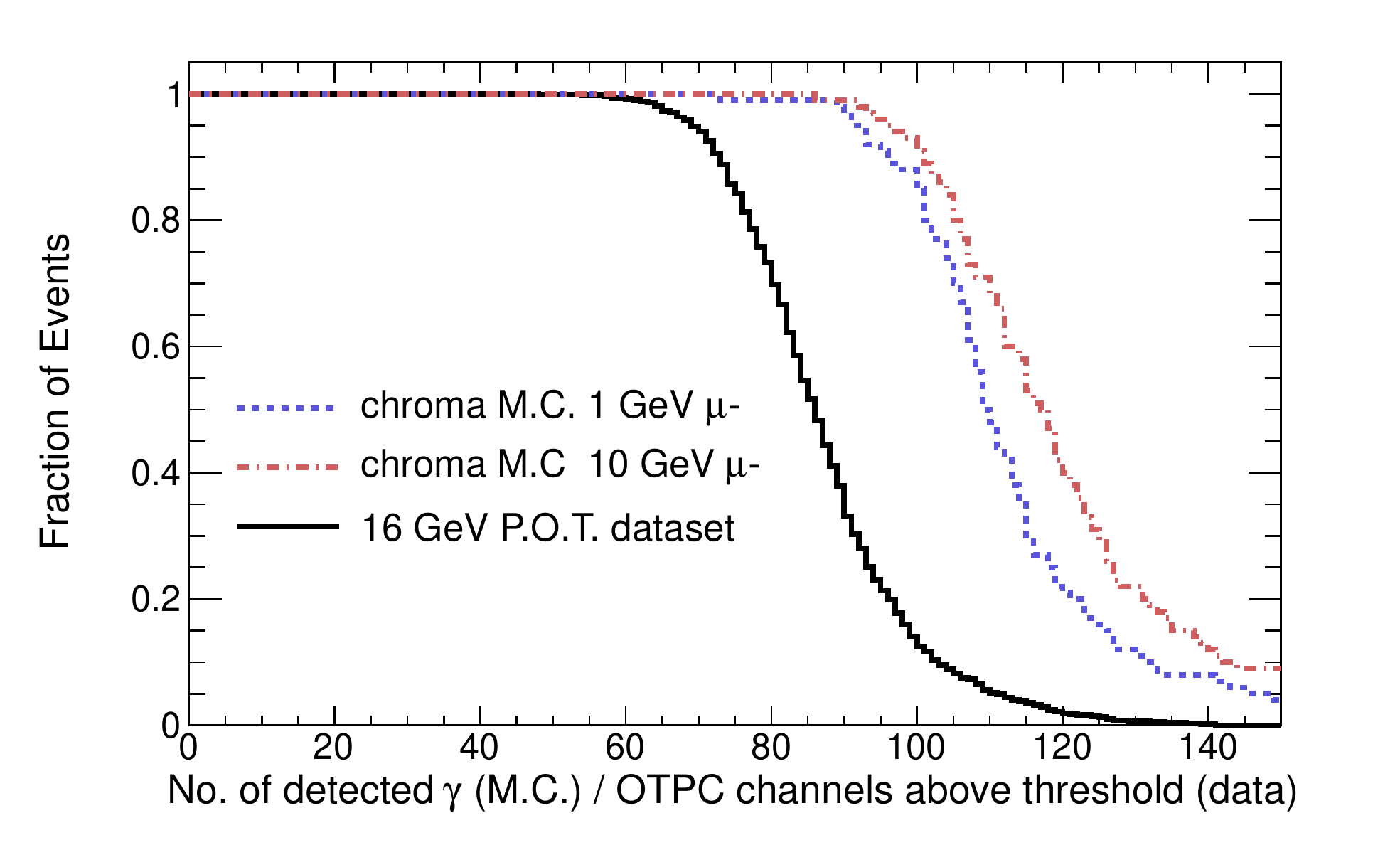}
\caption[Photo-detection efficiency]{Fraction of events in which
the number of OTPC channels is above a defined single
photo-electron threshold versus the number of channels required. }
\label{fig:Efficiency}
\end{figure}
The number of channels that passed the signal thresholds over the
1686 event dataset is shown in Figure~\ref{fig:Efficiency}.
Roughly 50\% of the events have 80 or more channels above
threshold per event. This is not a direct indication of the number
of individual photons detected per event, but should be a decent
proxy. Effects not included are over-estimation due to
charge-sharing between channels and under-estimation from channels
that may have two or more photon hits. From Poission statistics,
we estimate that 8$\%$ of strips will have more than one detected
photon.

\begin{table}
  \caption[Estimation of photo-detection inefficiencies]{Estimation of photo-detection inefficiencies compared to the detector Monte Carlo as shown in Fig.~\ref{fig:Efficiency}.  }
\vspace{5 pt}
  \centering
  \begin{tabular}{ p{2cm}| p{3.8cm}| p{8.5cm}  }
    \hline
    Source & Estimated effect & Explanation \\ \hline
    PM baffle & 6~$\pm$~3$\%$ loss of direct Cherenkov light &  PMs 0, 2, 3, and 4 show a smaller occupancy per event in the first several channels suggesting that the PM baffle (drawn in Fig.~\ref{fig:otpcray}) is
    intercepting some of the direct Cherenkov photons \\ \hline
    Reflection losses & 10~$\pm$~5$\%$ loss of direct Cherenkov light & Additional losses of the direct Cherenkov light due to areas of poor optical coupling at the fused-silica port window and MCP-PMT interface. For any air gaps in this interface, the critical angle for total internal reflection is $\sim$43$^\circ$.\\ \hline
    PM~1 & 5~$\pm$~3$\%$ total loss  & The detector Monte Carlo did not consider variations in the detection capabilities of the 5 MCP-PMTs. PM~1 has a notably lower efficiency than the other 4 PMs. \\ \hline
    Bad channels &  3~$\pm$~1$\%$ total loss  & Five channels in the OTPC DAQ were cut from the experimental datasets due to poor coupling. The detector Monte Carlo did not include the removal of these channels.
  \end{tabular}
  \label{tab:photoeff}
\end{table}

A comparison between data and the expectation from simulation shows a 70-80\% detection efficiency compared to the Monte Carlo.
Table~\ref{tab:photoeff} presents the sources and estimated detection losses that contribute to the observed photo-detection efficiency.

\subsection{Single photo-electron gain}
\label{subsec:gain}
\begin{figure}[]
\centering
\includegraphics[trim=1cm 11.9cm 6cm 1cm, scale=.42, clip=true]{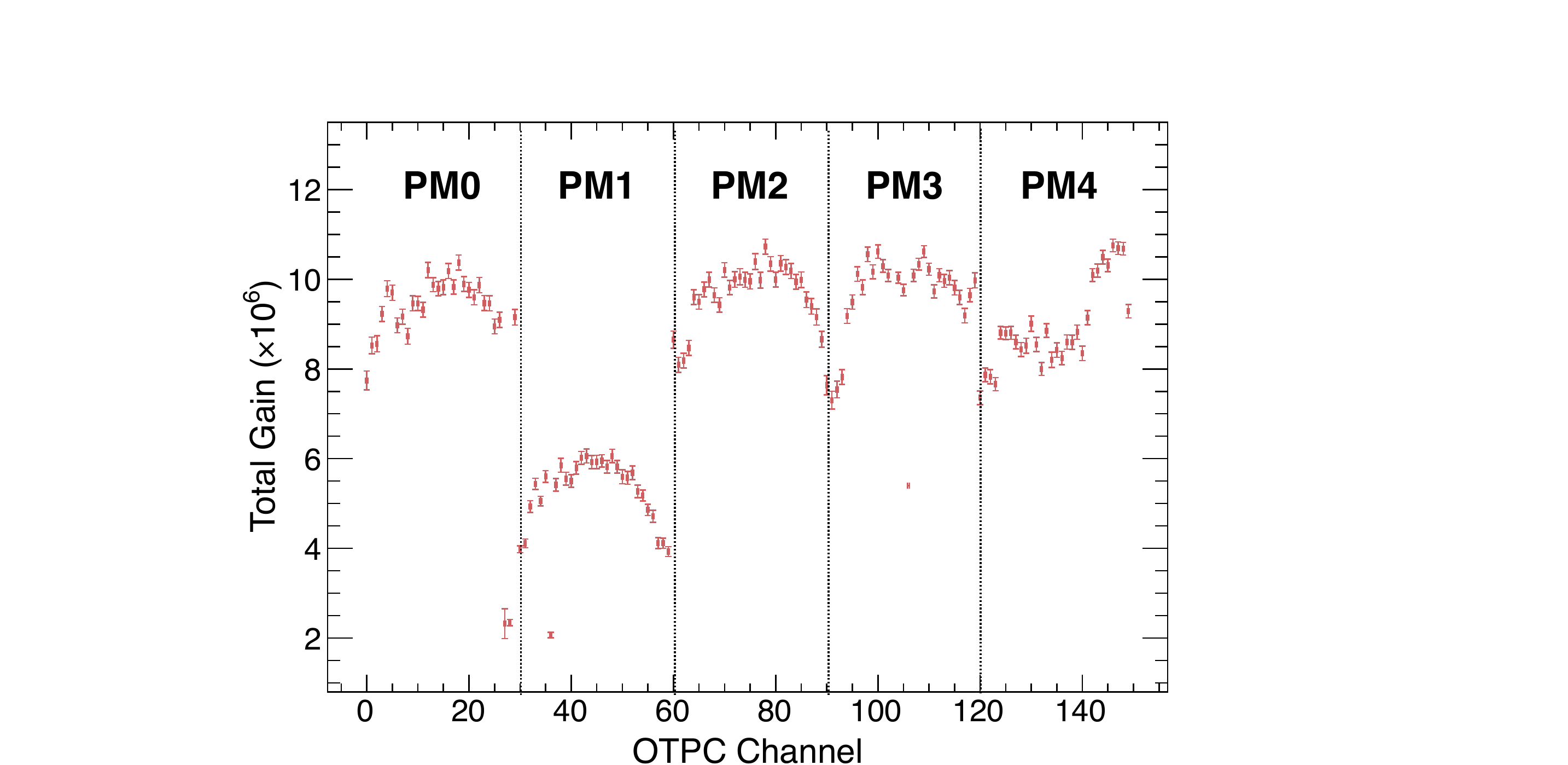}
\caption[OTPC channel gain]{The OTPC per-channel gain.  PM 0, 2, and 4 constitute the OTPC normal view; PM 1 and 3 are in the stereo view. The gain factor on each channel is comprised of three factors: the MCP-PMT~$\times$~electronics pre-amp gain~$\times$~variations in the PSEC4 count-to-voltage transfer function.}
\label{fig:stripgain}
\end{figure}

The gain per channel is shown in Figure~\ref{fig:stripgain}, which is the multiplicative
gain of the MCP-PMT, the 20~dB pre-amplifier board, and any relative gains between channels
in the PSEC4 ASIC\footnote{Since a full voltage calibration was not performed on these data,
there exist small gain differences between PSEC4 channels due to their relative transfer
functions from input voltage to output ADC counts. The vast majority of signals are small
($\textless$10$\%$ of the PSEC4 voltage range~\cite{psec4}), so this is not a large effect.}.
This is measured by taking the median integrated charge for signals above threshold over
the dataset. Error bars are taken to be the
RMS of the integrated charge distribution per channel.
MCP-PMT gains of about $10^6$ are found for PMs~0 and 2-4, who are measured to be about twice the gain of PM~1.
The gain on each channel is used to calibrate the number of detected photo-electrons per event.

\begin{figure}[]
\centering
\subfloat[]{\includegraphics[trim=0cm .9cm 0cm .8cm, scale=.42, clip=true]{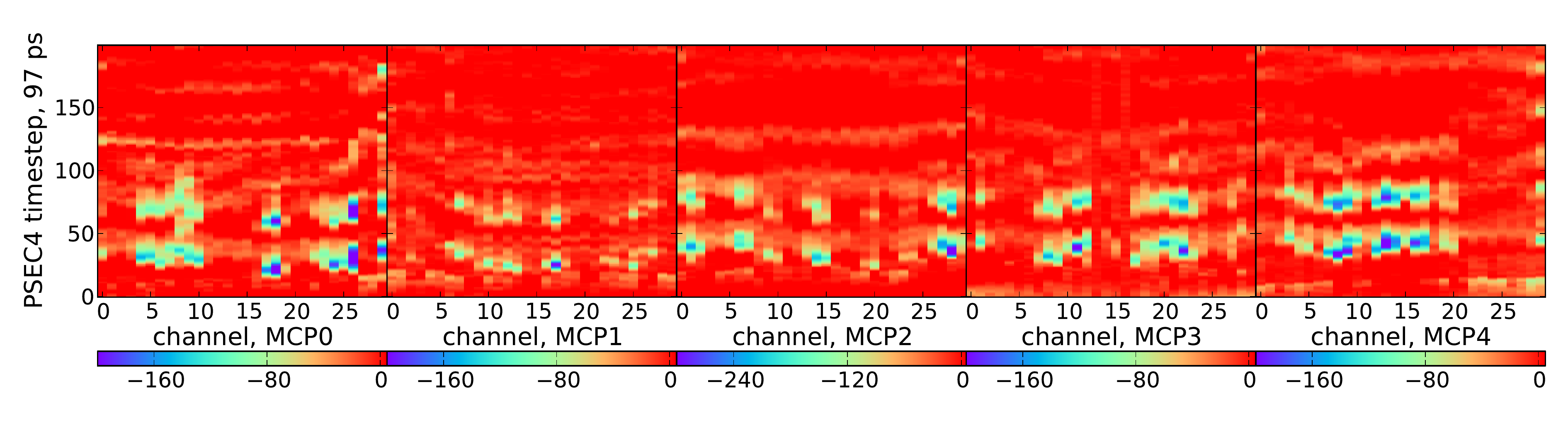}} \\
\vspace{-6 pt}
\subfloat[]{\includegraphics[trim=0cm .9cm 0cm .83cm,scale=.42, clip=true]{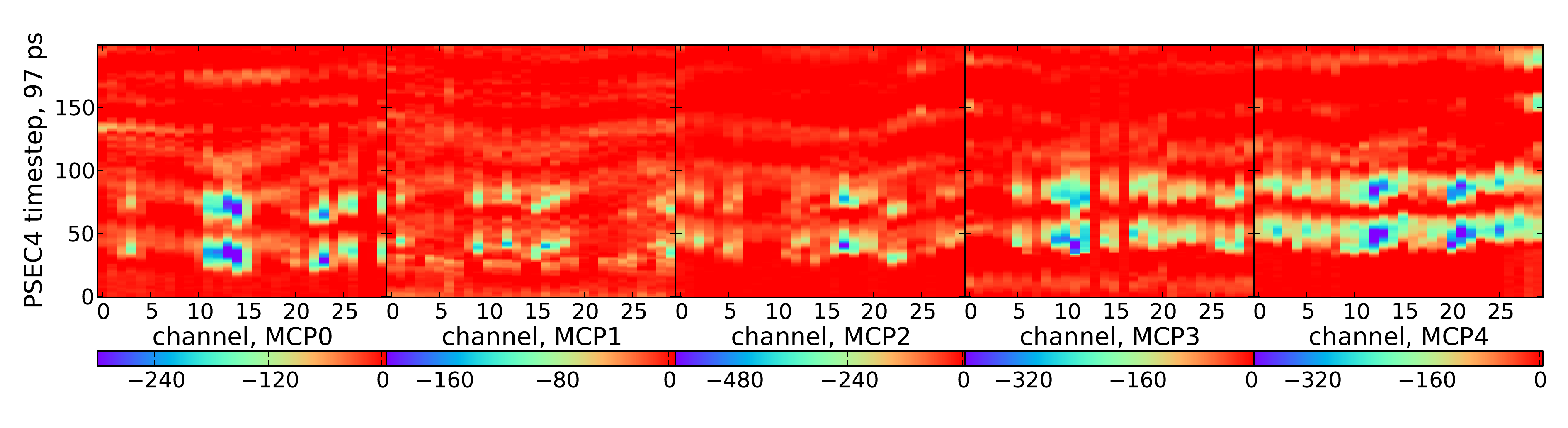}}\\
\vspace{-6 pt}
\subfloat[]{\includegraphics[trim=0cm .9cm 0cm .83cm,scale=.42, clip=true]{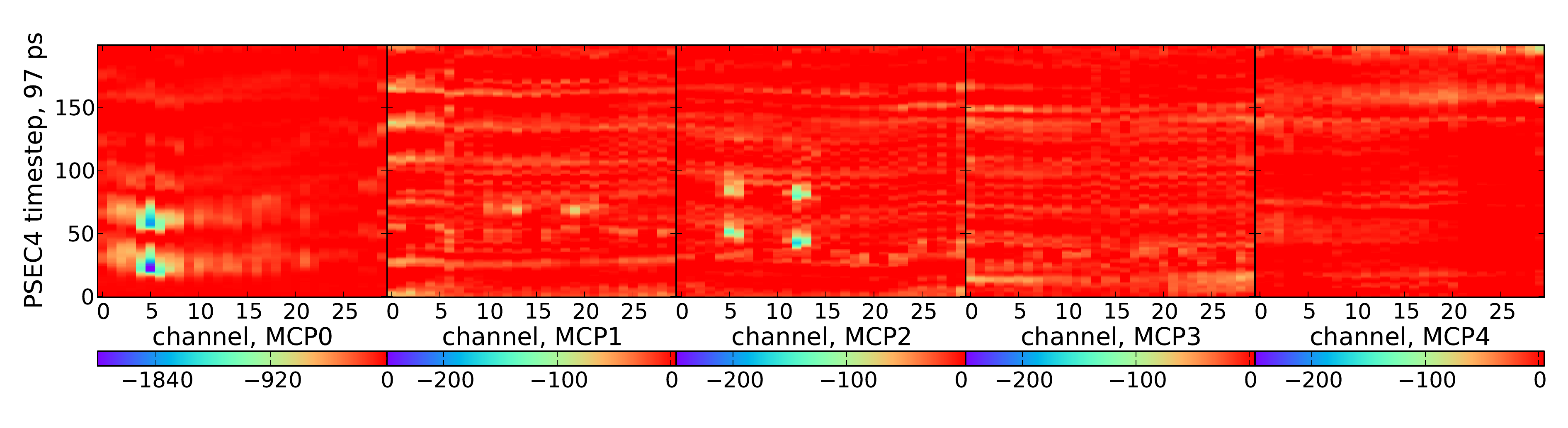}}\\
\vspace{-6 pt} \subfloat[]{\includegraphics[trim=0cm .9cm 0cm
.83cm,scale=.42,
clip=true]{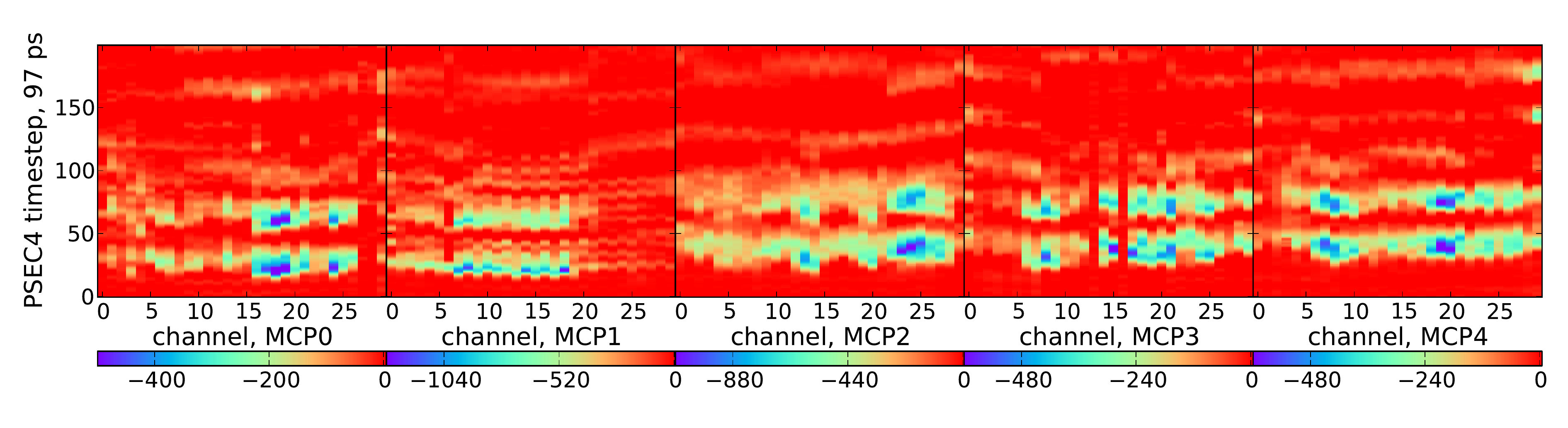}}
\vspace{-10 pt} \caption[Raw event displays]{Raw OTPC events:
time-step (97~ps per bin) vs. OTPC channel. The 5 panels in each
event are the individual OTPC PMs. PMs~0, 2, and 4 constitute the
OTPC normal view, PMs 1 and 3 are in the stereo view. Each panel
is scaled to the maximum signal value with the color value
indicative of the signal amplitude in PSEC4 ADC counts. (a)
Typical through-going event that is representative of most
recorded events. (b) Through-going event, in which the number of
detected photons increases near the right side. (c) Track with a
large number of photons over a short extent. (d) Large signal
amplitude along entire track, peaking in the 2nd panel. Events (c)
and (d) were recorded using trigger configuration 2
($\S$\ref{subsec:beamtrig}). Events (a) and (b) were recorded
using configuration 1, the through-going trigger mode.}
\label{fig:rawevents}
\end{figure}
\section{Photo-electrons along the track}
\label{sec:charge}

Four raw time-projected events are shown in Figure~\ref{fig:rawevents}. 
The PSEC4 time-step is plotted versus the OTPC channel, which is divided into 5 sub-panels representing each PM.
Each sub-panel is auto-scaled with the color value indicative of the signal amplitude in PSEC4 ADC counts.
The beam direction is from left-to-right.
The two time-separated pulses for every signal channel are the direct and anode-reflected waveforms. 
These data are reduced up to step~4 in $\S$\ref{sec:reduce}.

Events (a) and (b) in Fig.~\ref{fig:rawevents} are from the 16~GeV/c data that were recorded using the through-going trigger configuration. Events (c) and (d) are from the 8~GeV/c data, which  were recorded using trigger configuration 2. These two tracks left no signal in $R_2$, implying that these particles interacted in the volume, or else somehow missed the back trigger detector.

\begin{figure}[]
\centering
\includegraphics[scale=.45]{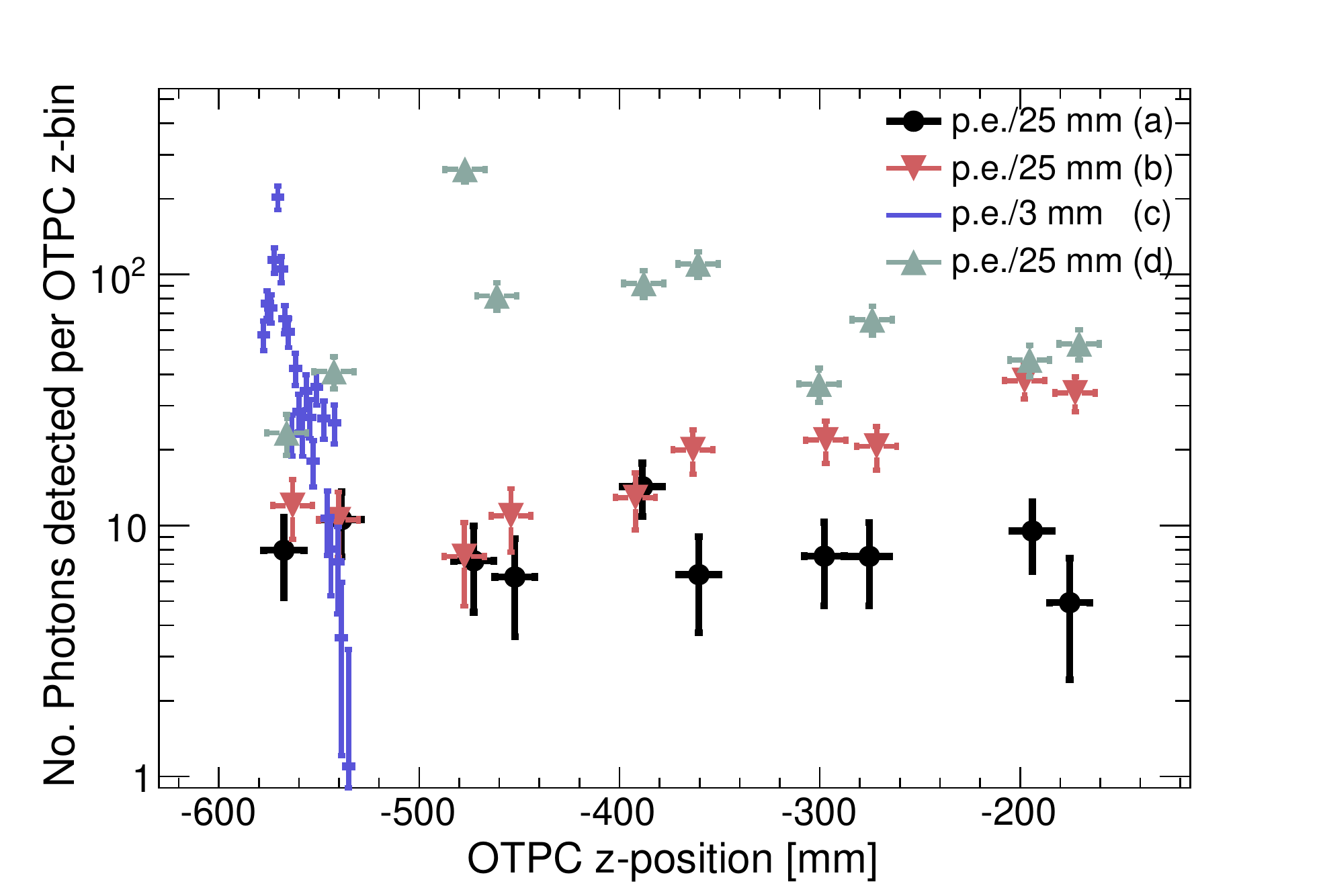}
\caption[Number of detected photons along particle track]{The number of photons detected along the OTPC z-axis, per z-bin defined by the horizontal error bars, for events presented in Fig.~\ref{fig:rawevents}. The gain calibrations shown in $\S$\ref{subsec:gain} were used to convert the waveform integrated charge to a number of photo-electrons (p.e.). Events a, b, and d show the number of p.e.'s per 2.5~cm interval along the OTPC z-axis. Event c shows the number of p.e.'s per 3~mm.}
\label{fig:chargealongtrack}
\end{figure}

By applying the gain calibration described in $\S$\ref{subsec:gain}, we can measure the effective number of photo-electrons along the track.
Figure~\ref{fig:chargealongtrack} shows the number of detected photons along the OTPC z-axis for the events in Fig.~\ref{fig:rawevents}. Event (a) has a roughly uniform number of photo-electrons along the $\sim$40~cm extent of the photo-detection coverage. Event (b) shows an increase in the detected photons in the last one-third of the track, possibly due to the creation of a $\delta$-ray along the track. Events (c) and (d) are distinct in that they exhibit
 a large localized peak in the number of
detected photons ($\textgreater$10$^2$ photons per $\sim$cm),
suggesting that these events may have an EM showering component.
Event (c) could also be explained by an wide-angled muon track
that exited the OTPC volume at PM 0.

\subsection{Comparison of datasets}
\label{subsec:compdataset} 
\begin{figure}[]
\centering \subfloat[]{\includegraphics[trim=0.5cm 0cm 0.99cm
.5cm, height=5.8cm, clip=true]{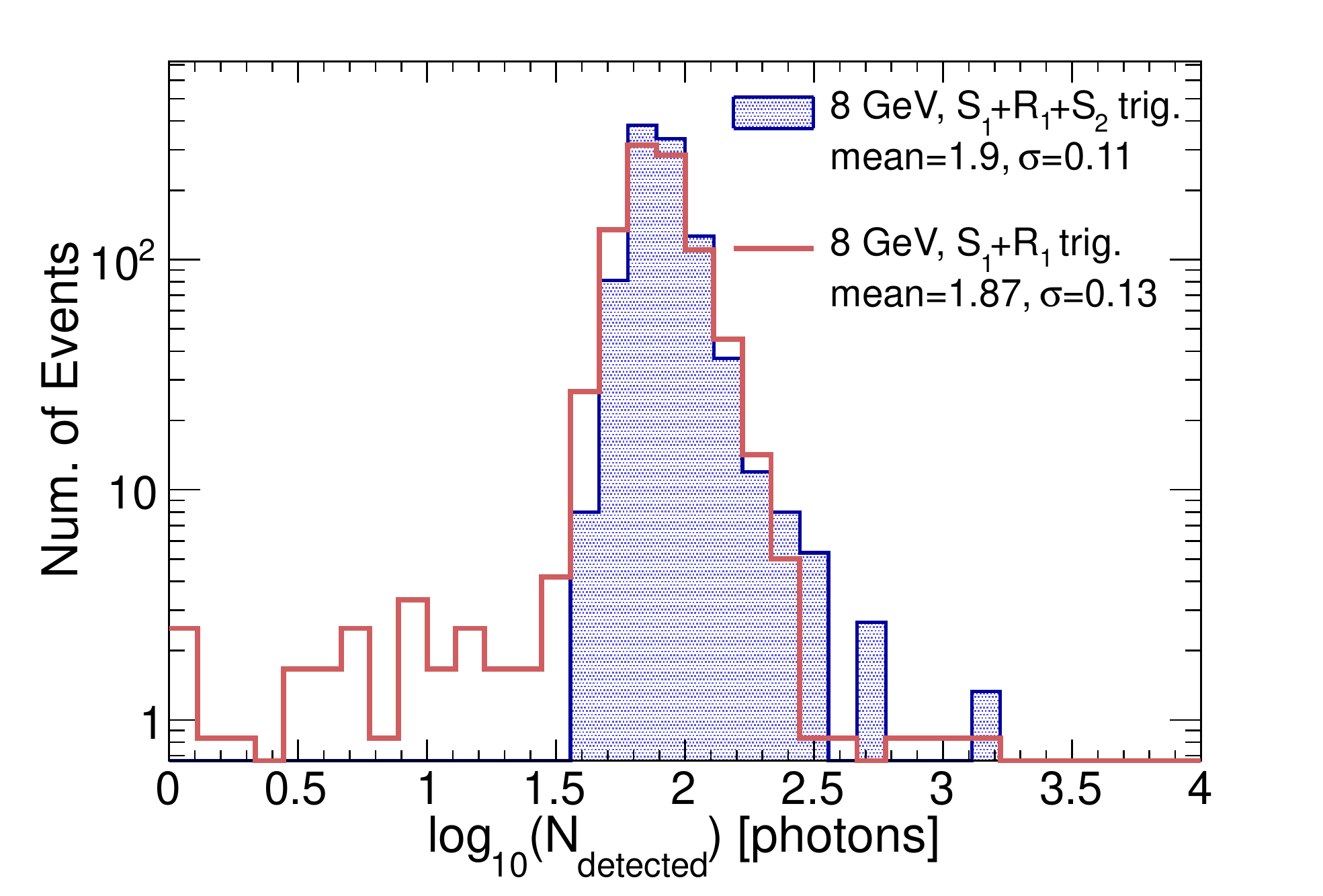}}
\subfloat[]{\includegraphics[trim=0.4cm 0cm 0.99cm
.5cm,height=5.8cm, clip=true]{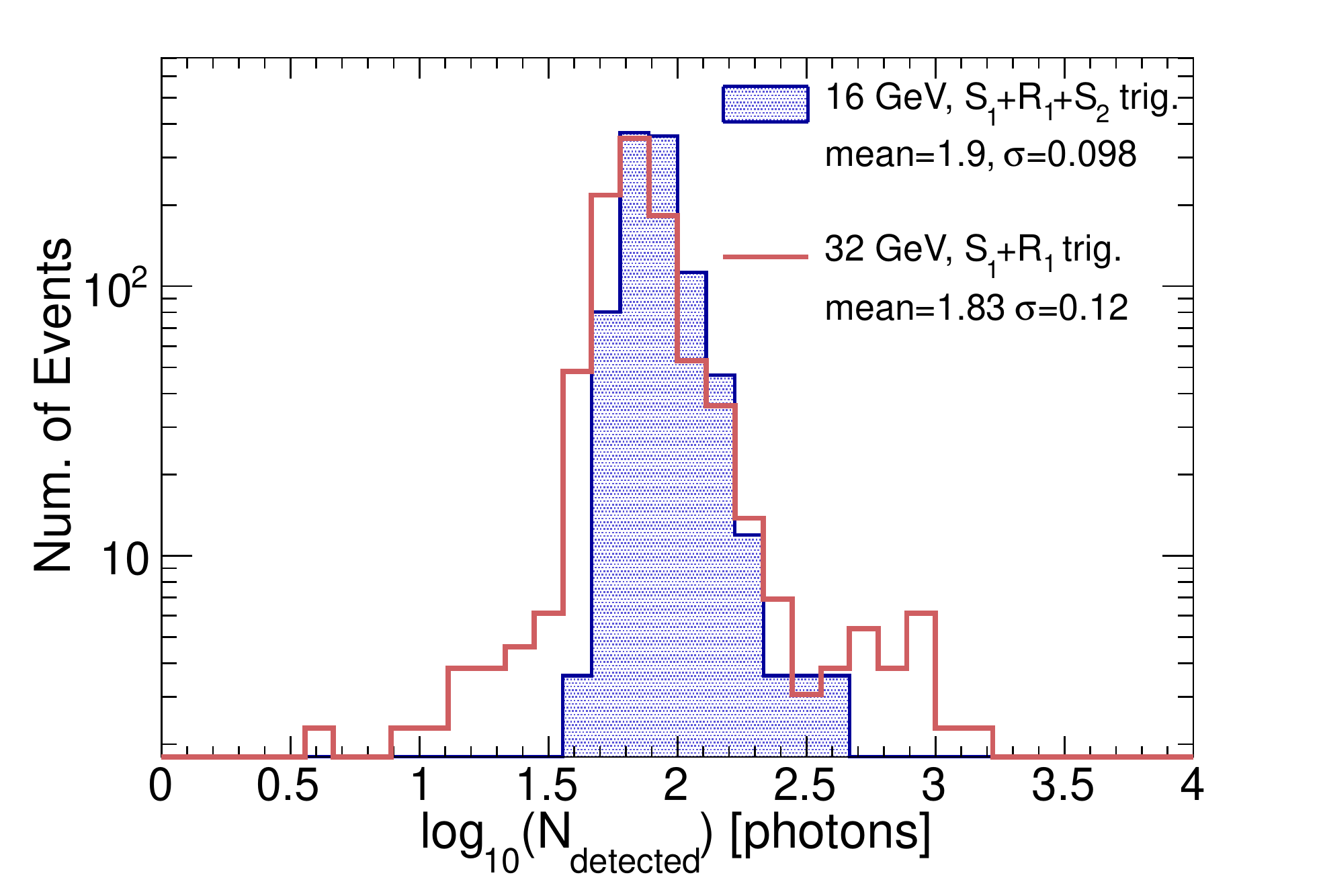}}
\caption[Number of photons detected, comparison between
datasets]{Dataset comparisons of the number of detected photons
per event. (a) The number of photons detected per event in the
8~GeV/c datasets are compared. The data recorded using the
$S_{1}$+$R_{1}$ trigger had poorer optical water quality given by
the 60-day curve in Fig.~\ref{fig:waterattn}. The through-going
triggered dataset was taken with optical water quality at or
better than the 6-hr curve. (b) The number of photons detected per
event in 16 and 32~GeV/c datasets are compared. The 16~GeV/c data
were recorded using the through-going trigger. The 32~GeV/c data
were recorded using the $S_{1}$+$R_{1}$ trigger and with poorer
optical water quality. } \label{fig:histophotons}
\end{figure}

A comparison of the number of detected
photons per event for datasets collected using the two trigger
configurations, at different secondary beam momenta, and at
varying optical water qualities is shown in
Figure~\ref{fig:histophotons}. The datasets that were acquired
using the through-going trigger ($S_{1}$+$R_{1}$+$S_{2}$,
configuration 1) have a well-defined distribution of detected
photons per event, which corresponds to the predominate flux of
multi-GeV muons that satisfy this trigger condition. Datasets
acquired using the second trigger configuration ($S_{1}$+$R_{1}$)
did not require a through-going particle. Though having a
different trigger, the majority of $S_{1}$+$R_{1}$ events do pass
through the OTPC and have a number of detected photons consistent
with the through-going trigger. There are also several events with
a large spread in the detected number of photons that are not seen
in the through-going triggered events.

Figure~\ref{fig:histophotons}a compares 8~GeV/c datasets that were recorded
using different trigger configurations (Datasets 1 and 2). Dataset 1 used the through-going trigger configuration.
 The number of photons per event in the dataset 1 is 79$\pm$20 at 1$\sigma$ from a
normal fit to the distribution.
The second 8~GeV/c dataset, triggered in the second configuration ($S_{1}$+$R_{1}$), has an expected number of photons of 74$\pm$25 with a fit over the same range.
The slight difference in the mean may be attributed to the poorer water quality
in the second dataset. Dataset 2 has a number of events with fewer than $\sim$30 detected
photons that are not observed in the through-going trigger dataset.

A similar comparison is made between a data collected at 16~GeV/c (Dataset 3) 
and 32~GeV/c (Dataset 4) in Fig~\ref{fig:histophotons}b. 
Dataset 3 was recorded using the through-going trigger. 
The number of photons per event in dataset 3 is 79$\pm$18 from a 
normal distribution fit, which corresponds to the photo-detection 
efficiency measurement in Figure~\ref{fig:Efficiency}. 
Dataset 1 has a core expectation of 67$\pm$22, but a larger 
overall variance from events that are unique to the $S_{1}$+$R_{1}$ 
trigger configuration. The downshift in the normal distribution 
between datasets may again be due to the poorer OTPC water quality when acquiring dataset 1.

\section{The time-projection}
\label{sec:timing}
In this and the following sections we use 1230 events from a secondary beam momentum of 16~GeV/c dataset,
which were triggered on through-going tracks at the best measured water quality.

\begin{figure}[]
\centering
\includegraphics[trim=0cm .25cm 0cm .4cm, clip=true, scale=.45]{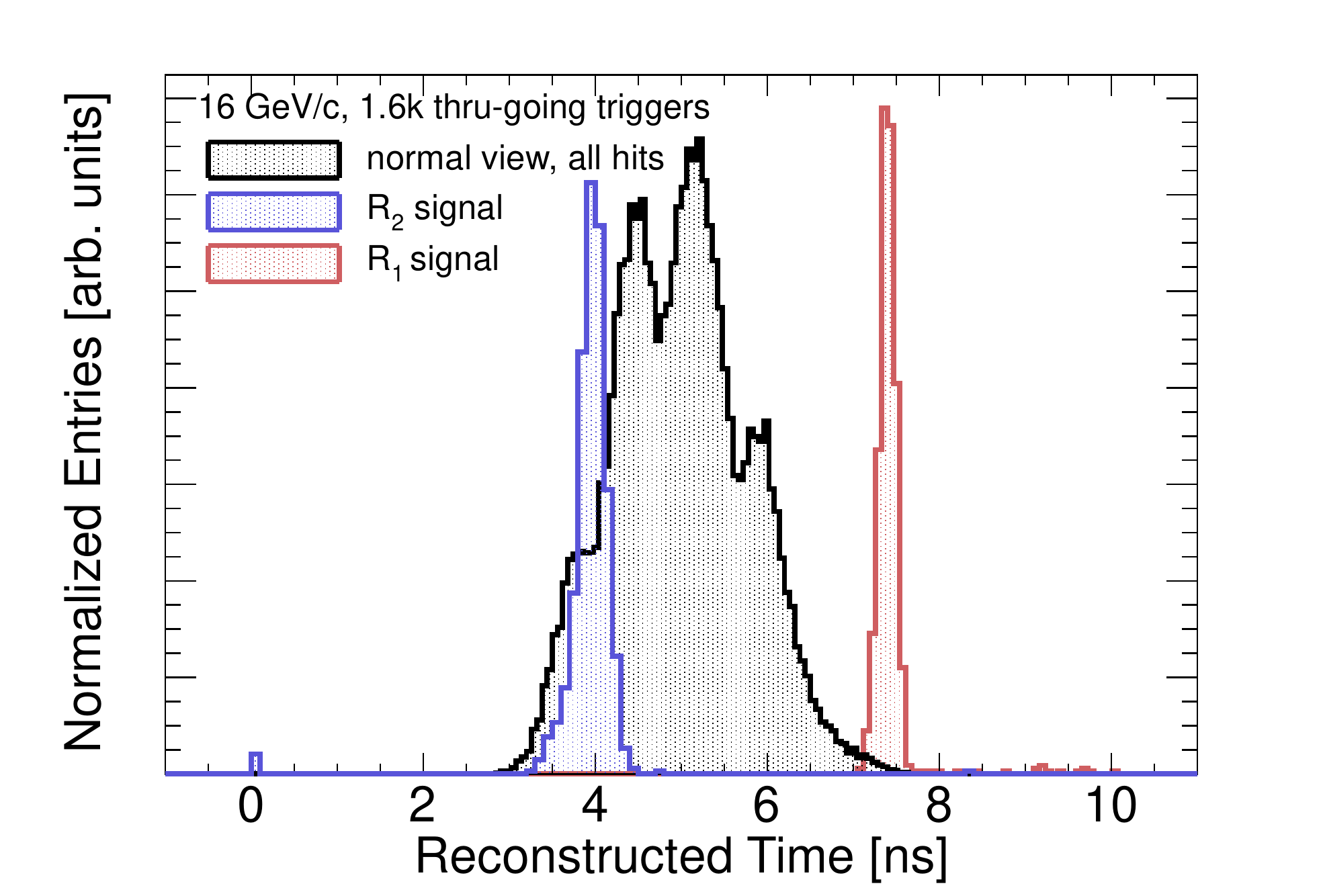}
\caption[Reconstructed event timing]{The times of the data and trigger signals from the reduced 16~GeV/c dataset. The reconstructed times of the detected Cherenkov photons (from the normal-view)  are shown by the black histogram. The `fast' trigger signals, $R_1$ and $R_2$ are shown in the red and blue, respectively. Each event is aligned with respect to the $R_1$ time. The time of $R_1$ is measured later than the $R_2$ signal due to cable delays. }
\label{fig:alltimes}
\end{figure}

We now add the time dimension of the detected photons, as
extracted from the waveforms in $\S$\ref{sec:features}, to the
data analysis. Figure~\ref{fig:alltimes} shows the measured time
structure of the events, in which the times-of-arrival of the
Cherenkov photons on the OTPC normal view are plotted with the
measured times of the beam trigger signals, $R_{1}$  and $R_{2}$.
The time structure seen in the Cherenkov photon data is due to the
three discrete PM locations in the OTPC normal view.

\subsection{Resolving the direct and mirror-reflected Cherenkov photons}
\label{subsec:resolvedm}

As shown in Equation~\ref{eqn:tproj}, the particle's time-projection along the longitudinal axis has two components: one from the particle's velocity and the other from its angle with respect to the OTPC z-axis. For through-going multi-GeV muons, we can assume the particle's velocity is constant along the OTPC with $\beta$=1. We introduce a rotated time basis, $t'$, which nulls out the velocity term in Eqn.~\ref{eqn:tproj}
\begin{subequations}
\begin{equation}
t'_{i} = t_{i} - \frac{z_{i}}{ c}
\end{equation}
\begin{equation}
\frac{dt'}{dz} = \frac{dt}{dz} - \frac{1}{c} = \frac{\tan\theta}{<v_{group}>}
\end{equation}
\label{eqn:rotatetime}
\end{subequations}
where $t_i$ and $z_i$ are the measured individual photon times and z-positions. As defined in $\S$\ref{sec:Detoptics}, $<v_{group}>$ is the weighted average of the group velocity dispersion over the optical efficiency range of the detector.
 In the $t'$ basis, the time-projection along the OTPC z-axis is a measure of the particle angle.

\begin{figure}[]
\centering
\includegraphics[scale=.45]{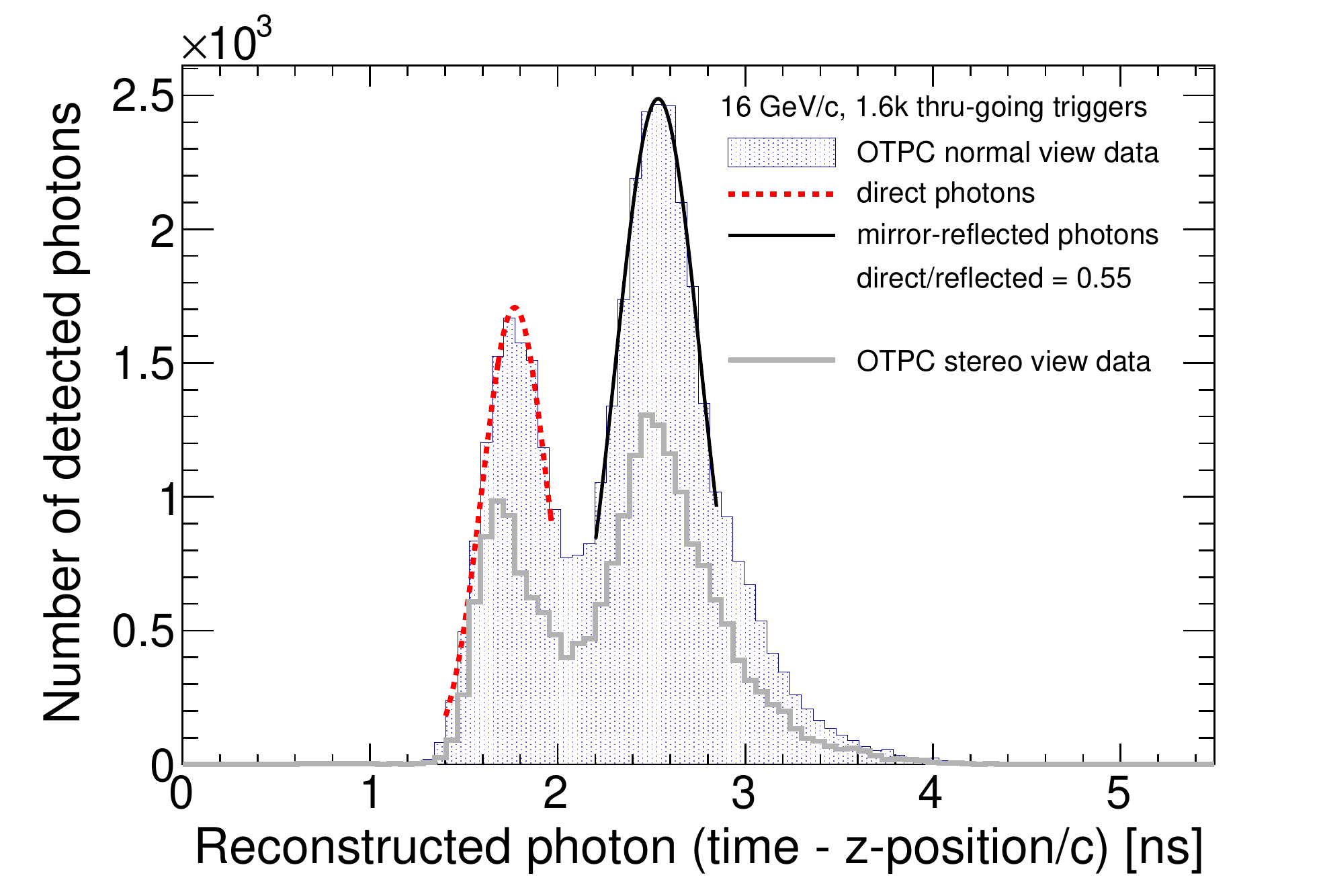}
\caption[Time-resolving the direct and mirror-reflected Cherenkov
photons]{Histogram of the inclusive dataset of detected photons in
a rotated time basis: \begin{math} t'_{i} = t_{i}-z_{i}/c
\end{math}. This removes the contribution to the time-projection
along the z-axis due to the particle's velocity (assuming
$\beta$=1). The direct and mirror-reflected Cherenkov photons are
distinctly visible and separated by 770~ps, as taken from the fit
parameters.} \label{fig:direct-mirror-all}
\end{figure}

Figure~\ref{fig:direct-mirror-all} shows the distribution of the
measured photon times in terms of their rotated time, $t'_{i}$.
The $R_2$ trigger time for each track serves as a fine
time-alignment between events. As expected, the $t'$ distribution
is bi-modal with peaks corresponding to the direct and
mirror-reflected photons. A distinct separation of 770~ps is
observed, which is consistent with the mean path length difference
between the direct and reflected photons divided by the photon
drift velocity, $<v_{group}>$. The path length difference equal to
the detector diameter, 18$\pm$1~cm ($\S$\ref{subsec:trackeqn}),
corresponds to an expected time difference between the direct and
mirror-reflected photons of 830$\pm$50~ps.


Figure~\ref{fig:direct-mirror-all} also shows a larger number of
mirror-reflected photons than direct photons. Using the relative
areas from the Gaussian fits to the distributions, we measure
fewer direct than reflected photons by a factor of 0.55 in the
normal view data. This was unexpected; the detector Monte Carlo
predicts approximately equal numbers of each for small-angle
tracks. The relative deficit of direct photons is most likely due
to reflections of the direct light at the fused-silica OTPC port
and MCP-PMT interface as discussed in Table~\ref{tab:photoeff}. It
is also likely that there is a small percentage of scattered
photons in the data; these would arrive later than the direct
photons and may be included in the mirror-reflected distribution
and also cause the observed tail in the $t'$ timing distribution
after 3~ns.

\subsection{An example event}
\label{subsec:timing}
\begin{figure}[]
\centering
\includegraphics[trim=0cm .25cm 1cm .4cm, clip=true,height=6.6cm]{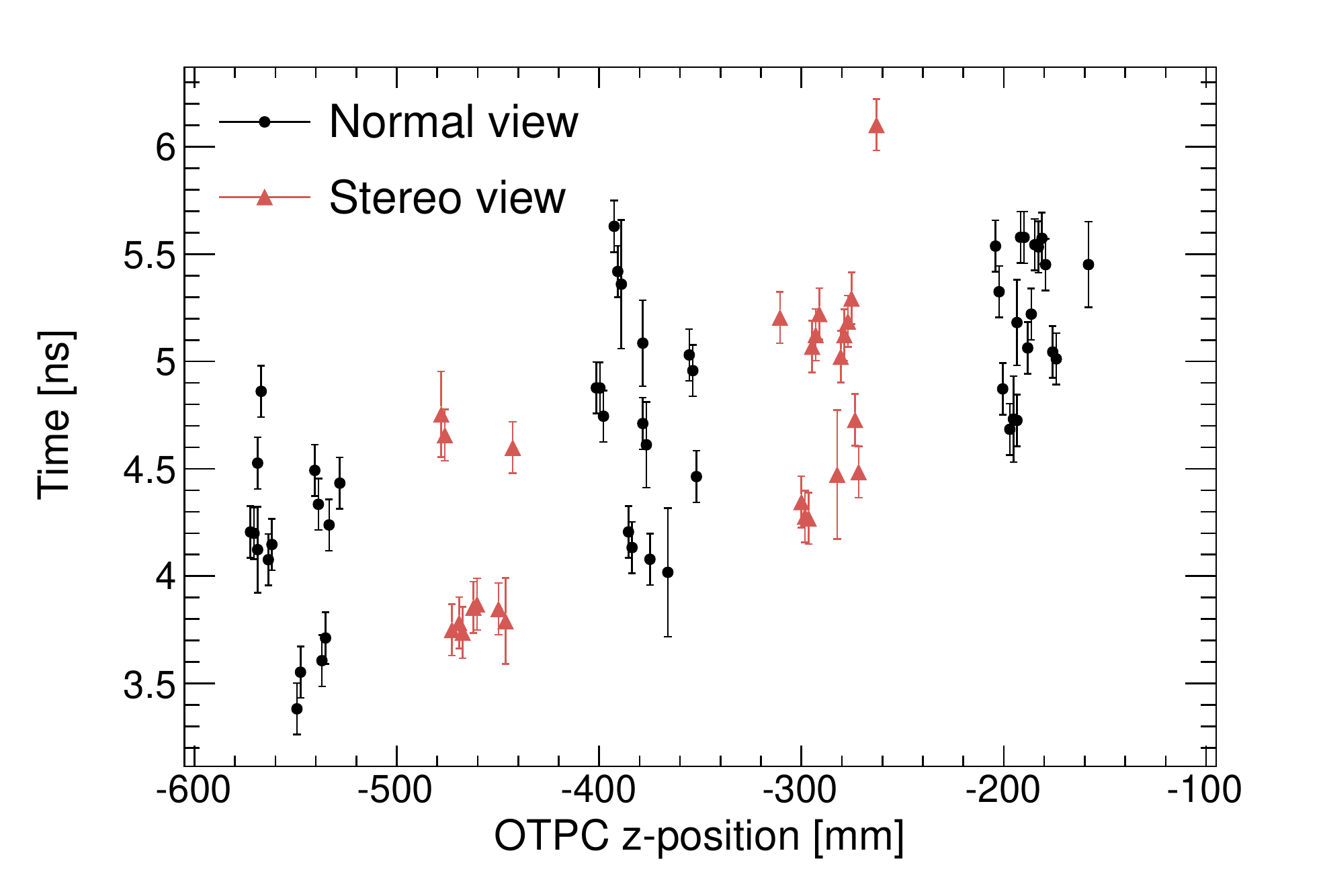}
\caption[Time projection on the OTPC z-axis, single event]{The time-projection along the beam axis for a single event. The time projection along the z-axis is shown for the event shown in Figure~\ref{fig:rawevents}a using 73 individually resolved Cherenkov photons.}
\label{fig:tprojection}
\end{figure}

\begin{figure}[]
\centering
\includegraphics[trim=6cm 4.8cm 7.8cm 5.5cm, clip=true,height=6.4cm]{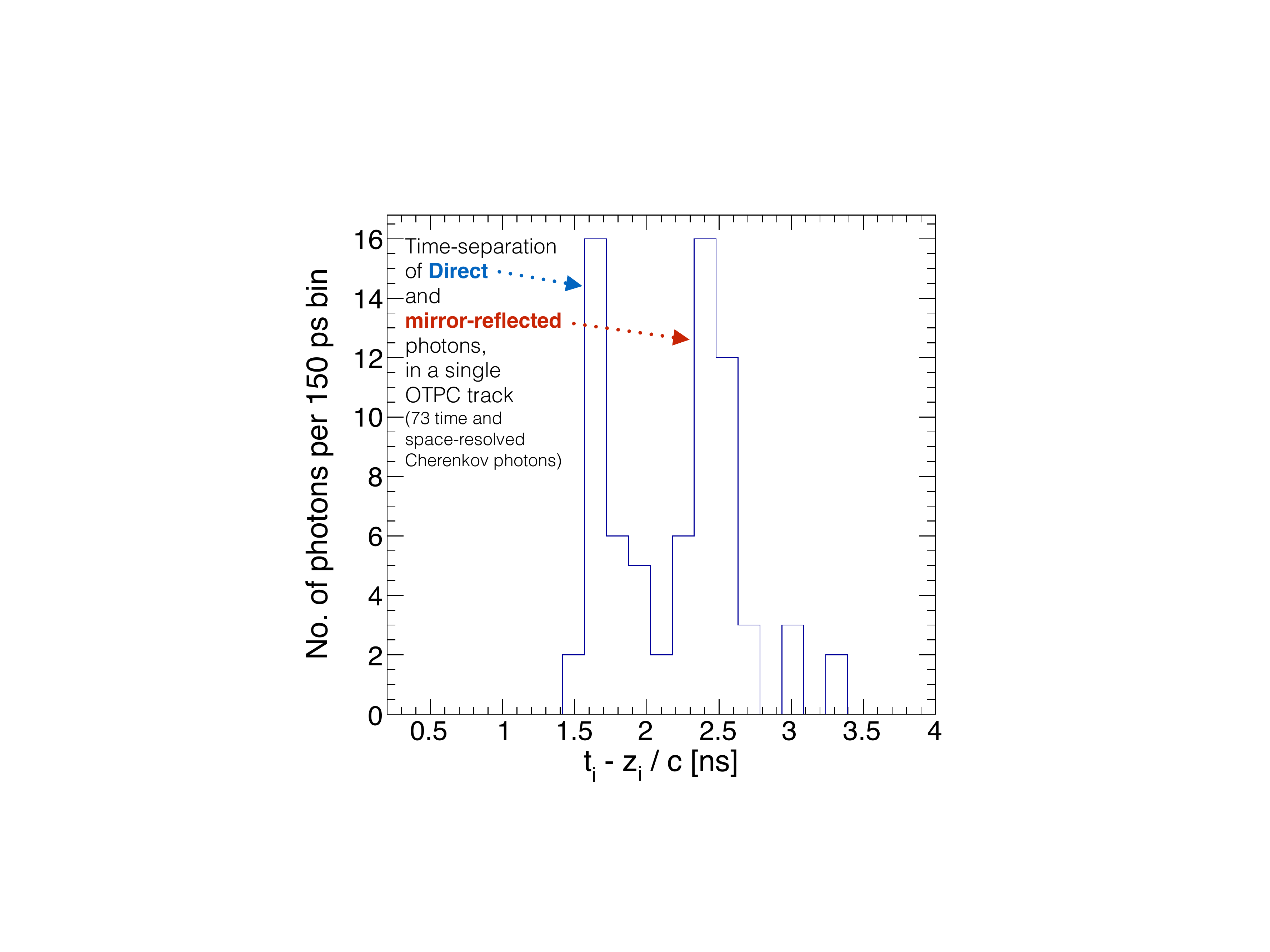}
\caption[Timing separation of direct and mirror-reflected light, single event]{Histogram of the detected photon times from the event in Fig.~\ref{fig:tprojection}, plotted in the rotated basis introduced in Fig.~\ref{fig:direct-mirror-all}. The separated peaks in the time distribution are from the direct and mirror-reflected Cherenkov photons.}
\label{fig:tprojection_sep}
\end{figure}

Figure~\ref{fig:tprojection}a shows the time-projection on the z-axis for the raw event shown in Fig.~\ref{fig:rawevents}a.
In this event, 73 individual photons were resolved in both time and space. Figure~\ref{fig:tprojection} shows their measured times, $t_i$, projected versus their detected positions, $z_i$.
The photons detected in the normal and stereo views are denoted by black and red data-points, respectively. Figure~\ref{fig:tprojection_sep} is a histogram of the rotated times,  $t'_i$, of this event. The peaks in the histogram are from the measured arrival-times of the direct and mirror-reflected photons.

\section{Angular resolution}
\label{sec:angle}

The direct Cherenkov photons can be used to measure the particle angle with respect to the OTPC z-axis as shown in Eqn.~\ref{eqn:rotatetime}. The most straightforward method of extracting this angle is to assume a linear track over the extent of the OTPC photodetector coverage. The slope of the linear fit, multiplied by the weighted average of the group velocity ($<v_{group}>$), is the tangent of the track angle.

\begin{figure}[]
\centering
\subfloat[]{\includegraphics[trim=0cm .1cm 1cm .4cm, clip=true,height=5.7cm]{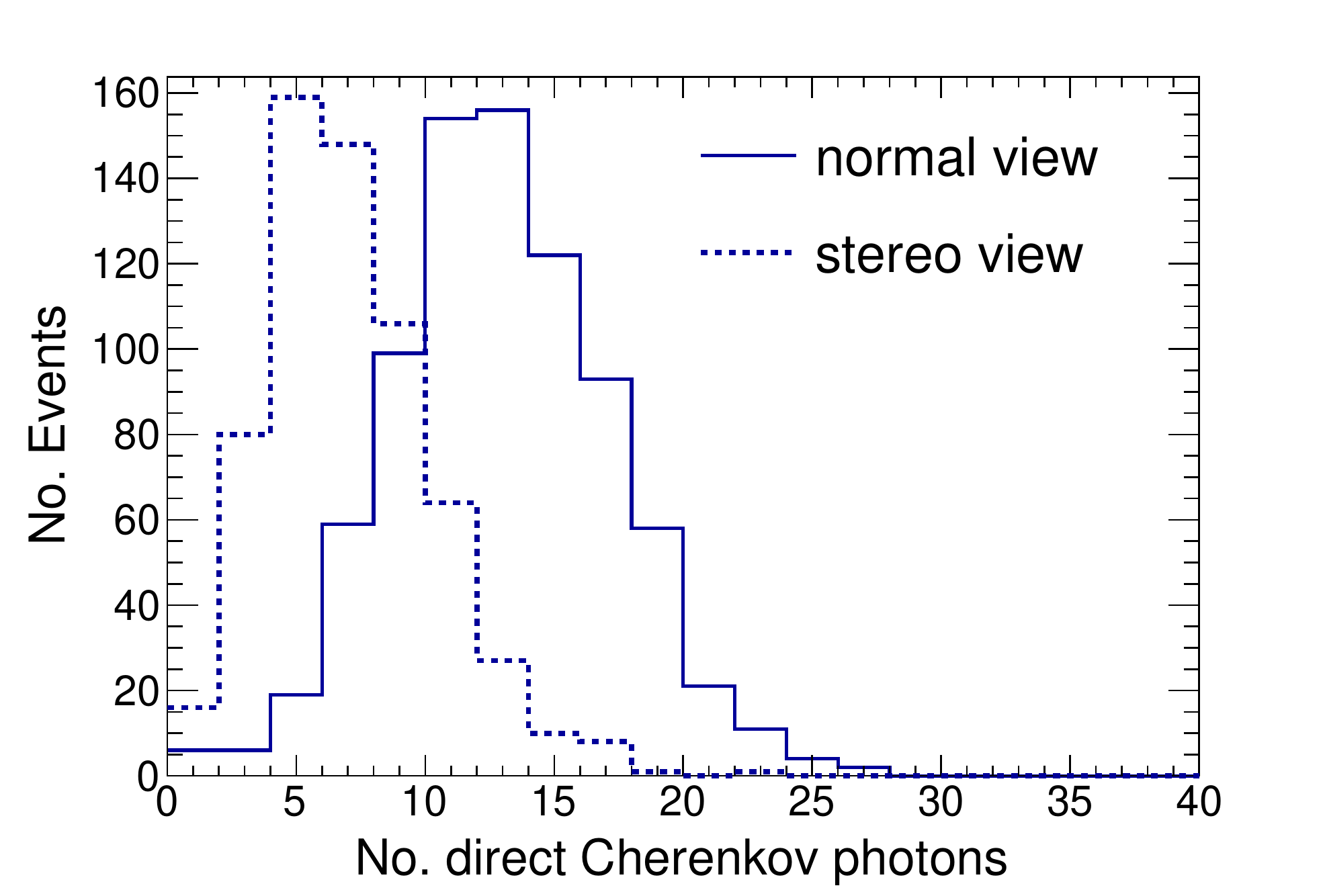}}
\subfloat[]{\includegraphics[trim=.5cm .1cm 1cm .4cm, clip=true,height=5.7cm]{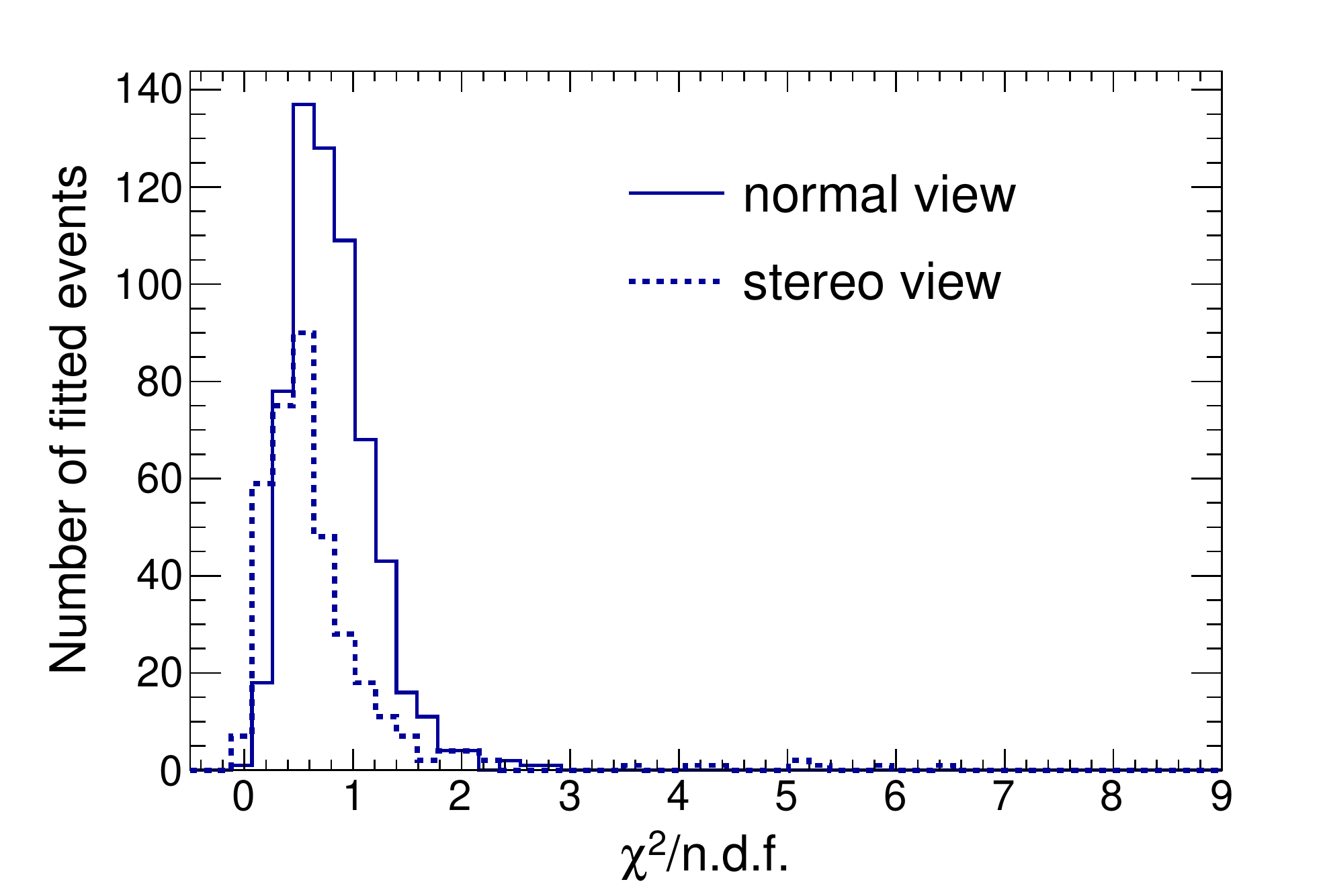}}

\caption[(a) Number of direct Cherenkov photons, (b) linear fit
$\chi$$^2$/n.d.f.]{ (a) Number of direct Cherenkov photons per
event.  The direct and mirror-reflected photons are separated by
applying a time-cut to the data presented in
Fig.~\ref{fig:direct-mirror-all}. (b) Goodness of fit as given by
the $\chi$$^2$ per number of degrees of freedom (n.d.f.). Events
from (a) with more than 8 normal-view and more than 4 stereo-view
direct photons are used in the fit. } \label{fig:angleinfo}
\end{figure}

It is possible to separate the direct and mirror-reflected photons
when the data are in the rotated time basis, $t'$, as shown in
Figure~\ref{fig:direct-mirror-all}. A normal view time-cut of
$\textless$1.95~ns and a stereo view time-cut of
$\textless$1.90~ns were used on the data to isolate the direct
photons. The number of direct Cherenkov photons per event is shown
in Figure~\ref{fig:angleinfo}a. The sample fitted consists of 522
events with more than 8 normal-view and more than 4 stereo-view
direct photons are fit, requiring that each PM has at least two
direct photons.  Separate linear fits to the direct stereo and
normal view photons are performed. The fits are reasonable, as
shown by the distributions of $\chi$$^2$ per degree of freedom
 in Figure~\ref{fig:angleinfo}b.

\begin{figure}[]
\centering
\includegraphics[scale=.46]{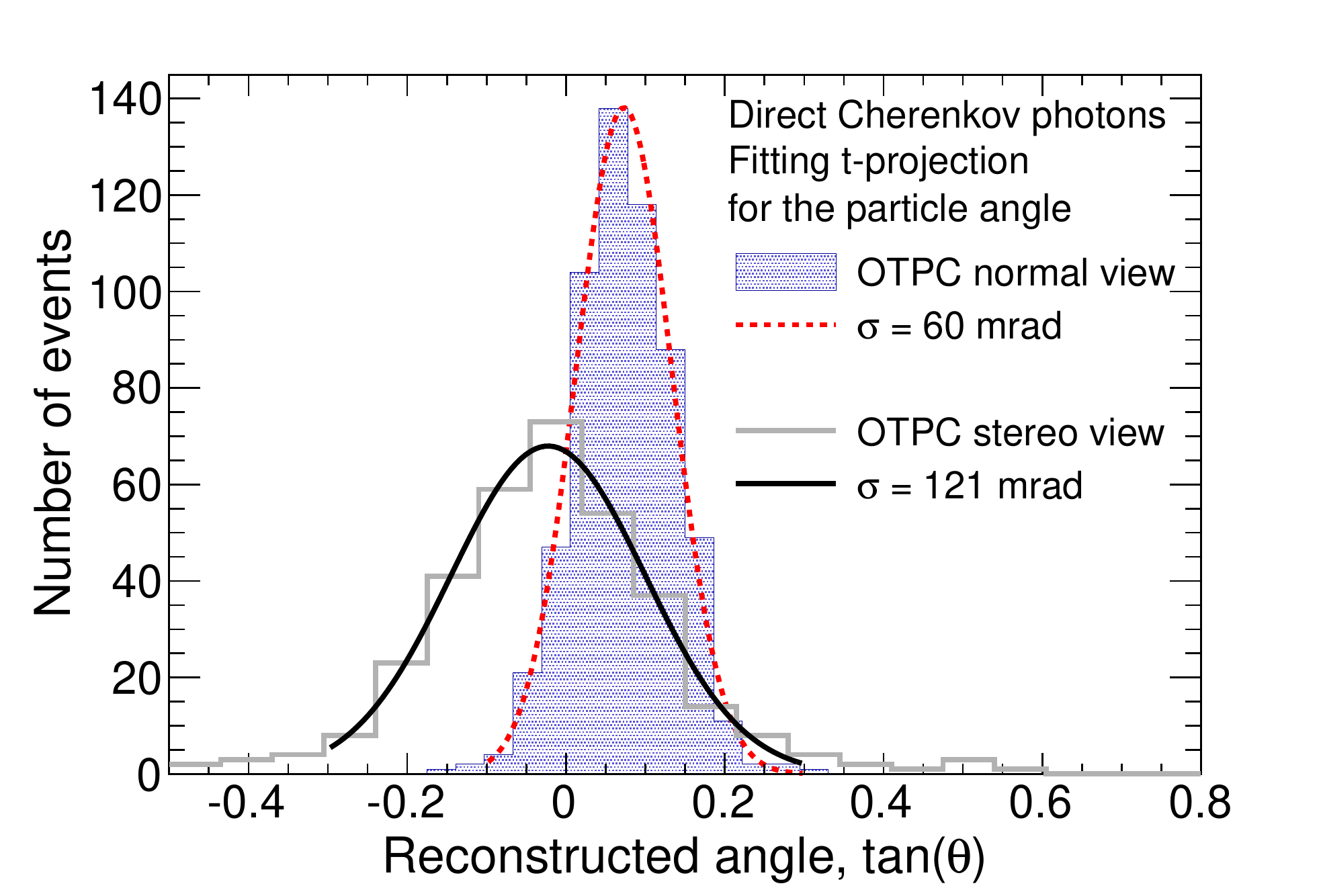}
\caption[Angular resolution]{Measured angle of tracks along the z-axis as seen by the stereo and normal views.
The direct Cherenkov light is fit with a straight line for events with more than 8 direct normal-view photons.
In the stereo view, at least 4 photons are required to fit the angle.
The measured angle distributions are: 73$\pm$60~mrad for the normal view, -20$\pm$121~mrad in the stereo view.
Fits that have a  $\chi$$^2$/n.d.f greater than 3, as shown in Fig.~\ref{fig:angleinfo}b, were not included in this measurement. }
\label{fig:angle}
\end{figure}

The returned slopes from the fits, \begin{math} \frac{dt'}{dz}
\end{math}, are multiplied by $<v_{group}>$=(c/1.38) and
histogrammed in Figure~\ref{fig:angle}. Linear fits to the photons
in the direct normal-view yield an angle relative to the nominal
beam axis over all the fitted tracks of 73$\pm$60~mrad at
1$\sigma$ resolution.  The stereo view shows a measurement of
-20$\pm$121~mrad. The stereo view measurement has a larger error
that is most likely due to its smaller z-coverage than the normal
view, which gives it a smaller lever arm on the angular
measurement. The normal view has 3 equally spaced PMs over a
z-range of $\sim$40~cm, but the stereo view has 2~PMs covering a
z-range of only 20~cm.

\begin{figure}[]
\centering
\includegraphics[height=5.5cm]{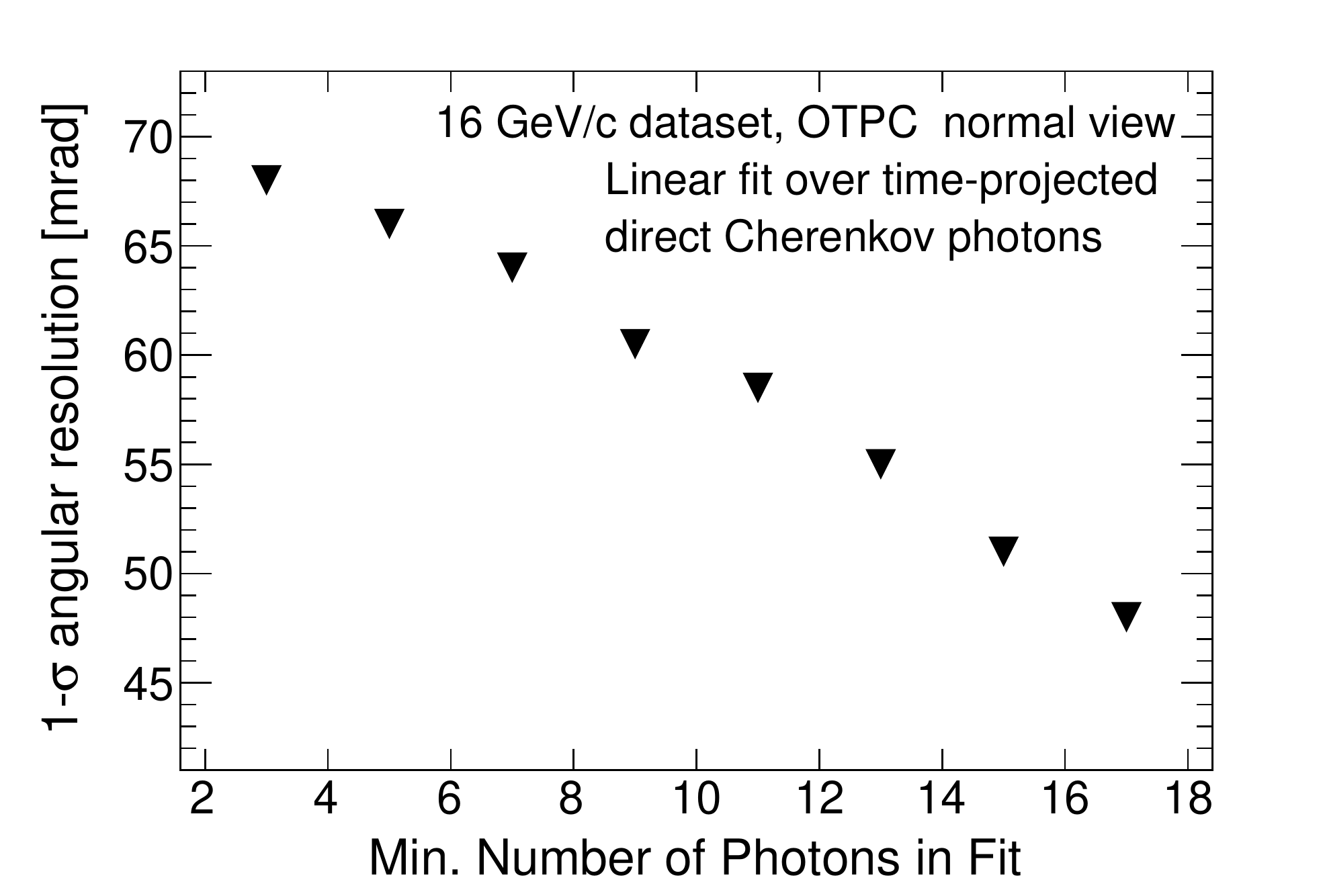}
\caption[1$\sigma$ angular resolution vs. num. of fitted photons]{The 1$\sigma$ angular resolution as a function of the minimum number of direct normal view photons included in the fit. The measurement in Fig.~\ref{fig:angle} is shown by the data-point with a minimum of 9 photons. An angular resolution of 48~mrad is achieved when using only events with greater than 17 direct photons in the normal view (99 events in the 16~GeV/c dataset). }
\label{fig:moreangle}
\end{figure}

A study of the angular resolution versus the number of direct
Cherenkov photons in the linear fit is shown in
Figure~\ref{fig:moreangle}. As a higher minimum number of photons
is required in the fit, the resolution is improved. Ninety-nine
events in the dataset have more than 17 discrete normal view
photons from the direct Cherenkov light. Fitting only these events
results in a 1$\sigma$ resolution of 48~mrad.

\section{Spatial resolution}
\label{sec:3d}

To fully reconstruct charged particle tracks, the mirror-reflected
photons are used in combination with the direct Cherenkov light. A
second time cut is made on the data in the rotated time basis,
$t'$ (Fig.~\ref{fig:direct-mirror-all}), to isolate the reflected
Cherenkov photons. A normal view time-cut of $\textgreater$2.05~ns
and a stereo view time-cut of $\textgreater$2.0~ns were used on
the data. A time-gap of 100~ps between the direct and reflected
photon time-cuts was used to separate the two regions.

\begin{figure}[]
\centering
\subfloat[]{\includegraphics[trim=0cm .1cm 1cm .6cm, clip=true,height=5.6cm]{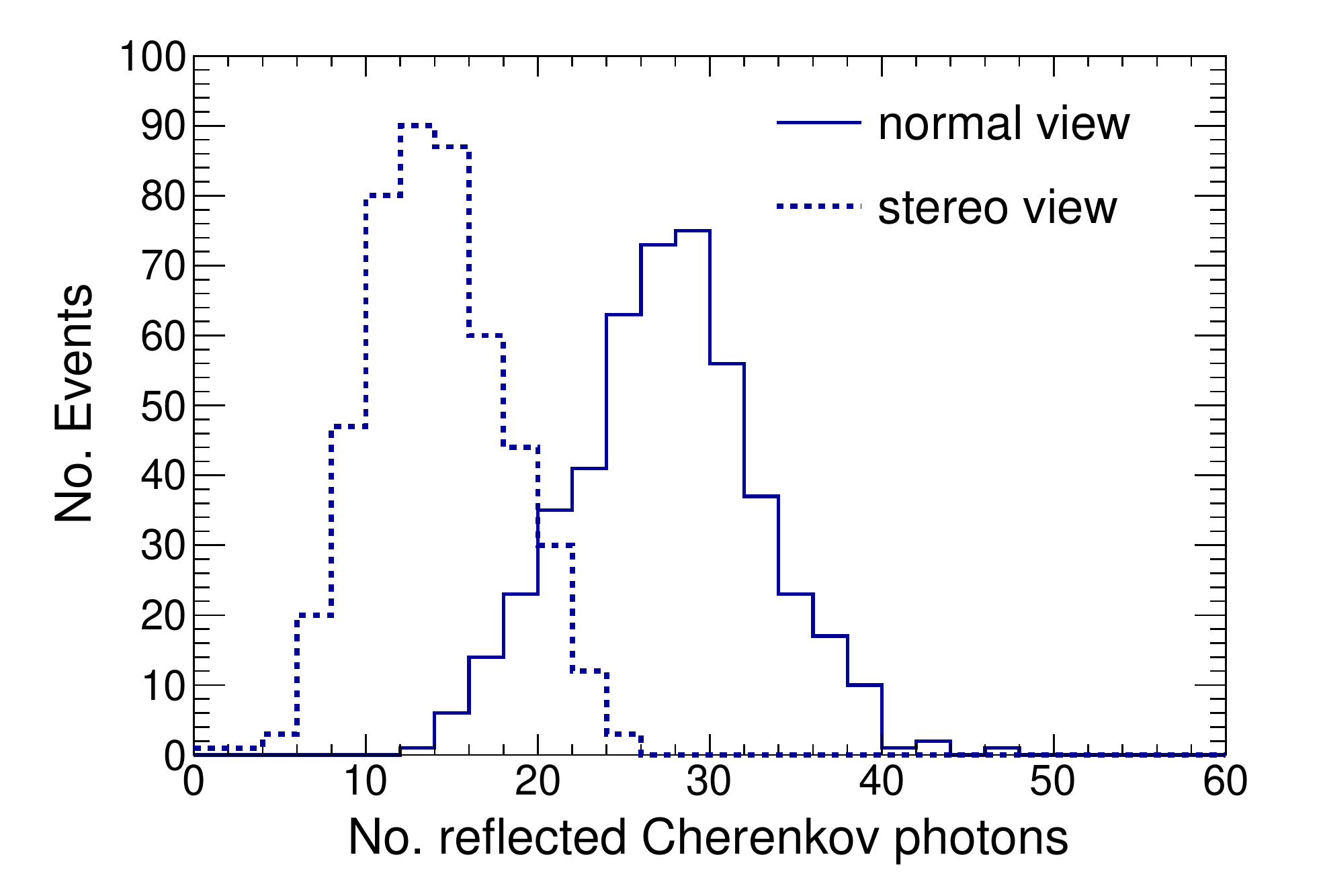}}
\subfloat[]{\includegraphics[trim=0cm .1cm 1cm .4cm,
clip=true,height=5.6cm]{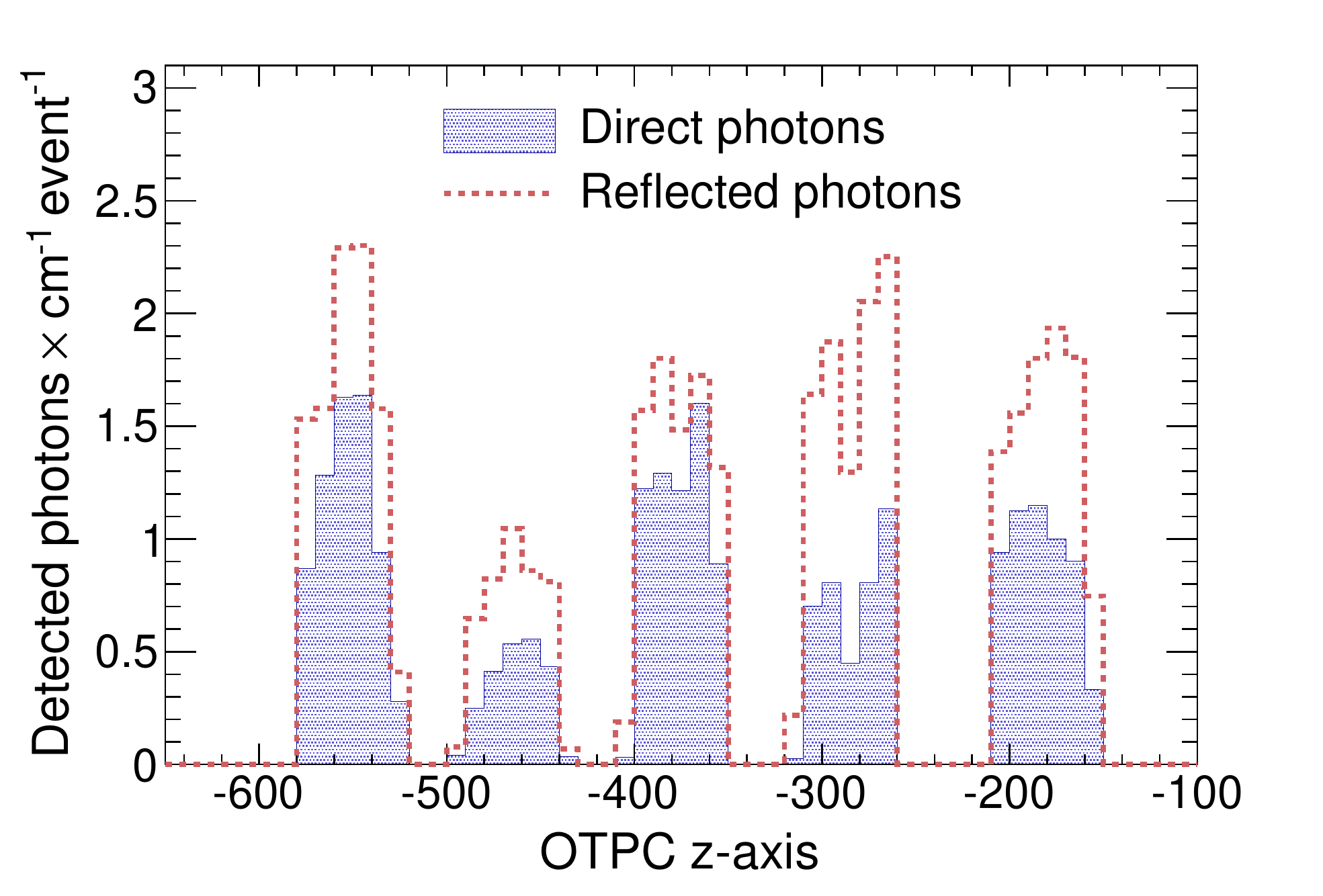}}
\caption[(a) Number of reflected Cherenkov photons (b) detected
photons per-z pos.]{(a) Number of reflected Cherenkov photons per
event.  The direct and mirror-reflected photons are separated by
applying a time-cut to the data presented in
Fig.~\ref{fig:direct-mirror-all}. The number of direct photons per
event is shown in Fig.~\ref{fig:angleinfo}a. (b) Number of direct
and reflected photons detected per event per cm. The five discrete
distributions along the OTPC z-axis are the 5~PM locations (left
to right, PM 0 to 4). } \label{fig:reflect}
\end{figure}

The number of reflected photons per event in the normal and stereo
views is shown in Figure~\ref{fig:reflect}a. The relative number
of direct and reflected detected photons at each PM per event per
cm is shown in Fig.~\ref{fig:reflect}b. As expected from
Fig.~\ref{fig:direct-mirror-all}, there is a preponderance of
reflected over direct photons.

A straight-forward way to reconstruct the spatial position of the
trajectory is to take the difference of the mean direct photon
arrival times and the mean reflected photon arrival times in the
$t'$ basis. Explicitly, the time difference at a PM for a given
event a, $\Delta t_{PM}$, using this method is
\begin{equation}
\Delta t_{PM} = \frac{1}{n}\sum_{i=1}^{n}t^{' mirror}_{i} - \frac{1}{m}\sum_{j=1}^{m}t^{' direct}_{j} - t_{0}
\label{eqn:dumb}
\end{equation}
where $n$ is the number of reflected photons per event per PM and  $m$ is the number of direct photons per event per PM. A time reference for the event, $t_{0}$ (the $R_2$ trigger signal time), is subtracted from each event for alignment. This time-difference measurement is inserted into Eqn.~\ref{eqn:mirror2} to obtain the lateral displacement of the particle track.

\begin{figure}[]
\centering
\includegraphics[scale=.5]{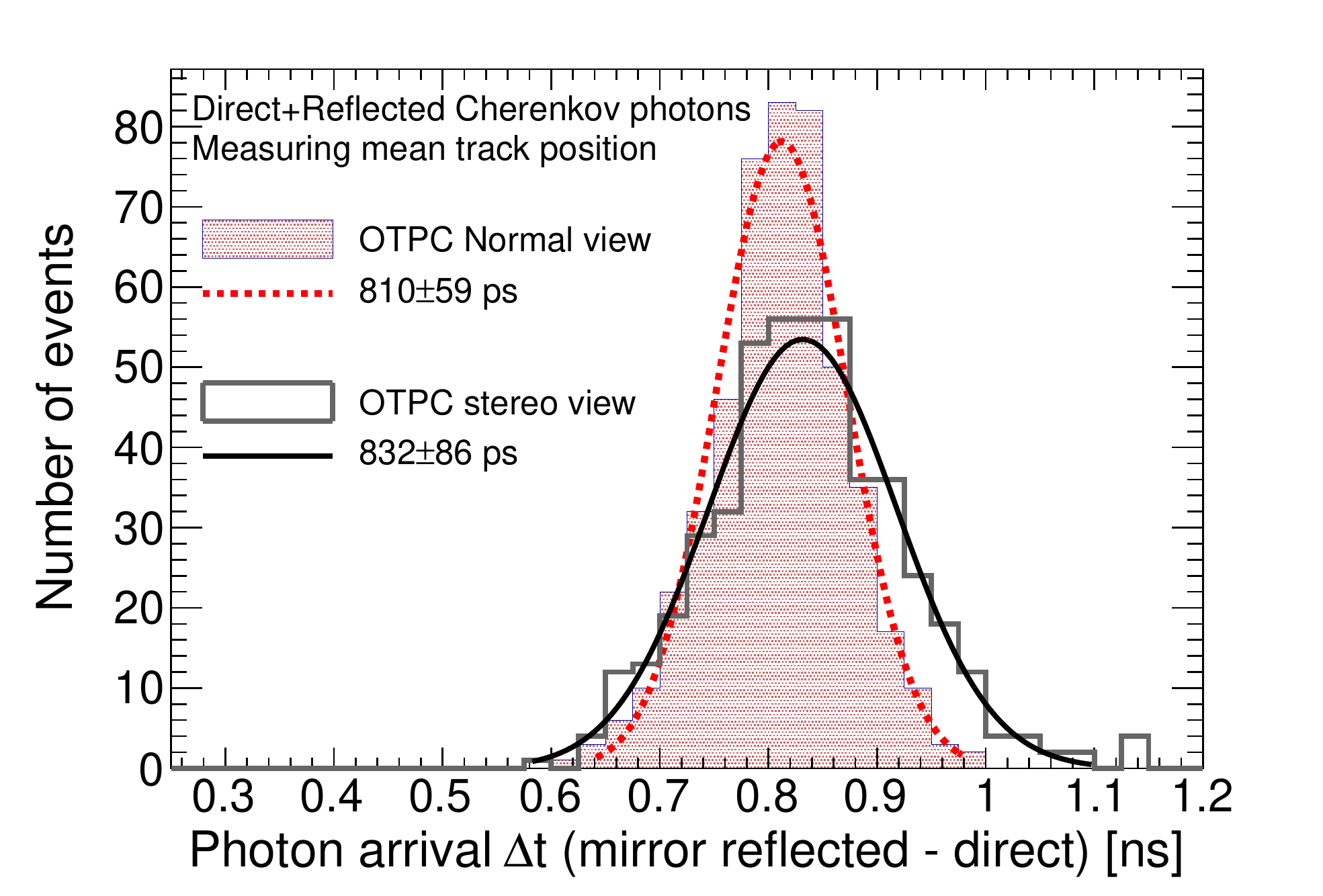}
\caption[Time resolving direct and mirror reflected photons for track reconstruction]{Time-of-arrival difference between the direct and mirror reflected Cherenkov photons for the normal and stereo views. The time differences were measured using Eqn.~\ref{eqn:dumb}. The normal view has an average time difference of 810 $\pm$ 59~ps
and the stereo data shows a time difference of 832 $\pm$ 86~ps. }
\label{fig:positionwtiming}
\end{figure}

The result of Equation~\ref{eqn:dumb} is shown in
Figure~\ref{fig:positionwtiming} when grouping together the normal
view data and stereo view data separately. This result
incorporates the same 522 events used in the measurement shown in
Fig.~\ref{fig:angle}. The normal view has an average time
difference of 810 $\pm$ 59~ps; the stereo data show a time
difference of 832 $\pm$ 86~ps\footnote{The mean time differences are slightly larger
than measured on the full 16~GeV/c dataset ($\S$\ref{subsec:resolvedm} and Fig.~\ref{fig:direct-mirror-all})
due to the event selection and the time cuts applied to separate the direct and mirror-reflected photons described
in the text}. 
These 1$\sigma$ uncertainties
correspond to spatial resolutions of 9.6~mm and 13.9~mm,
respectively, when combining data from the PMs.


\begin{figure}[]
\centering
\hspace{-17pt}
\subfloat[]{\includegraphics[trim=.23cm .1cm 1.2cm 1.0cm, clip=true,height=4.65cm]{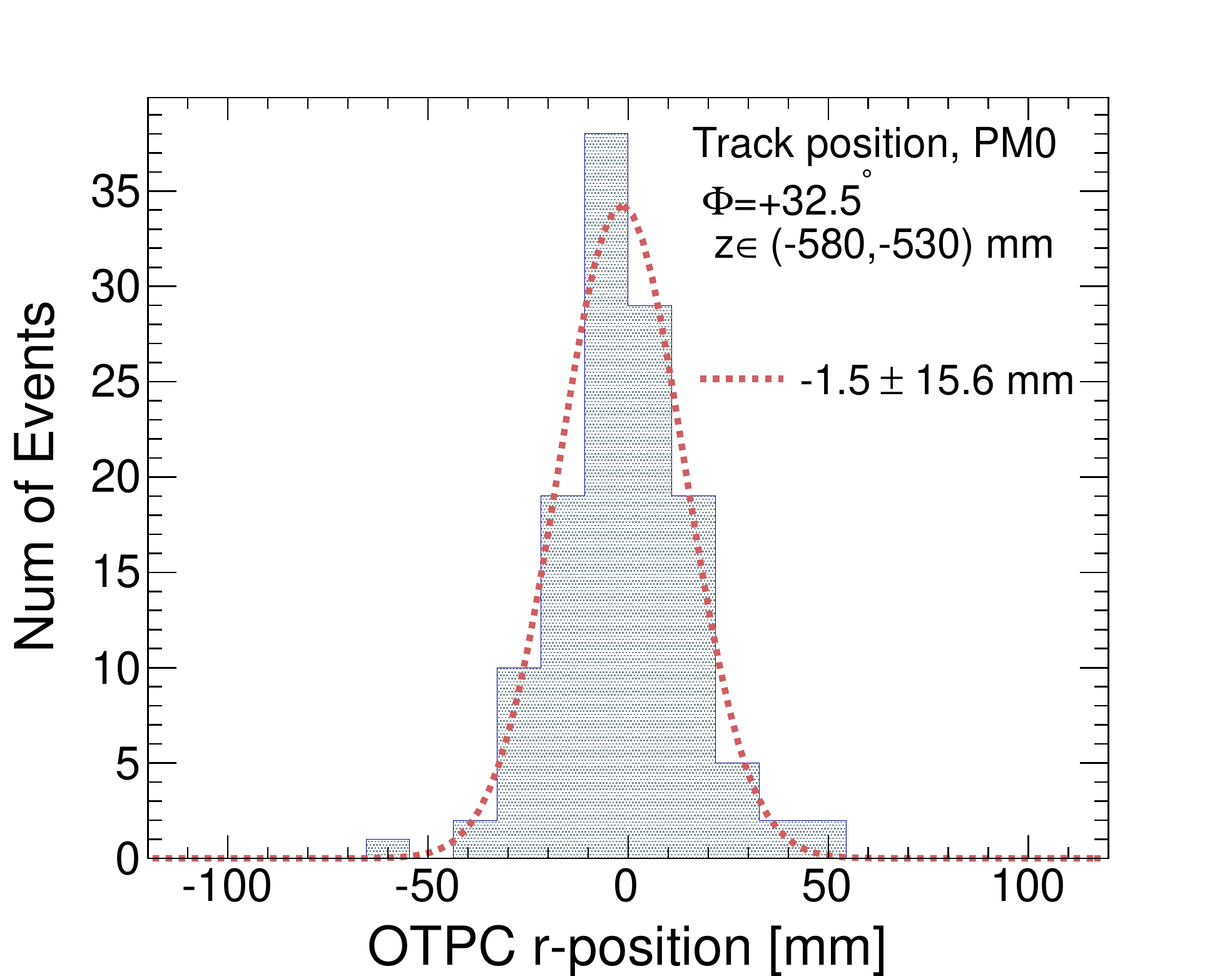}}
\subfloat[]{\includegraphics[trim=1.5cm .1cm 1.2cm 1.0cm, clip=true,height=4.65cm]{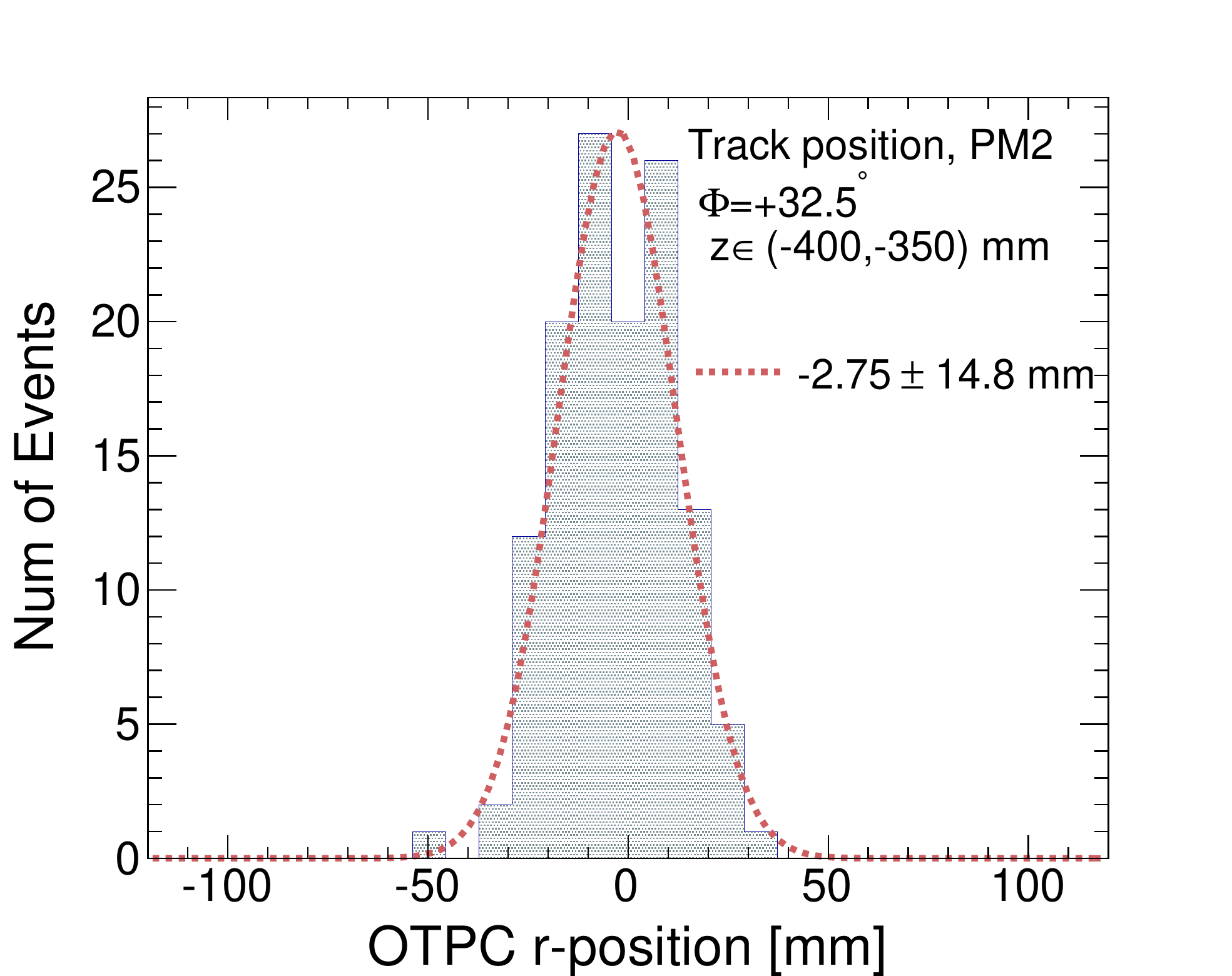}}
\subfloat[]{\includegraphics[trim=1.5cm .1cm 1.2cm 1.0cm, clip=true,height=4.65cm]{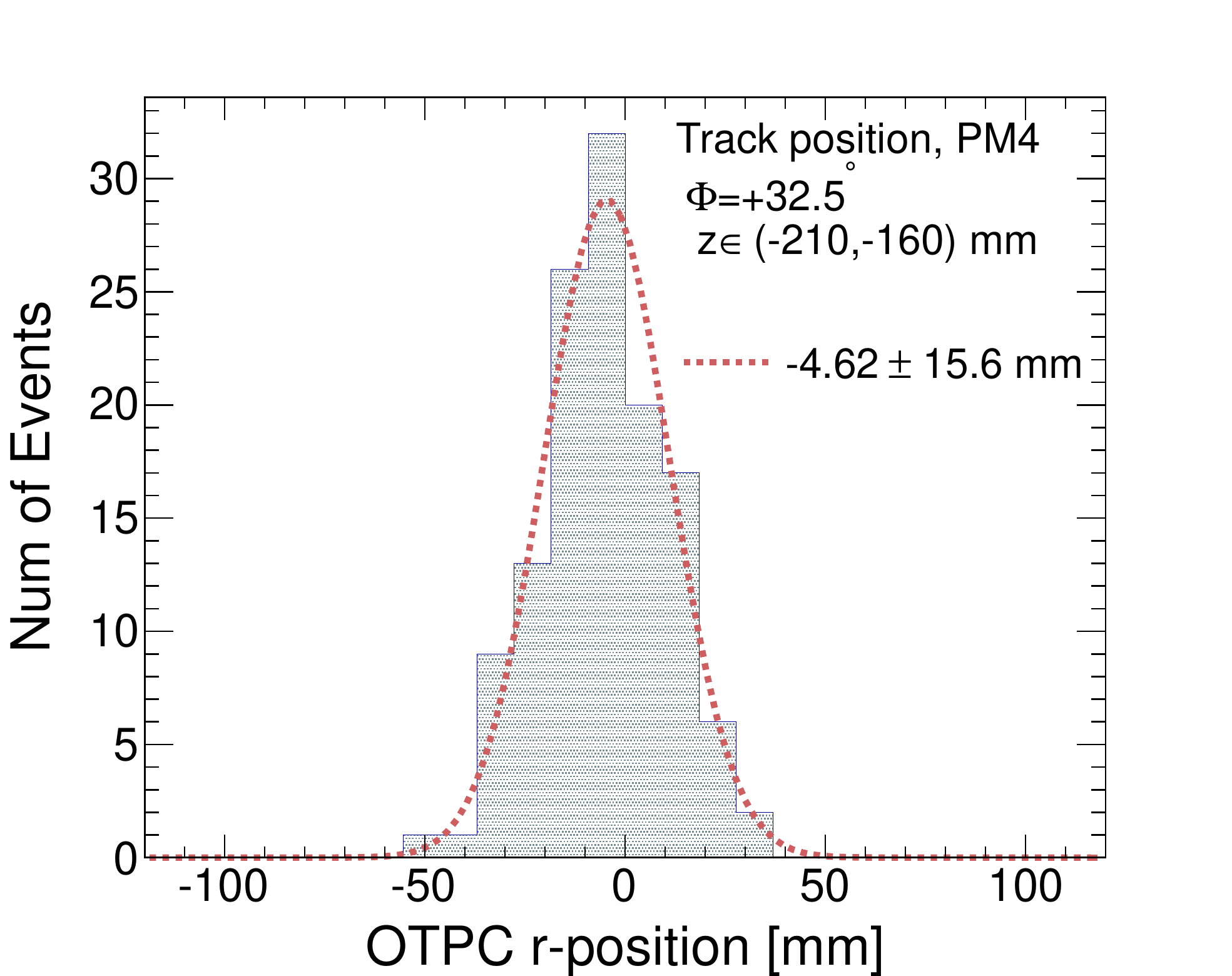}} \\
\subfloat[]{\includegraphics[trim=.25cm .1cm 1.2cm 1.0cm, clip=true,height=5cm]{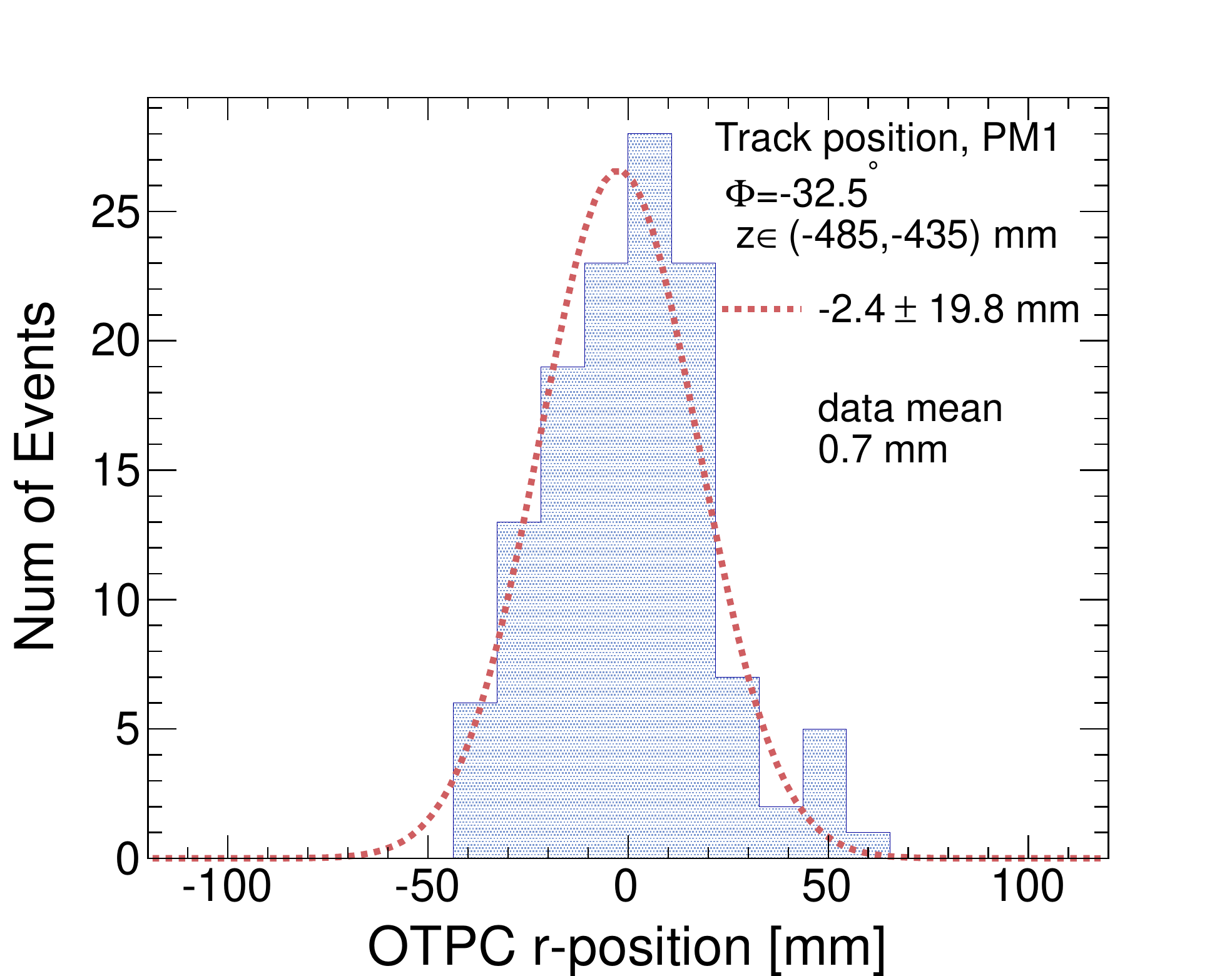}}
\subfloat[]{\includegraphics[trim=1.3cm .1cm 1.2cm 1.0cm, clip=true,height=5cm]{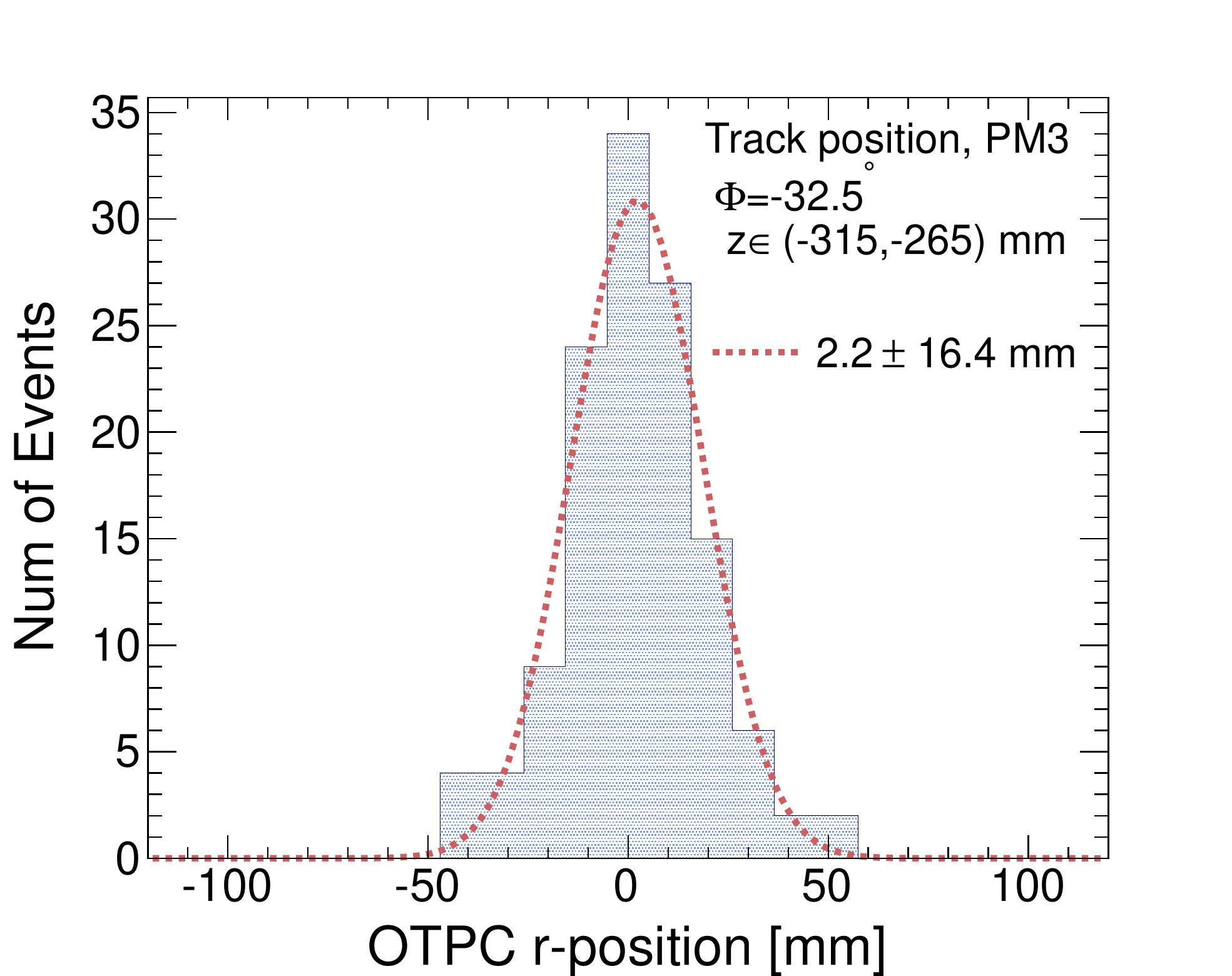}}
\caption[Reconstructed track position for each PM]{The reconstructed radial track position along the OTPC z-axis at each PM $\phi$-location.  }
\label{fig:trackreco}
\end{figure}

To utilize the OTPC as a tracking detector, we use each PM to
resolve the spatial coordinates along the z- and $\phi$-axes. We
require that an event has more than three direct and three
mirror-reflected photons for each PM. There are 127 through-going
events that satisfy this condition. As expressed in
Eqns.~\ref{eqn:dumb}, the radial position, r,  along the z and
$\phi$ axes of the OTPC is defined as the displacement from the
nominal beam axis. Using Eqns.~\ref{eqn:dumb} and
\ref{eqn:mirror2} we can reconstruct r at each of the five PM
positions in z- and $\phi$. The five distributions in the radial
position are shown in Figure~\ref{fig:trackreco}.

The properties of the data in Figure~\ref{fig:trackreco} are shown in Table~\ref{tab:recor}. The reconstructed radial position, r, at each z position is consistent with r=0, which is the expectation with the through-going trigger. At each normal view PM, the resolution is about 15~mm on this measurement. On the stereo view, the average error is 17.5~mm.
We expect the beam spatial location to have an RMS variation of $\sim$7~mm from the measured beam output position in the $R_2$ trigger (Fig.~\ref{fig:back-trig}).

\begin{figure}
\centering
\subfloat[]{\includegraphics[trim=.13cm .1cm 1.3cm 1.0cm, clip=true,height=5.3cm]{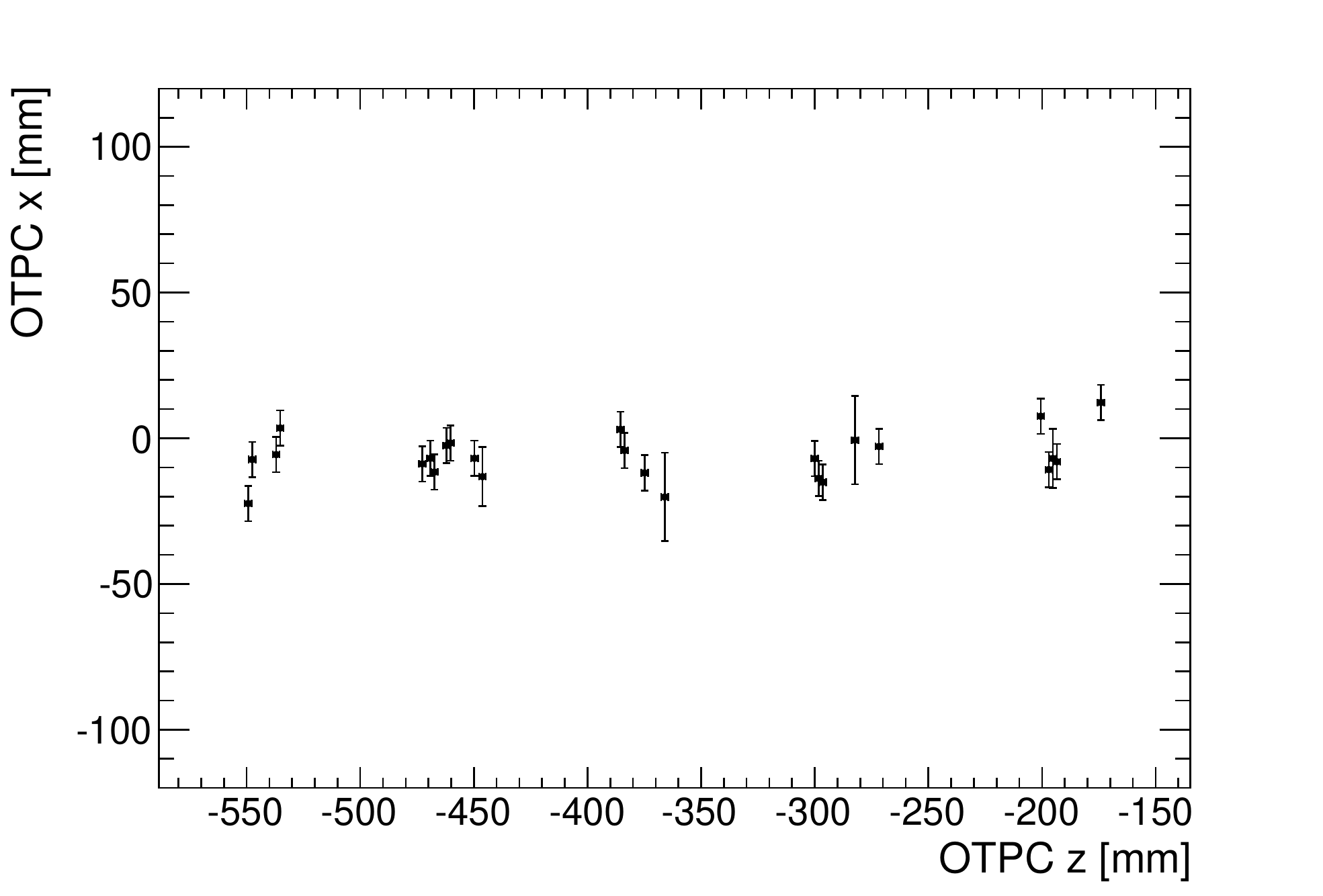}}
\subfloat[]{\includegraphics[trim=.13cm .1cm 1.3cm 1.0cm, clip=true,height=5.3cm]{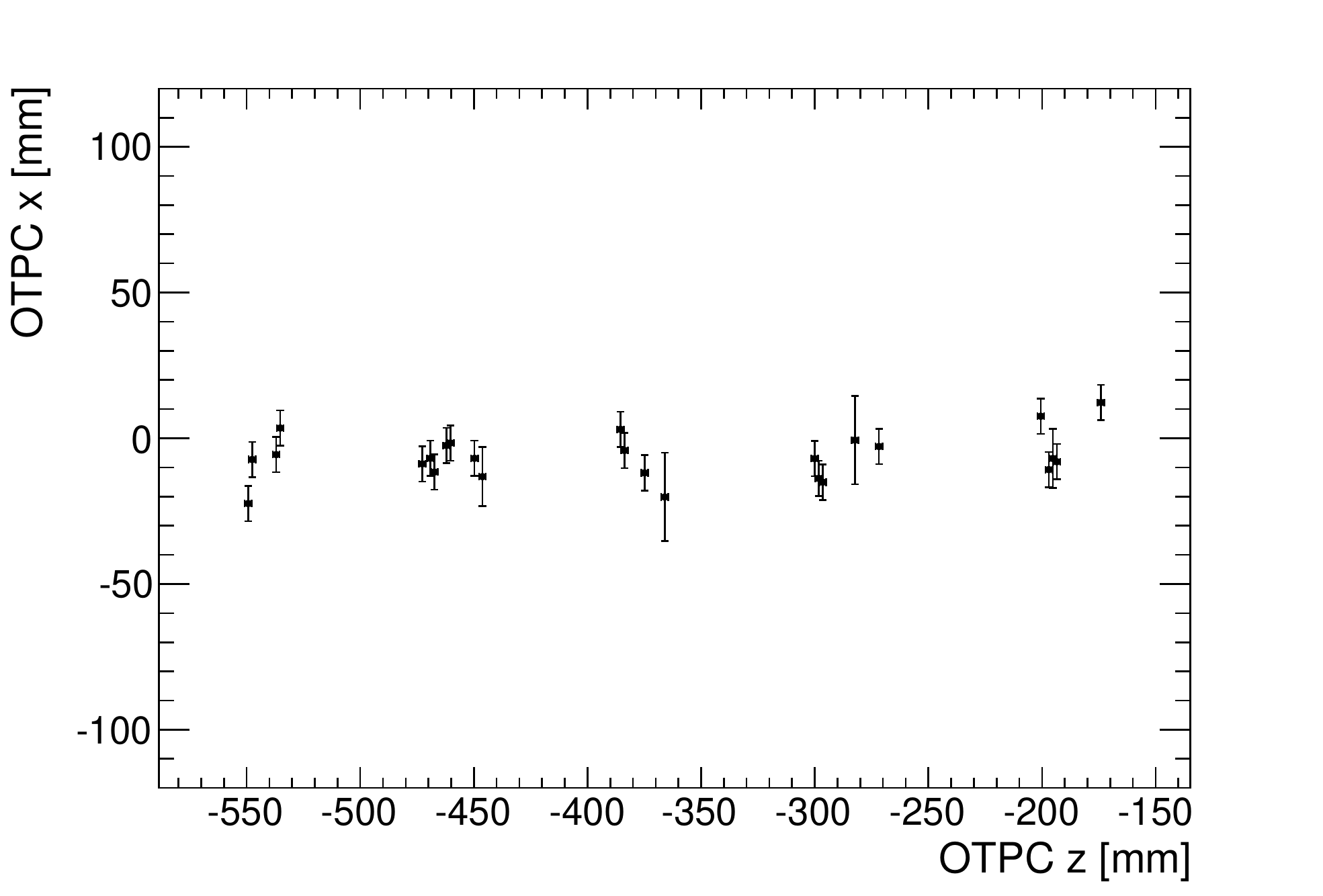}} \\
\subfloat[]{\includegraphics[height=9cm]{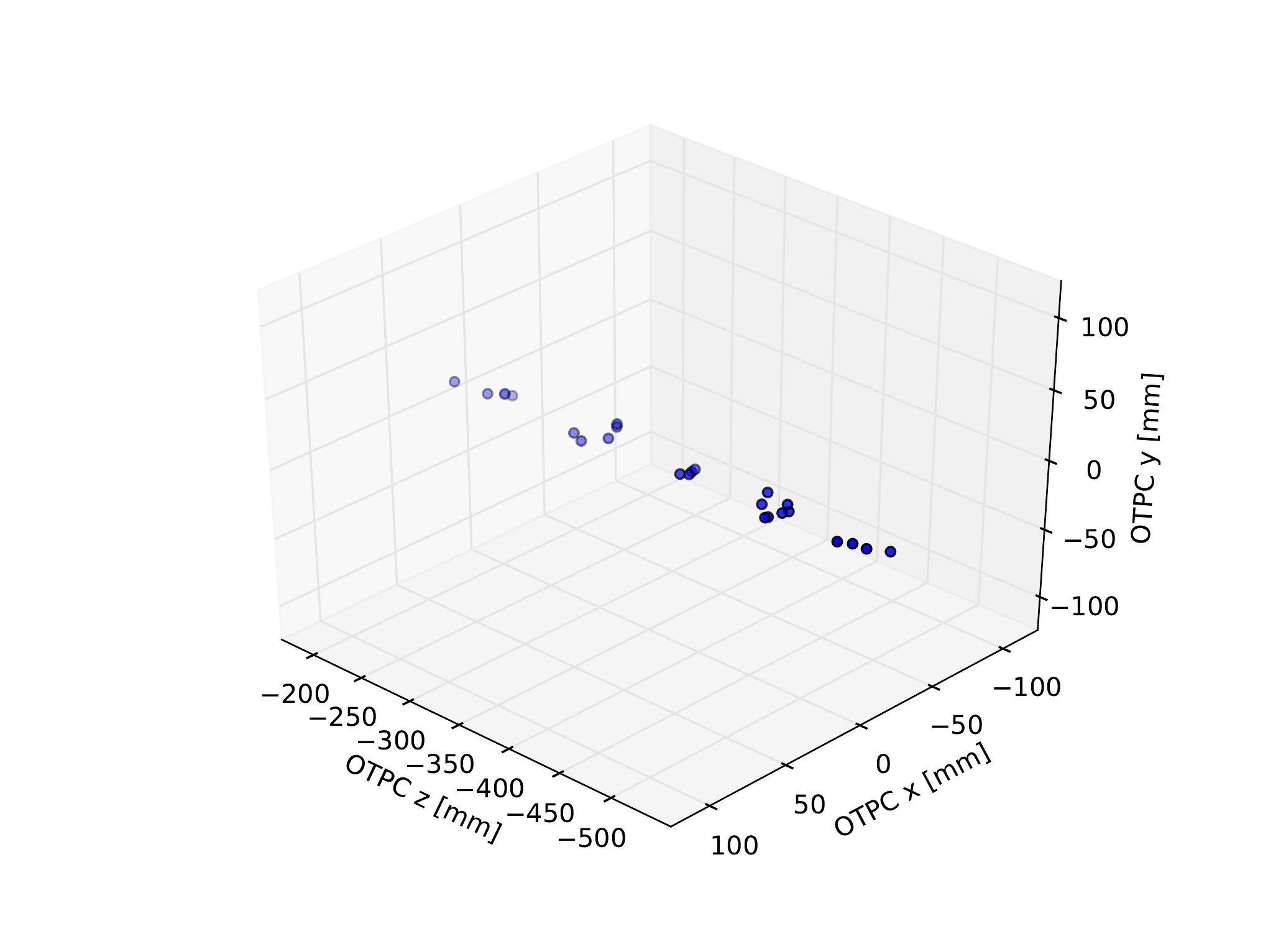}}
\caption[Reconstructed 3D track]{The 3D track reconstruction of the event shown in Figures~\ref{fig:rawevents}a and~\ref{fig:tprojection}.
The rotated times, $t'_i$, of the direct Cherenkov photons are projected on the reconstructed radial position (stereo, normal, z) and decomposed into Cartesian (x,y,z) coordinates.
(a) The track x vs. z position. (b) The track y vs. z position (c) The reconstructed x, y, and z of the track.}
\label{fig:3D}
\end{figure}

Figure~\ref{fig:3D} shows the 3D track reconstruction of the event displayed in Figs.~\ref{fig:rawevents}a and~\ref{fig:tprojection}. The direct Cherenkov photons are projected onto the reconstructed radial position, r, and the OTPC normal and stereo-view coordinates are decomposed into the Cartesian coordinate system defined in Fig.~\ref{fig:otpcdraw}.

\begin{table}
  \caption[Average reconstructed track positions along z and $phi$]{Average reconstructed radial positions of tracks along OTPC z-axis. The OTPC center-line is defined as r=0. These data are shown in Figure~\ref{fig:trackreco}}
\vspace{5 pt}
  \centering
  \begin{tabular}{ c c c  c c}
    PM  & $\bar{z}$ [mm] & $\phi$ [radians] & reconstructed r [mm]\\ \hline
    0  & -550 & 0.567 & -1.5 $\pm$ 15.6  \\
    2  & -375 & 0.567 & -2.75 $\pm$ 14.8  \\
    4  & -185 & 0.567 & -4.62 $\pm$ 15.6  \\ \\
    1  & -460 & -0.567 & 0.7 $\pm$ 19.8 \\
    3  & -290 & -0.567 & 2.2 $\pm$ 16.4  \\
  \end{tabular}
  \label{tab:recor}
\end{table}

\section{Conclusions}
\label{conclude}

We have constructed and characterized the performance of a small
prototype optical-TPC (OTPC), in which we measured the time and
position of arrival for each individual photon emitted by
Cherenkov radiation from a charged particle traversing the water
volume. The photon arrival times and 2D~spatial locations are
resolved to $\sim$75 ps and $\le 3\times3$~mm$^2$, respectively,
using the digitized waveforms. Given the arrival positions and the
drift times, the particle track
 position and direction can be reconstructed.

 While the present prototype is small and still
primitive, several of the techniques developed here may scale to
much large detectors capable of reconstructing events and vertices
in more complicated event topologies such as would be expected in
proton decay, neutrino-less beta decay, and neutrino interactions~\cite{ASDC, annie, elaginbb}.

The ability of the OTPC to resolve time differences much smaller
than 1~nanosecond lends itself to two strategies relevant to the
economics of scale-up for larger detectors. The first was to
cheaply multiply the $\sim$7$\%$ photocathode coverage on the
surface of the detector by using optical mirrors;  the mirrors
more than doubled the OTPC light collection. For each track, the
mirrors allowed for the added collection of Cherenkov photons
impinging on the detector wall opposing the MCP-PMT. The reflected
photons are clearly time-resolved in the data with an average
delay of 770~ps compared to the direct Cherenkov light.  The
second strategy is to digitize the
 microstrips  at one end of the strip with the opposing end open, so that direct pulse and the
reflection from the far end provides two distinct $\sim$ns-wide
pulses per detected photon. This cuts the electronics channel
count in half, and has the added advantage that the
times-of-arrival at each end are measured on the same electronics
channel.

For through-going tracks recorded at a secondary beam momentum of
16~GeV/c, we observed 80$\pm$20 detected Cherenkov photons per
event. Each photon is resolved in time, $z$, and $\phi$, allowing
a view of the 3D trajectory. We measured an angular resolution,
assuming a straight track, of 60~mrad when fitting events in which
more than 8 direct photons were detected among the three MCP-PMTs
in the OTPC normal view. By time-resolving the direct and mirror
reflected photons at 60~ps, we measure a transverse resolution on
the particle track of 15~mm.
We note that an entire reconstructed event
takes place over a 40-cm track length and in a duration of less than 3000
picoseconds.

\section{Acknowledgements}
We thank the staff of the Enrico Fermi Institute Electronics
Development Group for their essential technical and intellectual
contributions, in particular Mircea Bogdan for the design of the
Central DAQ printed-circuit card, Fukun Tang for first suggesting
the use of high-bandwidth microstrips for the photodetector
anodes, Mark Zaskowski for electronics assembly, and Mary Heintz
for her superb support of the infrastructure for electronics
design, computing, and web documentation. We are indebted to
Richard Northrop and Robert Metz, both of the University of
Chicago Physical Sciences Division Engineering Center,  for
expertise in mechanical design and skilled machining,
respectively, and  Justin Jureller of the UC Materials Research
Science Center for providing his time and the expertise to make
the OTPC water-absorbance measurements.

We express our gratitude to Jeffrey DeFazio, Raquel Ortega, and
Emile Schyns of PHOTONIS, Inc., for providing technical details
and support for the Planacon photodetectors. We thank Howard
Nicholson, then of the DOE, for suggesting the use of LAPPDs for
neutrino physics and for the name of the Optical Time Projection
Chamber.

This work could not have been done without the support of Fermilab
and, in particular, the Fermilab Test Beam operations group:
Eugene Schmidt Jr., Erik Ramberg, Mandy Rominsky, and Aria Soha.
Todd Nebel and Greg Sellberg provided excellent mechanical support. The
LArIAT collaboration provided technical assistance and knowledge
in welcoming the OTPC operation parasitically in their beam line.

This work was supported by  National Science Foundation grant
PHY-1066014 and Department of Energy award DE-SC0008172.
Part of this work was performed at the Fermilab Test Beam Facility,
which is operated by
Fermi Research Alliance, LLC under Contract No. De-AC02-07CH11359
with the United Stated Department of Energy.



\begin{thebibliography}{10}
\bibitem{spectro}
{A Cary 8500 dual-beam spectrophotometer was used for this measurement}.
\bibitem{pllchip}
{A Texas Instruments Clock Generator/Jitter Cleaner was used, model no.
  CDCE62005, and configured such that a 40MHz clock was distributed to each
  PSEC4 chip using LVCMOS signaling.}
\end{thebibliography}

\end{document}